\documentclass[10pt,reqno, usenames]{report}

\usepackage[a4paper, hmargin={2.8cm, 2.8cm}, vmargin={2.5cm, 2.5cm}]{geometry}
\usepackage{hyperref}
\usepackage{amsmath}
\usepackage{amssymb}
\usepackage[babel, farve, en]{ku-forside}% KU-forside
\usepackage[utf8]{inputenc}
\DeclareUnicodeCharacter{2014}{\dash}
\usepackage{IEEEtrantools}
\usepackage{comment}
\usepackage{lmodern}
\usepackage[auto]{microtype}
\usepackage{amssymb}%mathe symbole
\usepackage{amstext}
\usepackage{amsfonts}
\usepackage{mathtools}
\usepackage{amsxtra}
\usepackage{siunitx}
\usepackage{amsmath}%mathe
\usepackage{amsthm}%Theorems and proofs
\usepackage{amssymb}%mathe symbole
\usepackage{amstext}%mathe symbole
\usepackage{mathrsfs}
\usepackage{mathtools}
\usepackage{siunitx}
\usepackage{subcaption}
\usepackage[numbers,sort&compress]{natbib}
\usepackage{caption}
\usepackage{subcaption}
\usepackage{physics}
\newcommand{\eh}{\frac{1}{2}} %einhalb
\newcommand{\kl}[1]{\left(#1\right)}%automatische Klammern
\newcommand{\Bbra}[1]{\left\langle#1\right\vert}%automatische Bra []
\newcommand{\logv}[1]{\log\left\vert #1 \right\vert}
\newcommand{\h}{\hbar}
\DeclareSIUnit \parsec {pc}
\titel{\parbox{\linewidth}{General Relativity from\\ Intersection Theory and Loop Integrals
}} 
\undertitel{} %Skriv titlen på jeres projekt
\opgave{Master's thesis} 
\forfatter{Toni Teschke}%Skriv jeres gruppenummer
\dato{Niels Bohr Institute \today} %"today" skriver automatisk dagens dato. 
\vejleder{Academic advisers: Poul Henrik Damgaard and Hjalte Frellesvig}
\begin{document}
\maketitle
\newpage
\begin{center}
\chapter*{\centering Disclaimer}
\end{center}
The author of this work, it should be noted, grapples with a severe case of dyslexia. Consequently, it is imperative to recognize that certain elements within the text, such as spelling, punctuation, and the interpretation of words, may have been inadvertently affected due to the employment of dictation software. The author's earnest efforts have been invested in ensuring clarity and precision to the best of their ability within the limitations posed by their condition and the usage of such assistive technology.
\begin{center}
\chapter*{\centering Abstract}
\end{center}
In this work, we examine the gravitational scattering amplitude involving two Schwarzschild black holes as massive scalars in a $2\to2$ interaction. Specifically, we focus on the Second Post-Minkowski correction (2PM), considering only the contributions from box and cross-box diagrams, excluding the triangular contributions. We find that Feynman integrals, represented in the Baikov form, can be naturally interpreted as a pairing between a twisted co-cycle and a twisted cycle. We introduce the concept of twisted (co)-homology groups and explore the notion of an intersection number, which quantifies the pairing between elements of the twisted co-homology group and their dual counterparts. This insight leads to the development of a master integral decomposition formula, demonstrating that the intersection number acts as a scalar product in the space of Feynman integrals. We present a recursive algorithm for computing multivariate co-homology and provide a comprehensive example. Additionally, we offer codes for determining the dimension of each fiber, identifying their basis elements, and calculating their sizes. These codes are applied to both the box and cross-box diagrams, including their cut versions, to derive their minimal base elements and sizes. Furthermore, we use the method of multivariate intersection theory to determine the coefficients of the master integral bases for the cut gravity box and cross-box contributions. Finally, we evaluate the master integrals and their coefficients in the classical limit. Our results for the box and cross-box contributions of the master integral basis align with the existing literature on the 2PM correction. Additionally, we successfully compute the triangle contributions within the master integral basis of the box and cross-box contribution of the 2PM Correction. This demonstrates for the first time the applicability of intersection theory in the quantum field theoretic description of gravity.\\
\newpage
\tableofcontents
\newpage

\chapter{Introduction}
With upcoming experiments aiming to measure gravitational waves with unprecedented sensitivity, there is a growing demand for a high-precision understanding of general relativity \cite{Amaro-Seoane_2012}. These theoretical predictions are rooted in our modern understanding of general relativity, which involves describing gravity using quantum field theory. To achieve this goal, the development of new mathematical tools is essential for handling the complex loop integrals that arise. In this thesis, we introduce a technique known as intersection theory as a valuable addition to the toolkit for studying gravity.\\\\
Since Feynman introduced the concept of the graviton as a spin-2 particle in the 1960s to describe gravity within the framework of quantum field theory \cite{Feynman:1963ax}, and subsequent research by Gerardus 't Hooft and Martinus J.G. Veltman revealed the one-loop divergence of gravity \cite{tHooft:1974toh}, showing its renormalizability, as demonstrated in, and its ability to capture the essence of pure Einsteinian gravity, our comprehension of gravity has made substantial progress\cite{GOROFF1986709}.\\\\
Numerous thrilling results have been derived, in this way in recent decades, such as the classical description of a gravitational binary system and quantum corrections for gravitational processes at low energy.  Although quantum corrections for gravity are currently inaccessible to observation, we simply expect this to change in the coming decades. After all this is physics, not philosophy.\\\\
In recent years, the post-Minkowskian expansion of general relativity has garnered significant interest. This expansion has also been extended to include spinning objects. The primary rationale behind this approach stems from the observation that during the initial phases of a merger, when two objects are still separated by a considerable distance, gravitational interactions are relatively weak and can be effectively addressed using a weak coupling approximation. Where, the perturbation series organises the calculation of the scattering amplitude in a natural way and offers a convenient tool for studying the dynamics of such weak gravitational systems without the necessity to consider the small velocity limit, which is due to the Lorentz invariance of the amplitude. However, one must ensure a proper treatment of the classical limit of gravity as a trade-off for these advantages \cite{Cristofoli_2020}\\\\
%When one moves to a higher order of gravitational couplings in classical theory, as the reader already knows, one has to evaluate Feynman integrals with more and more loops. 
The set up that we are going to consider in this thesis, is the scattering amplitude involving two Schwarzschild black holes as massive scalars in a $2\to2$ interaction. Specifically, we focus on the Second Post-Minkowski correction (2PM) while considering only the contributions from box and cross-box diagrams, excluding the triangular contributions.\\\\
It's worth highlighting that the scattering amplitude includes not just classical and quantum components, both distinguishable by their behavior concerning $\h$, but also a superclassical component. The classical terms have a finite limit as $\h$ approaches zero, and the quantum terms vanish. On the other hand, superclassical contributions result in singular expressions, reflecting extremely rapid phase oscillations in the S-matrix. Hence, it's of paramount importance to ensure that superclassical terms cancel out when evaluating the classical limit of the scattering amplitude.\cite{Cristofoli_2020}\\\\
We will determine the classical limit using a method derived from the well-known method of regions in momentum space. This method involves evaluating the asymptotic expansion of the relevant Feynman integrals for small transferred momenta $q$. Specifically, the region characterized by the scaling relation $k\sim\mathcal{O}(q)$ produces the non-analytic term responsible for the emergence of the long-range potential. In general, the integral also receives a contribution from the highly relativistic hard region, where the scaling relation is $k\sim\mathcal{O}(1)$. This contribution consists of positive integer powers of $q^2$ and, as a result, does not influence the long-range behavior in position space. However, it plays a crucial role in the method of regions, ensuring the overall consistency of the small $q$ expansions. It's important to note that our main focus is on obtaining the classical limit of the master integral basis and their coefficients in the soft limit, so the hard region is not relevant to our current objective.\cite{Cristofoli_2020}\\\\
%With increasing physical precision of the scattering amplitude or, in other words, with increasing loop order, the complexity of the Feynman integrals also increases. Understanding the underlying structure of these integrals is crucial. On the one hand, this can improve our ability to predict and thus advance the interplay between theory and experiments; on the other hand, it can improve and at best modify our understanding of quantum field theory. Therefore, it is of great importance that we find new ways to perform computation and thus explain the surprising simplifications that one regularly encounters.\cite{Weinzierl:2022eaz}\\\\
As the physical precision of the scattering amplitude increases, or, in other words, as we consider higher loop orders, the complexity of Feynman integrals also increases. Gaining a deep understanding of the underlying structure of these integrals is paramount. This understanding can yield two significant benefits: firstly, it enhances our predictive capabilities, facilitating progress in the interplay between theory and experiments. Secondly, it has the potential to improve and even reshape our comprehension of quantum field theory. Therefore, it is crucial that we explore innovative methods for computation, allowing us to elucidate the unexpected simplifications that frequently arise.
When working with Feynman integrals, there is one source that causes the most inconvenience and the greatest computational effort. Namely, the complexity of their evaluation. In this respect, it is intuitively clear that one should disregard as many redundancies as possible and re-express the Feynman integrals in terms of the base elements of the system. These are the master integrals which, roughly speaking, form vector spaces for a given scattering amplitude in which, the accompanying Feynman integrals can be expressed in terms of the master integrals and accompanying coefficients. This is conventionally achieved through integration by parts (IBPs) identities and a glorified Gaussian elimination process, more on which later.\cite{Gasparotto:2023cdl}\\\\
In recent years, it has been found that many essential aspects of Feynman integrals are captured by the Intersection theory of twisted co-homologies, succinctly "Intersection theory", and accompanying hypergeometric functions. Notably, these theories were originally developed by Japanese mathematicians in the final quarter of the previous century. In their pioneering paper, "Feynman Integrals and Intersection Theory," Mastrolia and Mizera made a groundbreaking contribution. They were the first to demonstrate that the hypergeometric properties of Feynman integrals enable the formation of a co-homology intersection number, succinctly termed the "intersection number." This intersection number concept can be intuitively understood as the scalar product between Feynman integrals. Within this theoretical framework, it is therefore possible to express any Feynman integral in terms of the master integrals by projecting them onto the basis. In this way, the brute force solution characteristic of the use of integration by part identities is avoided.\cite{Mastrolia:2018uzb}\\\\
At the time of writing, intersection theory is not yet competitive in terms of computational power when compared with more established methods such as integration by parts identities. Nevertheless, we trust in the continued innovation and a growing understanding of the twisted co-homology group as it provides a new and comprehensive framework that will uncover a novel understanding of Feynman integrals. These new structures may be invisible to the more traditional approaches, and it is almost certain that we will see computational improvements in the coming years.\\\\
In Chapter \ref{GRnQFT} we will delve into the pertinent elements of classical general relativity and quantum field theory. Our aim here is to establish a shared foundation and forge a connection between these two theories. Subsequently, we will construct a quantum field theory description of gravity and execute a gauge fixing procedure for the graviton. To wrap up, we will derive the Feynman rules for gravitons by expanding the mathematical components of general relativity, in the the graviton field and the gravitational constant, in the context of flat spacetime.\\\\
In Chapter \ref{FIE} we will review the essential notions of Feynman integrals such as performing a derivation of the Baikov representation, this representation has been chosen as it offers the most transparent link to the Intersection theory of twisted cohomologies, writing down the graviton box and cross-box, in the Baikov representation, introduce integration by parts relations and the concept of master integrals and finally relate integration by parts relations to the Baikov representation.\\\\
In chapter \ref{ISTU} we provide an overview of the univariate case of Intersection theory. We will first introduce the most essential mathematical ideas to understand Intersection theory. We will then show how the pairing between the Co-Homology group and its dual leads to the concept of intersection numbers. Further, we introduce the concept of linear relation as well as the master decomposition formula 
and the Feynman integral decomposition formula, this it is possible to obtain the coefficients associated with the master integrals.\\\\

In Chapter \ref{Intersection_Theory_II}, our focus shifts to the examination of multivariate Intersection theory. The primary challenge in this context revolves around creating a multivariate extension of the univariate intersection number algorithm. To achieve this, we must establish new connection coefficients throughout the entire manifold and gain a comprehensive understanding of the layered integration variables. By the end of this chapter, we will delve into a detailed example, breaking down each step of the algorithm. This approach aims to provide clarity, especially for readers who may be new to the concepts.\\\\ 

In chapter \ref{GravBox} we apply the concepts developed in the last two chapters to the box and cross-box Contributions of the 2PM scattering amplitude and find their master integral coefficients, which, after the mathematical arc that came before it, is almost an afterthought. \\

In Chapter \ref{Soft}, we undertake a soft limit expansion of both the master integrals and their corresponding coefficients. This process leads us to acquire a complete master integral representation within the classical limit of the box and cross-box contributions of the 2PM scattering amplitude.

\graphicspath{{./images/}}
\chapter{General relativity form a field-theoretic perspective}\label{GRnQFT}
In this chapter, we refresh the reader's understanding of general relativity and the path integral approach to quantum field theory and lay the necessary groundwork for quantizing general relativity. For this, it is necessary to formulate the Einsteinian field equations as a path integral. This is done via the Einstein-Hilbert action. 
Finally, we will discuss how to interpret the classical limit $\h\to0$ of quantum field theory.
As particle physicists, we will primarily work in the negative convention for the metric and set, $c=G=\h=1$.
\section{General Relativity}
In general relativity, gravity is a consequence of the geometry of space-time, which is described mathematically by the metric tensor and its derivatives. Physical laws are generally covariant, i.e. they are invariant under general coordinate transformations, which means they are the same for different observers in different reference frames \cite{Dirac}\cite{carroll2019}.%, this is achieved by tensor fields that by definition fulfil the necessary transformation laws. 
The equivalent of the physical laws for different observers is achieved by introducing tensor fields that by definition fulfil the necessary transformation laws.
Following the intuition that Gauge symmetry is, very roughly speaking, a coordinate transformation, one realises that the general covariance is nothing other than the Gauge symmetry of a spin-2 particle, the infamous graviton, which we will treat later via the Faddeev-Popov procedure. We follow the calculation and convention laid out in \cite{Dirac}. These conventions include the mostly negative metric as well as the definition of the Ricci tensor via the curvature tensor. For matters of convenience and generality, we will work in an arbitrary space-time dimension $D$, where the gravitational coupling constant $G_N$ has the mass dimensionality of 
\begin{equation}
\left[G_N\right]=[\mathrm{mass}]^{-(D-2)}
\end{equation}
However, it will often be convenient for us to work with $\kappa$ rather than $G_N$:
\begin{equation}
    \kappa^2=32 \pi G_N
\end{equation}
This definition of $\kappa$ agrees with  \cite{donoghue1995introduction} 
As physicists, we work in the explicit coordinate form using tensor index notation. The Riemann curvature tensor then takes the form:
\begin{equation}
R_{\mu\nu \rho \sigma}=\partial_\rho\Gamma_{\mu\nu\sigma}-\partial_\sigma\Gamma_{\mu\nu\rho}-\Gamma_{\mu\beta\rho} \Gamma_{\nu \sigma}^\beta+\Gamma_{\beta\mu\sigma} \Gamma_{\nu\rho}^\beta
\end{equation}
As the reader surely remembers, the Christoffel symbols of the first kind are
\begin{equation}
\Gamma_{\rho\mu\nu}=\frac{1}{2}\left(\partial_\nu
g_{\rho\mu}+\partial_\mu g_{\rho\nu}-\partial_\rho g_{\mu\nu}\right)\label{Chris}
\end{equation}
The Ricci tensor, $R_{\mu\nu}$ and Scalar $R$, can be obtained from the Riemann curvature tensor in the following manner:
\begin{align}
& R_{\mu \nu}=g^{\rho \sigma} R_{\rho \mu \nu \sigma}\label{ric} \\
& R=g^{\mu \nu} R_{\mu \nu}\label{rics}
\end{align}
The Einstein-Hilbert action can then be defined in terms of the Ricci scalar and the square root of the negative of the determinant of the metric tensor $g=\det[g_{\mu\nu}]$. It takes the form:
\begin{equation}
S_{\mathrm{EH}}=\int d^D x \sqrt{-g} R\label{EH}
\end{equation}
If we have a Lagrangian density or short Lagrangian $\mathcal{L}$ that describes the matter distribution, we can add a matter term to the Einstein Hilbert action
\begin{equation}
S_\phi=\int d^D x \sqrt{-g} \mathcal{L}_\phi\label{phi}
\end{equation}
In this work we will restrict ourselves to the case of a massive scalar field, hence the Lagrangian takes the form:
\begin{equation}
\mathcal{L}_\phi=\frac{1}{2}\left(g^{\mu \nu} \partial_{\mu}\phi\partial_{\nu}\phi-m^2 \phi^2\right)
\end{equation}
If we combine both expressions, Eq. \eqref{EH} and Eq. \eqref{phi}, we get the total action:
\begin{align}
    S_{\mathrm{c}}
    =&\frac{2}{\kappa^2} S_{\mathrm{EH}}+S_\phi\\
    =&\int d^D x \sqrt{-g}\left(\frac{2 R}{\kappa^2}+\frac{1}{2}\left(g^{\mu \nu} \partial_\mu \phi \partial_\nu \phi-m^2 \phi^2\right)\right)\label{Sc}
\end{align}
\section{Quantum Field Theory}
In this section, we will apply the path integral quantisation and gauge theoretic description to the gravitational field $g_{\mu\nu}$, which we will linearise and then treat, like any other quantum field, using the minimal coupling of the Einstein-Hilbert actions to a scalar field in the path integral. This leads to a non-renormalizable quantum theory of gravity. However, in the classical low-energy limit, we can ignore the diversion terms safely. %We have already discussed that the general covariance of the gravitational action translates into a gauge theory of a spin-2 particle. 
The following discussion of quantum field theory is based on the work of \cite{Peskin:1995ev},\cite{Srednicki2006},\cite{Schwartz2014} and \cite{Jakobsen:2020ksu}.
As for the spin-1 particle, it is necessary to fix the gauge in the path integral. We employ the Faddeev–Popov gauge fixing procedure, which involves the use of a gauge fixing function $G_{\sigma}$ %and gauge fixing parameters denoted by $\epsilon$.
By employing a delta function, we single out a specific gauge choice, allowing the path integral to encompass only independent field configurations. The perturbation function associated with the action written down in equation Eq. \eqref{Sc} takes the form: 
\begin{equation}
Z_\omega=\int \mathcal{D} g_{\mu \nu} \mathcal{D} \phi \operatorname{det}\left(\frac{\delta G}{\delta \epsilon}\right) \delta\left(G_\sigma-\omega_\sigma\right) e^{i S_c}
\end{equation}
However, it is important to note that the gauge fixing function $G_{\sigma}$ leads to the breaking of general covariance in the Einstein Hilbert action. To address this, we introduce ghosts via the expansion of the Jacobian. Then, we integrate the arbitrary field $\omega_{\sigma}$ using a Gaussian weight function in the partition function $Z_{\omega}$. This integration results in the final expression for the partition function $Z$ in the covariant gauge, utilising the gauge fixing function $G_{\sigma}$:
\begin{align}
Z & =\int \mathcal{D} \omega_\sigma Z_\omega \exp \left(i \int d^D x \frac{1}{\kappa^2 \varrho} \eta^{\sigma\rho} \omega_\sigma \omega_\rho\right) \\
& =\int \mathcal{D} g_{\mu \nu} \mathcal{D} \phi \mathcal{D} c \mathcal{D} \bar{c} \exp\kl{i S_{\mathrm{c}}+i \frac{2}{\kappa^2} S_{\mathrm{gf}}+i S_{\mathrm{gh}}}\label{Z1}
\end{align}
In Eq. \eqref{Z1}, there are three different types of fields: the gravitational field $g_{\mu\nu}$, the Scalar field $\phi$, and the two ghost fields $c$ and $\bar{c}$. Along with the existing $S_{c}$, two new actions are introduced in the partition function. The first one is the gauge fixing action $S_{\mathrm{gf}}$, derived from the Gaussian weight function. The second one is the ghost term $S_{\mathrm{gh}}$, obtained by expanding the Jacobian determinant. The gauge fixing term is expressed as follows:
\begin{equation}
S_{\mathrm{gf}}=\frac{1}{2 \varrho} \int d^D x \eta^{\sigma \rho} G_\sigma G_\rho
\end{equation}
The ghost term is given by: 
\begin{equation}
S_{\mathrm{gh}}=\int d^D x \bar{c}^\alpha \frac{\partial G_\alpha}{\partial \epsilon_\beta} c^\beta
\end{equation}
Since our main focus is on the classical low-energy limit of quantum gravity, we can disregard the ghost field and get: 
\begin{align}
S & =S_{\mathrm{c}}+\frac{2}{\kappa^2} S_{\mathrm{gf}} \\
& =\frac{2}{\kappa^2}\left(S_{\mathrm{EH}}+S_{\mathrm{gf}}\right)+S_\phi\label{Ssum}
\end{align}
\subsection{The classical limit of QFT and Gravity}
The classical limit of quantum field theory can be defined as the limit where $\h\to0$. The common interpretation of this limit is that even a slight variation in the classical field configuration will cause the integrand $e^{iS}$ to oscillate rapidly so that only configurations with $\delta S=0$ contribute to the classical field. The equations $\delta S=0$ are then the classical equations of motion. The classical limit of quantum field theory is analysed in detail, for example, in \cite{Kosower:2018adc} and for gravity in \cite{PRLDamgaard1}\cite{Bjerrum-Bohr:2019kec}\cite{Cheung:2020gyp} are recommended. Counter to the whitetail belief that loops are purely quantum mechanical processes, contribute loop Feynman diagrams to the classical scattering amplitude of the field. This is the consequence of several cancellations of $\h$. An essential distinction in classical physics is between waves and particles, which does not exist in a quantum mechanical description. In a quantum mechanical framework, the wavenumber $l^{\mu}$ and the particle momenta $p^{\mu}$ are related by: 
\begin{equation}
p^{\mu}=\h l^{\mu}\label{peqhl}  
\end{equation}
If we want to introduce $\h$ explicitly as a dimensional quantity in a Feynman diagram, we should consider each "quantum" momentum, whether it is in the classical limit interpreted as a wave or a particle. This distinction is of importance since in the classical limit the particle has a finite momentum, while a wave has a finite wavenumber. The question arises now, whether a given quantum mechanical excitation $p^{\mu}$ or $l^{\mu}$ in Eq. \eqref{peqhl} should be finite. Let's consider now the convention $\h=1$, which at first seems contradictory to the limit $\h\to0$. In this case, the classical limit rather means that the action $S$ is much greater than $\h$, i.e. much greater than unity.  If we assume, as before, that momenta are by default "particle momenta", we see from Eq. $\eqref{peqhl}$ that the momentum of classical particles should remain finite, while the momentum of classical waves should be set to 0. Thus, if $p^{\mu}$ is the momentum of a quantum excitation in a given Feynman diagram, then $p^{\mu}=\h l^{\mu}$ is the wavenumber, which should remain finite. Then $a^{\mu}$ must be small in comparison.  We have two types of particles, massive scalars and gravitons. Massive scalars are interpreted as point particles and a finite momentum On the other hand, gravitons in the classical limit behave as waves and their momentum is set to 0. These conclusions apply to both external and internal loop moments. In our later analysis, the classical limit is to some extent synonymous with a long-range limit.  As more wavenumbers in momentum space correspond to long distances in position space.  A rigorous discussion may be found in \cite{Kosower:2018adc}. In Classical Limit, these effects can also be thought of as descriptions of extended objects in \cite{PhysRevD.73.104029}. Finite-size effects are described by including non-minimal terms in the action.
.\section{Gauge-Fixing}
Besides the already mentioned standard textbooks of quantum field theory \cite{Peskin:1995ev} \cite{Srednicki2006}\cite{Schwartz2014}, is this section largely based on \cite{Jakobsen:2020ksu}, as well as \cite{rafiezinedine2018simplifying} and a more detailed in \cite{Jakobsen:2020diz}. 
As mentioned before, we are using the covariant gauge with an arbitrary gauge parameter $\varrho$. This results in the gauge fixed action from Eq.\eqref{Ssum}:
\begin{equation}
S_{\mathrm{EH}}+S_{\mathrm{gf}}=\int d^D x\left(\sqrt{-g} R+\frac{1}{2 \varrho} \eta^{\rho \sigma} G_\rho G_\sigma\right)\label{SEHgf1}
\end{equation}
In order to ensure that the classical limit of the action in Eq. \eqref{Ssum} corresponds to general relativity, we must determine a suitable gauge condition that satisfies the coordinate conditions dictated by the principle of general covariance. One commonly used coordinate condition is known as the harmonic gauge.
\begin{equation}
g^{\mu \nu} \Gamma_{\mu \nu}^\sigma=0
\end{equation}
However, since it is not linear, it is not suitable for studying the quantum theory of gravity. Therefore, we linearise the harmonic gauge with the following parametrisation.
\begin{equation}
g_{\mu \nu}=\eta_{\mu \nu}+h_{\mu \nu}\label{glin}
\end{equation}
This is then known as the de Donder gauge:
\begin{align}
    \partial_\mu\left(h_\sigma^\mu-\frac{1}{2} \eta_\sigma^\mu h_\nu^\nu\right)=0
\end{align}
For generality, we can replace $h_{\mu\nu} $  with $g_{\mu\nu} $ and get:
\begin{equation}
\eta^{\mu \nu} \partial_\mu g_{\sigma \nu}-\frac{1}{2} \partial_\sigma \eta^{\mu \nu} g_{\mu \nu}=0
\end{equation}
Where the gauge fixing function takes the form
\begin{equation}
    G_\sigma=\partial_\mu\left(h_\sigma^\mu-\frac{1}{2} \eta_\sigma^\mu h_\nu^\nu\right)
\end{equation}
\section{Linearisation of the Action}
Having determined all the quantities in the action equation Eq. \eqref{SEHgf1}, we follow the idea of \cite{Dirac} in rewriting the Einstein-Hilbert action \eqref{EH} such that it solely depends on the metric and its first derivatives and proceed from there, in our derivation of the Feynman rules for gravity from $S_c$ as \cite{Jakobsen:2020ksu} laid out. So let's begin:
\begin{align}
S_{\mathrm{EH}} & =\int d^D x \sqrt{-g}R\\
&=\int d^D x \sqrt{-g} g^{\mu \nu}\left(\Gamma_{\rho \mu, \nu}^\rho-\Gamma_{\mu \nu, \rho}^\rho-\Gamma_{\mu \nu}^\rho \Gamma_{\rho \sigma}^\sigma+\Gamma_{\mu \sigma}^\rho \Gamma_{\nu \rho}^\sigma\right)\label{EHPI2}\\
&=\int d^D x \sqrt{-g} g^{\mu \nu}\left(\Gamma_{\mu \nu}^\rho \Gamma_{\rho \sigma}^\sigma-\Gamma_{\mu \sigma}^\rho \Gamma_{\nu \rho}^\sigma\right)\label{EHPI3}
\end{align}
In Eq. \eqref{SEHgf1} we inserted the definition of the Ricci scalar given in terms of three Riemann curvature tensors given in Eq. \eqref{rics}, performing partial integration gives us \eqref{EHPI3}. Inserting the definition of the Christoffel symbol \eqref{Chris}, performing partial integration one more time, gives us the  desired expression:
\begin{equation}
S_{\mathrm{EH}}=\int d^D x \sqrt{-g} \frac{1}{4}\left(2 g^{\sigma \gamma} g^{\rho \delta} g^{\alpha \beta}-g^{\gamma \delta} g^{\alpha \beta} g^{\rho \sigma}-2 g^{\sigma \alpha} g^{\gamma \rho} g^{\delta \beta}+g^{\rho \sigma} g^{\alpha \gamma} g^{\beta \delta}\right) g_{\alpha \beta, \rho} g_{\gamma \delta, \sigma}\label{SEHg}
\end{equation}
We recognise that for Eq.  \eqref{SEHg} it follows by the parameterisation, Eq.\eqref{glin}, that $\partial_{\sigma}g_{\mu\nu}=\partial_{\sigma}h_{\mu\nu}$. Therefore are both actions $S_{\mathrm{EH}}$ and $S_{\mathrm{gf}}$ of quadratic order in $\partial_{\sigma}h_{\mu\nu}$, thus we only have to expand $\sqrt{-g}$ and $g^{\mu\nu}$ in $S_{\mathrm{EH}}$ and $S_{\mathrm{gf}}$. The expansion of $g^{\mu\nu}$ follows the permitrisation of Eq. \eqref{glin} and recognises that $g^{\mu\nu}=\kl{\eta_{\mu\nu}+h_{\mu\nu}}^{-1}$, which we can expand into:
\begin{equation}
    g^{\mu \nu}=\eta^{\mu \nu}-h^{\mu \nu}+h_\rho^\mu h^{\rho \nu}-h_\rho^\mu h_\sigma^\rho h^{\sigma \nu}+\ldots\label{gmnexp}
\end{equation}
Further, we expand $\sqrt{-g}$ in the usual way, and get:
\begin{align}
\sqrt{-g} & =\exp \left(\frac{1}{2} \operatorname{tr} \ln \left(\eta_\nu^\mu+h_\nu^\mu\right)\right) \\
& =\exp \left(\frac{1}{2}\left(h_\mu^\mu-\frac{1}{2} h_\nu^\mu h_\mu^\nu+\frac{1}{3} h_\nu^\mu h_\rho^\nu h_\mu^\rho-\frac{1}{4} h_\nu^\mu h_\rho^\nu h_\sigma^\rho h_\mu^\sigma+\ldots\right)\right) \\
& =1+\frac{1}{2} h-\frac{1}{4} \mathcal{P}_{\mu \nu}^{\rho \sigma} h^{\mu \nu} h_{\rho \sigma}+\ldots\label{sqgexp}
\end{align}
With
\begin{align}
I_{\alpha \beta}^{\mu \nu} & =\frac{1}{2}\left(\delta_\alpha^\mu \delta_\beta^\nu+\delta_\beta^\mu \delta_\alpha^\nu\right)\label{IT}\\
\mathcal{P}_{\alpha \beta}^{\mu \nu} & =I_{\alpha \beta}^{\mu \nu}-\frac{1}{2} \eta^{\mu \nu} \eta_{\alpha \beta}
\end{align}
The action can then be rewritten into a form more suitable for deriving Feynman's rules for gravity. As we have bigger fish to fry, we will not go through the calculation together. However, if the reader is interested in how it is done, the calculation can be found in \cite{Jakobsen:2020diz}.
The final result one obtains is:
\begin{equation}
\left(S_{\mathrm{EH}}+S_{\mathrm{gf}}\right)_{h^2}=\frac{1}{4} \int d^D x \partial^{\rho}h_{\mu \nu}^\rho\left(\delta_\sigma^\rho \mathcal{P}_{\alpha \beta}^{\mu \nu}-2\left(1-\frac{1}{\varrho}\right) \mathcal{P}_{\rho \kappa}^{\mu \nu} \mathcal{P}_{\alpha \beta}^{\sigma \kappa}\right)\partial_{\sigma}h^{\alpha \beta}\label{SEHpgfI}
\end{equation}
As we want to derive the Feynman rules for gravity, we will use the rescaled gravitational field $\mathfrak{h}_{\mu\nu}=\frac{h_{\mu\nu}}{\kappa}$ and the action to $\frac{2}{\kappa^2}\kl{S_{\mathrm{EH}}+S_{\mathrm{gf}}}$. Thus, the action Eq. \eqref{SEHpgfI} takes  the form:
\begin{equation}
\frac{2}{\kappa^2}\left(S_{\mathrm{EH}}+S_{\mathrm{gf}}\right)_{h^2}=\frac{1}{2} \int d^D x \partial_{\rho}\mathfrak{h}_{\mu \nu}\left(\delta_\sigma^\rho \mathcal{P}_{\alpha \beta}^{\mu \nu}-2\left(1-\frac{1}{\varrho}\right) \mathcal{P}_{\rho \kappa}^{\mu \nu} \mathcal{P}_{\alpha \beta}^{\sigma \kappa}\right) \partial_{\rho}\mathfrak{h}^{\alpha \beta}\label{SEHSgfrs}
\end{equation}
\section{Covariant Gauge Gravity Propagator}
Following 
\cite{Jakobsen:2020ksu}, we will go into momentum space in the derivation of the graviton propagator. Hence, Eq. \eqref{SEHSgfrs} has the following form:
\begin{equation}
\frac{2}{\kappa^2}\left(S_{\mathrm{EH}}+S_{\mathrm{gf}}\right)_{h^2}=\frac{1}{2} \int \frac{d^D p}{(2 \pi)^D} \tilde{\mathfrak{h}}_{\mu \nu}^{\dagger} p^2\left(\mathcal{P}_{\alpha \beta}^{\mu \nu}-2\left(1-\frac{1}{\varrho}\right) \mathcal{P}_{\rho \kappa}^{\mu \nu} \frac{p^\rho p_\sigma}{p^2} \mathcal{P}_{\alpha \beta}^{\sigma \kappa}\right) \tilde{\mathfrak{h}}^{\alpha \beta}
\end{equation}
As the graviton propagator is defined as the inverse of $G^{\mu\nu}_{\alpha\beta}=\Delta_{\alpha\beta}^{\mu\nu}$ s.t.:
\begin{equation}
\Delta_{\alpha \beta}^{\mu \nu}=\mathcal{P}_{\alpha \beta}^{\mu \nu}-2\left(1-\frac{1}{\varrho}\right) \mathcal{P}_{\rho \kappa}^{\mu \nu} \frac{p^\rho p_\sigma}{p^2} \mathcal{P}_{\alpha \beta}^{\sigma \kappa}
\end{equation}
When analysing the tensor $\Delta_{\alpha \beta}^{\mu \nu}$, it is important to note that this structure depends on two key factors: the momentum of the graviton $p^{\mu}$, and the covariant gauge parameter $\varrho$. When it comes to inverting 
$\Delta_{\alpha\beta}^{\mu\nu}$, there exist multiple approaches. We will write an Ansatz for the most general covariant quadratic operator that depends on the graviton momentum. There are five such independent operators for this:
\begin{align}
I_{\alpha \beta}^{\mu \nu} & =\frac{1}{2}\left(\delta_\nu^\mu \delta_\beta^\alpha+\delta_\beta^\mu \delta_\alpha^\nu\right)\label{Bes1}\\
T_{\alpha \beta}^{\mu \nu} & =\frac{1}{4} \eta^{\mu \nu} \eta_{\alpha \beta}\label{Bes2}\\
C_{\alpha \beta}^{\mu \nu} & =\frac{1}{2}\left(\eta^{\mu \nu} \frac{p_\alpha p_\beta}{p^2}+\frac{p^\mu p^\nu}{p^2} \eta_{\alpha \beta}\right)\label{Bes3}\\
\mathcal{J}_{\alpha \beta}^{\mu \nu} & =I_{\rho \kappa}^{\mu \nu} \frac{p_\sigma p^\rho}{p^2} I_{\alpha \beta}^{\sigma \kappa}\label{Bes4}\\
K_{\alpha \beta}^{\mu \nu} & =\frac{p^\mu p^\nu}{p^2} \frac{p_\alpha p_\beta}{p^2}\label{Bas5}
\end{align}
Any quadratic operator that involves $p^{\sigma}$ and $\eta^{\mu\nu}$ can be expressed as a linear combination of these operators. To simplify the notation, we will adopt an index-free approach.
\begin{equation}
    \mathcal{P}=I-2 T,\quad\text{and}\quad\Delta=\mathcal{P}+2\left(1-\frac{1}{\varrho}\right) \mathcal{P} \mathcal{J} \mathcal{P}
\end{equation}
Since our objective is to determine $G=\Delta^{-1}$, we will represent $G$ as a linear combination of the five operators introduced earlier in equations \eqref{Bes1} to \eqref{Bas5}.
\begin{equation}
G=\alpha_1 I+\alpha_2 T+\alpha_3 C+\alpha_4 \mathcal{J}+\alpha_5 K\label{ansatz}
\end{equation}
The coefficients $\alpha_n$ are then determined via the equation.
\begin{equation}
\Delta G=I\label{DGeq1}
\end{equation}
If we multiply $G$ and $\Delta$, we get:
\begin{align}
\Delta_{\alpha \beta}^{\mu \nu} G_{\gamma \delta}^{\alpha \beta}&=\alpha_1 I_{\gamma \delta}^{\mu \nu}\label{eqsbas1} \\
&+\left(\alpha_1\left(-4+\frac{2}{\varrho}\right)+\alpha_2\left(-1+\frac{1}{2 \varrho}\right)(D-2)-\alpha_3 \frac{1}{\varrho}\right) T_{\gamma \delta}^{\mu \nu}\label{eqsbas2}\\
&+\Bigg(\alpha_1\left(2-2 \frac{1}{\varrho}\right)+\alpha_2 \frac{D-2}{4}\left(1-\frac{1}{\varrho}\right)+\alpha_3\left(-\frac{D-2}{2}+\frac{D}{4 \varrho}\right)-\left(\alpha_4+\alpha_5\right) \frac{1}{2 \varrho}\Bigg) C_{\gamma \delta}^{\mu \nu}\label{eqsbas3}\\
&+\left(\alpha_1 \left(-2+\frac{2}{\varrho}\right)+\alpha_4 \frac{1}{\varrho}\right) \mathcal{J}_{\gamma \delta}^{\mu \nu}\label{eqsbas4}\\
& +\left(\alpha_3(D-2)\left(\frac{1}{2}-\frac{1}{\varrho}\right)+\alpha_5 \frac{1}{\varrho}\right) K_{\gamma \delta}^{\mu \nu} \\
& +\left(\alpha_2 \frac{D-2}{4}\left(-1+\frac{1}{\varrho}\right)+\alpha_3\left(-\frac{D-2}{2}+\frac{D-4}{4 \varrho}\right)-\left(\alpha_4+\alpha_5\right) \frac{1}{2 \varrho}\right) A_{\gamma \delta}^{\mu \nu}\label{eqsbas5}
\end{align}
By comparing the obtained result with Eq. \eqref{DGeq1}, where $\Delta G=I$, we can derive a set of six equations that determine the value of $\alpha_n$. However, it is important to note that not all these equations are independent, and we can determine the value of $\alpha_1$ to $\alpha_5$ using only the first five. Specifically, from the first  Eq. \eqref{eqsbas1}, we find that $\alpha_1=1$. Moreover, we obtain an additional result from the fourth  Eq. \eqref{eqsbas4}.
\begin{align}
    0&=2\left(-1+\frac{1}{\varrho}\right)+\alpha_4 \frac{1}{\varrho}\\
    \alpha_4&=-2\kl{1-\varrho}
\end{align}
We can then determine the coefficients $\alpha_2$, $\alpha_3$, and $\alpha_5$ by analyzing Equations \eqref{eqsbas2}, \eqref{eqsbas3}, and \eqref{eqsbas4}. Finally, we obtain the following values for the coefficients:
\begin{align}
\alpha_1&=1\label{cof1} \\
\alpha_2&=-\frac{4}{D-2}\label{cof2} \\
\alpha_3&=0\label{cof3}\\
\alpha_4&=-2(1-\varrho)\label{cod3} \\
\alpha_5&=0\label{cof4}
\end{align}
For a quick verification, we can substitute the determined coefficients into equations \eqref{eqsbas1} to \eqref{eqsbas5}. As a result, all lines except the first one will vanish, yielding $G\Delta=I$. To obtain the tensor structure of the graviton propagator, we can incorporate the coefficients from line \eqref{cof1} to \eqref{cof4} into the expression for $G^{\mu\nu}_{\alpha\beta}$ given in equation \eqref{ansatz}. This will yield the following result:
\begin{equation}
    G=\mathcal{P}^{-1}-2(1-\varrho) \mathcal{J}
\end{equation}
Where $\mathcal{P}^{-1}$ is the inverse propagator to $\mathcal{P}$ and is given by :
\begin{equation}
\mathcal{P}^{-1}=I-\frac{4}{D-2} T \label{P-1}
\end{equation}
When $\varrho=1$, the covariant gauge operator simplifies to $G=\mathcal{P}^{-1}$, which is commonly referred to as the de Donder propagator. In four space-time dimensions ($D=4$) $\mathcal{P}^{-1}=\mathcal{P}$.
Ultimately, both $G$ and $\Delta$ exhibit certain similarities:
\begin{equation}
\Delta  =\mathcal{P}-2\left(1-\frac{1}{\varrho}\right) \mathcal{P} \mathcal{J} \mathcal{P},\quad
G =\mathcal{P}^{-1}-2(1-\varrho) \mathcal{J},
\end{equation}
If we write down $G$ with explicit index notation using the definitions of $\mathcal{P}^{-1}$, Eq. \eqref{P-1} and $\mathcal{J}$ Eq. \eqref{Bes4}, the graviton propagator in the covariant de Donder-type gauge:
\begin{equation}
G_{\alpha \beta}^{\mu \nu}=I_{\alpha \beta}^{\mu \nu}-\frac{1}{D-2} \eta^{\mu \nu} \eta_{\alpha \beta}-2(1-\varrho) I_{\rho \kappa}^{\mu \nu} \frac{p^\rho p_\sigma}{p^2} I_{\alpha \beta}^{\kappa \sigma}
\end{equation}
If we set $\varrho=1$, then the graviton propagator reduces to the de Donder type graviton propagator:
\begin{equation}
G_{\alpha \beta}^{\mu \nu}=I_{\alpha \beta}^{\mu \nu}-\frac{1}{D-2} \eta^{\mu \nu} \eta_{\alpha \beta}\label{GPROP}
\end{equation}
Diagrammatic this looks as follows:

\begin{figure}[h]
\centering
\includegraphics[width=0.66\textwidth]{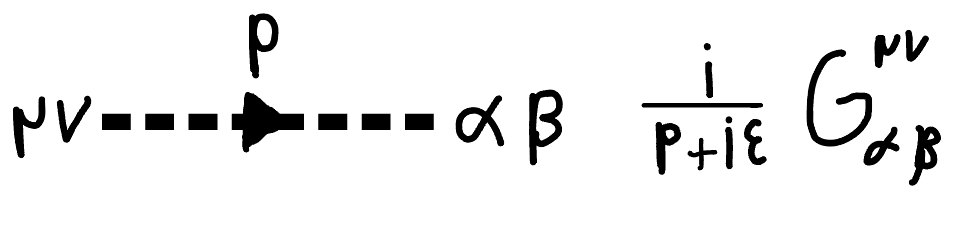}
\label{fig:figure1}
\end{figure}

\section{Scalar Gravity Vertex}
This section again follows the work \cite{Jakobsen:2020ksu}
as well as \cite{rafiezinedine2018simplifying}. Since the matter in question is described by a scalar field, its interaction with the gravitational field can be described by a scalar gravitational vertex. The rules for scalar gravitational Vertices can be derived by expanding the matter term of the action, denoted as $S_\phi$, which takes the following form:
\begin{equation}
S_\phi=\frac{1}{2} \int d^D x \sqrt{-g}\left(g^{\mu \nu} \phi_{, \mu} \phi_{, \nu}-m^2 \phi^2\right)
\end{equation}
If we recall that the number of gravitational interactions corresponds, roughly speaking, to the order of expansion of the action $S_{\phi}$ with respect to the metric $h_{\mu\nu}$, we can observe that the simplest vertex involving one graviton and one scalar $\phi^2h$ corresponds to the first order of expansion of $S_{\phi}$ in terms of $h_{\mu\nu}$. This expansion, in turn, relies on the first-order expansion of $g^{\mu\nu}$ (as given by Eq. \eqref{gmnexp}) and $\sqrt{-g}$ (as given by Eq. \eqref{sqgexp}) in terms of $h_{\mu\nu}$. If we perform this expansion, we obtain the following expression:
\begin{align}
S_\phi & \approx \frac{1}{2} \int d^D x\left(\eta^{\mu \nu} \phi_{, \mu} \phi_{, \nu}-m^2 \phi^2\right) \\
& -\frac{1}{2} \int d^D x\left(h^{\mu \nu} \phi_{, \mu} \phi_{, \nu}-\frac{1}{2} h_\mu^\mu\left(\phi^{, \nu} \phi_{, \nu}-m^2 \phi^2\right)\right)
\end{align}
In order to obtain the $\phi^2h$-vertex in a form more convenient for our purposes, we can transform the action into momentum representation and perform a rescaling of the form $\mathfrak{h}_{\mu\nu}=\frac{h{\mu\nu}}{\kappa}$. This leads us to the Feynman rules for the $\phi^2h$-vertex function, which reads as follows:
\begin{equation}
V^{\mu\nu}_{\phi^2h}(q,l,m)=
I_{\alpha \beta}^{\mu \alpha} q^\alpha l^\beta-\frac{q l-m^2}{2}\eta^{\mu\nu}\label{Verrul} 
\end{equation}
and 
\begin{figure}[h]
\centering
\includegraphics[width=0.66\textwidth]{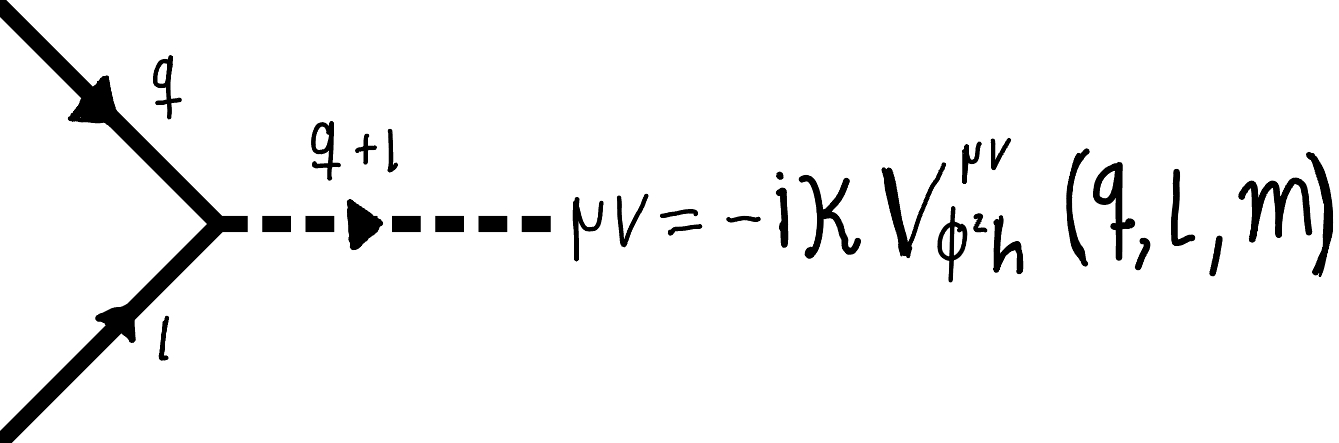}
\label{VerrulD}
\end{figure}
furthermore, we have the well-known scalar propagator:
\begin{equation}
    S(p,m)=\frac{i}{p^2-m^2+i \epsilon}\label{SPROP}
\end{equation}
\section{Feynman Integral of the Gravity-Box and Cross Box}
From quantum field theory we know that any scattering amplitude can be described as an expansion of the interacting fields in the coupling constant, for gravity this is $G_N$, where the resulting loops are the off-shell interaction of the mediating boson with the interacting fields. Such an expansion is valid in the low-energy limit of gravity, where the spacetime curvature is small and special relativity applies.
\begin{figure}[h]
\centering
\includegraphics[width=0.9\textwidth]{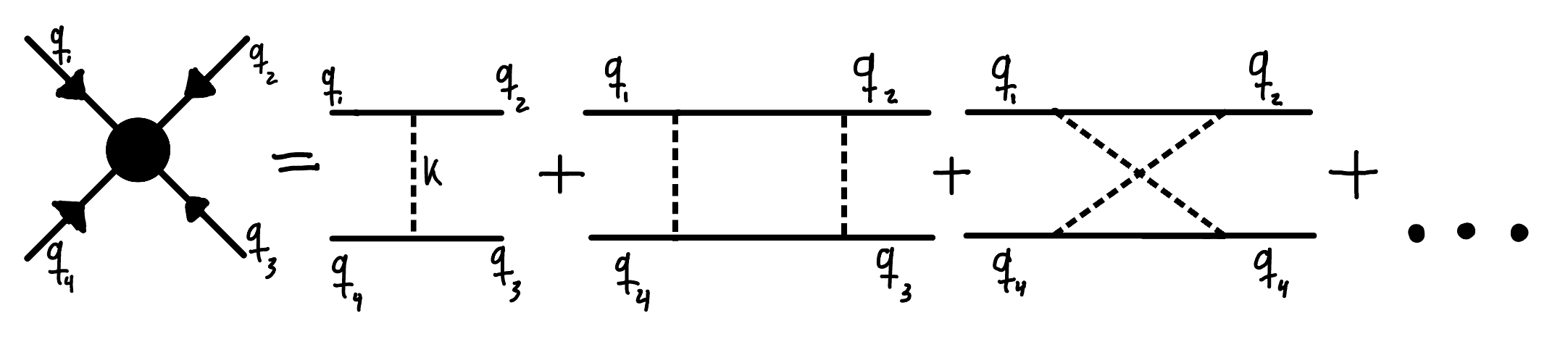}
\label{VerrulD2}
\end{figure}
In our particular case, we consider the gravitational scattering amplitude $2\to2$ with momentum $p_1$,$p_2$, with $m_1$ and $p_3$ and $p_4$ with $m_2$. The scalar gravitational scattering amplitude is expanded to 2nd order in $G_N$. giving us what is commonly known in the literature as the second post-Minkowskian expansion (2PM), which in addition to the first-order tree-level contribution includes a gravitational box contribution, a cross-box contribution and triangular contributions, which we will not investigate further in this paper, we will focus exclusively on the box and cross-box contributions, we will cover the full 2PM contribution with intersection theory in a forthcoming paper.
First, we have the tree-level interaction between the two scalars with the vertex rule Eq. \eqref{Verrul} and the de Donder type graviton propagator Eq.\eqref{GPROP}, which results in: 
\begin{equation}
\mathcal{M}=\kappa^2 V^{\mu}_{\phi^2h}(p_1,p_2,m_1)G_{\mu\nu}^{\alpha\beta}(p_1+p_2)V^{\alpha\beta}_{\phi^2h}(p_3,p_4,m_1)
\end{equation}
Interestingly, performing the calculation purely at the tree-level contribution would yield Newtonian gravity.\cite{Feynman:1963ax}. 

Secondly, we have the box and Crossbox contribution of the full 2PM contribution, applying the Feynman rules for the scalar graviton vertex Eq. \eqref{Verrul}, the de Donder type graviton propagator Eq. \eqref{GPROP}, as well as the scalar propagator Eq. \eqref{SPROP}, we can derive, the following two Feynman integrals.
\begin{align}
&I_{B}\kl{q_1,q_2,q_3,q_4,k,m_1,m_2}\\
=&\int_{\mathbb{R}^{D}}\frac{d^{D}l}{\kl{i\pi}^{D/2}}\kappa V^{\alpha\beta}(q_1,-q_1-l,m)S(q_1+k,m_1)\kappa V^{\epsilon\zeta}(q_2,q_1+k,m_1)\\
&\frac{1}{(Q+k)^2}G_{\epsilon\zeta\iota\mu}(h)\kappa V^{\iota\mu}(q_4-k,q_3,m_2)S(q_4-k,m_2)\kappa V^{\xi\rho}(q_4,k-q_4,m_3)\frac{1}{k^2}G_{\xi\rho\alpha\beta}(h)\nonumber
\end{align}
with $Q=q_1+q_2$. Writing the Feynman rules for the scalar graviton vertex Eq. \eqref{Verrul}, the de Donder type graviton propagator Eq. \eqref{GPROP}, as well as the scalar propagator Eq. \eqref{SPROP} explicitly will lead to:
\begin{align}
&I_{B}\kl{q_1,q_2,q_3,q_4,k,m_1,m_2}\label{FIGB}\\
=&\kappa^4\int_{\mathbb{R}^{D}}\frac{d^{D}l}{\kl{i\pi}^{D/2}}\kl{I^{\alpha\beta}_{\gamma\delta}(-k-q_1)^\gamma q_1^\delta-\frac{\kl{-q_1-k}q_1-m^2_1}{2}\eta^{\alpha\beta}}\frac{1}{\kl{k+q_1}^2-m_1+i\varepsilon}\nonumber\\
&\kl{I^{\epsilon\zeta}_{\eta\theta}(q_1+k)^{\eta}q_2^\theta-\frac{\kl{q_1+k}q_2-m_1^2}{2}\eta^{\epsilon\zeta}}\frac{1}{(Q+k)^2}\kl{I_{\epsilon\zeta\iota\mu}-\frac{1}{D-2}\eta_{\epsilon\zeta}\eta_{\iota\mu}}\nonumber\\
&\kl{I^{\iota\mu}_{\kappa\lambda}(q_4-k)^{\kappa}q_3^{\lambda}-\frac{(q_4-k)q_3-m_2^2}{2}\eta^{\iota\mu}}\frac{1}{\kl{q_4-k}^2-m_2^2+i\varepsilon}\nonumber\\
&\kl{I^{\xi\rho}_{\pi\sigma}(k-q_4)^\pi q_4^\sigma-\frac{(k-q_4)q_4-m^2_2}{2}\eta^{\xi\rho}}\frac{1}{k^2}\kl{I_{\alpha\beta\xi\rho}-\frac{1}{D-2}\eta_{\alpha\beta}\eta_{\xi\rho}}\nonumber
\end{align}
with
\begin{equation}
I_{\alpha \beta}^{\mu \nu}=\frac{1}{2}\left(\delta_\alpha^\mu \delta_\beta^\nu+\delta_\beta^\mu \delta_\alpha^\nu\right)
\end{equation}
see Eq. \eqref{IT}. 
%Performing not-so-simple calculations that can be found under $TBI$, we can increase the prettiness of the result and arrive at the following expression for the box:
%\begin{align}
%&I_{B}\kl{q_1,q_2,q_3,q_4,l,m_1,m_2}\\
%=&\frac{\kappa^4}{16}\int_{\mathbb{R}^{D}}\frac{d^{D}l}{\kl{2\pi}}\left(\frac{1}{2} \left(s-m_1^2-m_2^2\right)^2-\frac{2 m_1^2 m_2^2}{D-2}\right)^2\\
%&\frac{1}{l^2 (l+Q)^2 \left((l+q_1)^2-m_1^2\right) \left(l-q_4)^2-m_2^2\right)}
%\end{align}
with the Mandelstam variables:
\begin{align}
s=-\kl{q_1+q_4}^2&=-\kl{q_2+q_3}^2\nonumber\\
u=-\kl{q_1+q_3}^2&=-\kl{q_2+q_4}^2\\
t=-\kl{q_1+q_2}^2&=-\kl{q_3+q_4}^2\nonumber
\end{align}
Secondly, for the cross-box contribution we get the:
\begin{align}  
&I_{CB}\kl{q_1,q_2,q_3,q_4,k,m_1,m_2}\\
=&\int_{\mathbb{R}^{D}}\frac{d^{D}l}{\kl{i\pi}^{D/2}}\kappa V^{\alpha\beta}(q_2,-q_2-k,m)S(q_2+k,m_1)\kappa V^{\epsilon\zeta}(q_1,q_2+k,m_1)\nonumber\\
&\frac{1}{(Q+k)^2}G_{\epsilon\zeta\iota\mu}(h)\kappa V^{\iota\mu}(q_4-k,q_3,m_2)S(q_4-k,m_2)\kappa V^{\xi\rho}(q_4,k-q_4,m_3)\frac{1}{l^2}G_{\xi\rho\alpha\beta}(h)\nonumber
\end{align}
with $Q=q_1+q_2$. 
Writing the Feynman rules for the scalar graviton vertex Eq.\eqref{Verrul}, the de Donder type graviton propagator Eq.\eqref{GPROP}, as well as the scalar propagator Eq. \eqref{SPROP} explicitly gives us:
\begin{align}
&I_{CB}\kl{q_1,q_2,q_3,q_4,l,m_1,m_2}\label{FIGCB}\\
=&\kappa^4\int_{\mathbb{R}^{D}}\frac{d^{D}l}{\kl{i\pi}^{D/2}}\nonumber
\kl{I^{\alpha\beta}_{\gamma\delta}(-q_2-k)^\delta q_2^\gamma-\frac{\kl{-q_2+k}q_2-m^2_1}{2}\eta^{\alpha\beta}}\frac{1}{\kl{k+q_2}^2-m_1+i\varepsilon}\nonumber\\
&\kl{I^{\epsilon\zeta}_{\eta\theta}(q_2+k)^{\eta}q_1^\theta-\frac{\kl{q_2+l}q_1-m_1^2}{2}\eta^{\epsilon\zeta}}
\frac{1}{(Q+l)^2}\kl{I_{\epsilon\zeta\iota\mu}-\frac{1}{D-2}\eta_{\epsilon\zeta}\eta_{\iota\mu}}\nonumber\\
&\kl{I^{\iota\mu}_{\kappa\lambda}\kl{q4-k}^{\kappa}q_3^{\lambda}-\frac{\kl{q_4+k}q_3-m_2^2}{2}\eta^{\iota\mu}}
\frac{1}{\kl{q_4-k}^2-m_2^2+i\varepsilon}\nonumber\\
&\kl{I^{\xi\rho}_{\pi\sigma}(k-q_4)^\pi q_4^\sigma-\frac{(q_4+k)q_4-m^2_2}{2}\eta^{\xi\rho}}\frac{1}{k^2}\kl{I_{\alpha\beta\xi\rho}-\frac{1}{D-2}\eta_{\alpha\beta}\eta_{\xi\rho}}\nonumber  
\end{align}
with
\begin{equation}
I_{\alpha \beta}^{\mu \nu}=\frac{1}{2}\left(\delta_\alpha^\mu \delta_\beta^\nu+\delta_\beta^\mu \delta_\alpha^\nu\right)
\end{equation}
see Eq. \eqref{IT}.  

%Performing not-so-simple calculations that can be found under $TBI$, we can increase the prettiness of the result and arrive at the following expression for the corss-box:
%\begin{align}
%&I_{CB}\kl{q_1,q_2,q_3,q_4,l,m_1,m_2}\\
%=&\frac{\kappa^4}{16}\int_{\mathbb{R}^{D}}\frac{d^{D}l}{\kl{2\pi}}\left(\frac{1}{2} \left(u-m_1^2-m_2^2\right)^2-\frac{2m_1^2 m_2^2}{D-2}\right)^2\\
%&\frac{1}{l^2(-m1^2+(l+q_1)^2)(-m_2^2+(l+q_3)^2)(l+Q)^2)}
%\end{align}
With the Mandelstam variables:
\begin{align}
s=-\kl{q_1+q_4}^2&=-\kl{q_2+q_3}^2\nonumber\\
u=-\kl{q_1+q_3}^2&=-\kl{q_2+q_4}^2\label{MSV}\\
t=-\kl{q_1+q_2}^2&=-\kl{q_3+q_4}^2\nonumber
\end{align}
\newpage

\chapter{Feynman Integral essentials}
In this section \label{FIE}, we will outline the properties of Feynman integrals and diagrams that are essential for our later work. 
This section is based on the works of \citep{Gasparotto:2023cdl}\citep{Grozin:2011mt}\citep{Weinzierl:2022eaz}.  Let's examine a diagram with $p_1,...,p_E$ external momenta. In a general kinematic setup means that the diagram $n_{Ex}=E+1$ has outer edges with momentum $p_i$,% $n_I$ internal edges with momenta $k_i$,
and $l$ loops with momentum $k_i$. The total number of momenta denoted as $q_i$ spans from $i=\{1,...,l-1,l,l+1,...,l+E\}$.% For a given diagram there are $s_{ij},j\geq i$ scalar products of number: 
%\begin{equation}
%    N_E=\frac{E(E+1)}{2}
%\end{equation}
%of the external kinematic quantities such that $i>L$, and
\begin{equation}
    n=\frac{l(l+1)}{2}+lE\label{n}
\end{equation}
scalar products of the internal variables, such that $i\leq l$.
Each Feynman diagram, is characterised by $E$ external edges, and $l$ loops. The Feynman integral of a specific dimensional regularized space-time dimension $D$ is given by:
\begin{equation}
I=\int \prod_{j=1}^{l} \frac{d^D k_j}{ i\pi^{\frac{D}{2}}}
\frac{1}{\mathsf{D}_1^{a_1} \ldots \mathsf{D}_n^{a_n}}\label{FI} 
\end{equation}
Here $n$ is defined by Eq. \eqref{n} and $a_j$ denotes the power of the propagator. 
\begin{equation}
    \mathsf{D}_j=\left(-q_j^2+m_j^2\right)^{a_j}\quad\text{with}\quad q_j=\sum_{r=1}^{l} \mathbb{A}_{j r} k_r+\sum_{r=1}^E \mathbb{B}_{j r} p_r\quad\text{such that}\quad j\in\{1,...,n\}
\end{equation}
This is known as the inverse propagator. The propagators have a momentum $k$. Notably, when $l\geq2$, the set $(\mathsf{D}_1,...,\mathsf{D}_n)$ are in one-to-one correspondence with a set of scalar products that involve the loop momentua $k$. Notably, when $l\geq2$, the set $(\mathsf{D}_1,...,\mathsf{D}_n)$ is, in general, larger than the set of denominators appearing in the Feynman diagram. These elements are referred to in the literature as irreducible scalar products.
The power $a_j$ is an integer such that $a_j\in\mathbb{Z}$ for $j\in\{1,...,n\}$ and $|a|=\sum_{j=1}^{n}a_j$. The entire set of possible integrals is collectively referred to as the integral family. Its elements can be organised in sectors known as topologies. For any given Feynman integral within the integral family, we can assign an integer $\mathrm{ID}$ based on the following rule.
\begin{equation}
    I_{a_1, a_2, \ldots, a_n} \rightsquigarrow \mathrm{ID}=\sum_{j=1}^n 2^{j \cdot \Theta\left(a_j-\frac{1}{2}\right)}-1\label{ID}
\end{equation}
Integers sharing the same $\mathrm{ID}$ are grouped together in a sector within the integral family. Hence, a sector is nothing but a collection of integrals characterised by having propagators with the same powers, which are strictly positive integers, see Eq. \eqref{ID}. These powers are often called denominators. Subsectors or topologies are characterised by a smaller collection of the above-mentioned subset of propagators. 
In addition, it is useful to assign a specific characterisation to any given Feynman integral, as follows:
\begin{equation}
I_{a_1, a_2, \ldots, a_n} \rightsquigarrow\left\{\begin{aligned}
r & =\sum_{i=1}^n a_i \Theta\left(a_i-\frac{1}{2}\right), \\
s & =-\sum_{i=1}^n a_i \Theta\left(-a_i+\frac{1}{2}\right) .
\end{aligned}\right.\label{sr}
\end{equation}
The quantities presented in Eq.\eqref{sr}, $\Theta$ 
quantify the complexity of the integral, larger values of $r$ and $s$ imply an expected increase in the complexity of the corresponding Feynman integral. If we are only interested in the values of $\mathrm{ID}$ and $r,s$, the notation $I_{\mathrm{ID},r,s}$ or $I_{r,s}$ will be used. The equation \eqref{FI} is not the only possible representation of a Feynman integral. In this thesis we will mostly work with the Baikov representation.

\section{Baikov Representation}

In this section \label{SECBR}, we will introduce the Baikov representation for the one-loop case. The derivation is based on  \citep{Weinzierl:2022eaz} as well as \citep{Frellesvig:2021vdl}. The Baikov representation applies to a certain subset of the Feynman diagram where the number of internal edges $n$ is equal to the number of independent scalar products which involves the loop momenta. Earlier we denoted the external momenta with $p_1,p_2,\dots,p_{E}$ and their dimension, $E=\dim\kl{p_1,p_2,\dots,p_{E}}$. For generic external momenta and $D\geq n_{Ex}-1$ we have $E=n_{Ex}-1$. The Lorentz invariant quantities for the one-loop case that involves the loop momenta are: 
\begin{equation}
    k^2,\quad k_i\cdot p_j
\end{equation}
%In total, for the one-loop case 
%\begin{equation}
%N_E=1+E
%\end{equation}
linear independent scalar products which involve the loop-momenta.
We denote the linear independent scalar products involving the momenta by:
\begin{equation}
\sigma=\left(k^2, k\cdot p_1,\dots, k\cdot p_E\right), \quad\sigma=\left(\sigma_0, \sigma_i\right)
\end{equation}
with  $\sigma_0=K^2$ and $\sigma_i=K, \cdot P_i\quad i\in\{0,...,E\}$. 
Later during the derivation of the Baikov representation, we will use them as our new coordinate system.

A Feynman diagram $G$ has a Baikov representation if $N_V=n$
and if we may express any internal inverse propagator as a linear combination of the linear independent scalar products which involve the loop momenta and terms that are independent of the loop momenta. The second condition states, that there is an invertible $(N_V+1)\cross(N_V+1)$ dimensional matrix $C$ and a loop momentum independent $N_V$ dimensional vector $f$ such that:
\begin{equation}
-q_a+m^2_a=C_{ab}\sigma_b+f_{a}\label{BC2}
\end{equation}
for all $a\in\{1,...,n\}$. The loop momenta $k$ can be decomposed into a parallel $k_{\|}$ and a perpendicular component $k_{\perp}$:
\begin{equation}
k=k_{\|}+k_{\perp} 
\end{equation}
where the parallel component $k_{\|}$ lives inside the space of the external momenta,
\begin{equation}
k_{\|}\in\left\{ p_1, p_{2,},\dots,p_{n}\right\}\\
\end{equation}
while the perpendicular component lives inside the orthogonal space. 
It further follows for the integration measure:
\begin{equation}
    d^{D}k=d^{E}k_{\|}d^{D-E}k_{\perp}\label{Imea1}
\end{equation}
%as usual, we go into the complex Euclidean space where the measure reads:
%\begin{equation}
%   d^D K=i d^D k, \quad d^E K_{\|}=i d^E k_{\|}, \quad d^{D-e} K_{\perp}=d^{D-E} k_{\perp}
%\end{equation}
By the usual argumentation follows for the volume integration measure: 
\begin{equation}
d^{D-E} k_{\perp}=k_{\perp}
^{D-1-E} d\vert k_{\perp}\vert d^{D-E-1}\Omega
\end{equation}
with
\begin{align}
\int d^{D-E-1}\Omega=\Omega_{D-E}=\frac{2\pi^{{D-E} / 2}}{\Gamma\kl{\frac{D-E}{2}}}   
\end{align}
and therefore is expressed as:
\begin{equation}
I=\frac{2}{\Gamma\kl{\frac{D-E}{2}}i\pi^{\frac{E}{2}}}\int\frac{N(k)\left|k_{\perp}\right|^{D-E-1} d\left|k_{\perp}\right|d^E k_{\|}}{\mathsf{D}_1(k)^{a_1}\cdots \mathsf{D}_{n}(k)^{a_{n}}}\label{Inc1}
\end{equation}
As $k_{\|}$ lives  inside the space of external momentum, it can be expressed as a linear combination of the external momentum $p_i$ times some scaling factor $r_i$:
\begin{equation}
k_{\|}=\sum_{i=1}^E r_i p_i
\end{equation}
This now implies the following
\begin{align}
k\cdot p_j=k_{\|} \cdot p_j=\sum_{i=1}^E r_i p_i \cdot p_j  
\end{align}
We recognise that $p_i \cdot p_j$ forms in a general matrix $G$ whose entries are scalar products $G_{ij}=p_i \cdot p_j$. This is known as a Gram matrix. 
Thus, the scaling factor $r_i$ can be expressed as:
\begin{equation}
r_i=\sum_{j=1}^E G_{i j}^{-1}\left(k \cdot p_j\right)
\end{equation}
From this follows: 
\begin{align}
k_{\|}^2 & =\sum_{i j=1}^E r_i r_j G_{i j} \\
& =\sum_{i j=1}^E\left(k \cdot p_i\right) G_{i j}^{-1} G_{i j} G_{i j}^{-1}\left(k \cdot p_j\right) \\
& =\sum_{j=1}^E\left(k \cdot p_i\right) G_{i j}^{-1}\left(k \cdot p_j\right)\label{kpa}
\end{align}
This identity will later come in quite handy. Following our integration measure Eq. \eqref{Imea1}, we will first calculate $d^E k_{\|}$ in our new coordinates $\sigma$, as $k_{\|}$ lies within the span of the external momenta $p_i$, and this only involves $\sigma_i$. The associated Jacobian matrix between the loop momenta and the external momenta reads: 
\begin{align}
J=\frac{\partial\left(\sigma_1,\dots,\sigma_E\right)}{\partial\left(K_1^0,\dots,K^E\right)}
=\left(\begin{array}{cccc}
P_1^0 & P_1^1 & \ldots & P_1^{e-1} \\
P_2^0 & P_2^1 & \ldots & P_2^{e-1} \\
& & \ldots & \\
P_e^0 & P_e^1 & \ldots & P_e^{e-1}
\end{array}\right)
\end{align}
As $JJ^{T}=p_i\cdot p_j$ we see that the Jacobian is nothing else than the square root of the Gram determinant of the external momenta:
\begin{equation}
\operatorname{det}(J)=\left(\operatorname{det}\left(G\left(P_{1,}, P_E\right)\right)\right)^{\frac{1}{2}}
\end{equation}
Therefore the $d^{E}k_{\|}$ follows the integration measure as seen below: 
\begin{align}
d^E k_{\|} =\left|\frac{\partial k_l}{\partial\sigma_i}(P)\right|^{\frac{1}{2}} \prod_{i=1}^E d \sigma_i =\sqrt{\operatorname{det}\left(G\left(p_1,\dots,p_E\right)^{-1}\right)}\prod_{i=1}^E d \sigma_i\label{dkpa}
\end{align}
Next, we are going to calculate $d k^2_{\perp}$ in our new coordinates $\sigma$, as this only involves the internal momentum, the Jacobian matrix simply reads:
\begin{equation}
\frac{\partial \sigma_0}{\partial k_{\perp}^2}=\frac{\partial\left(k^2\right)}{\partial k_{\perp}^2}=\frac{\partial\left(k_{\|}^2+k_{\perp}^2\right)}{\partial k_{\perp}^2}
\end{equation}
For the integration measure it follows:
\begin{align}
d k_{\perp}^2 & =d \sigma_0 \\
2 k_{\perp} d k_{\perp} & =d \sigma_0 \\
d k_{\perp} & =\frac{1}{2}\left|k_{\perp}\right|^{-1} d \sigma_0\label{dkper1}
\end{align}
In order to express $k_{\perp}$ in our new coordinates $\sigma$, preferably a Gram matrix, we first write-down its norm $k_{\|}^2
=k^2-k_{\perp}^2$. If we then include the expression for $k_{\|}$ Eq. \eqref{kpa}, which we derived earlier, we get: 
\begin{equation}
k_{\perp}^2 =k^2-k_{\|}^2=k^2-\sum_{j=1}^E\left(k\cdot p_i\right) G_{ij}^{-1}\left(k\cdot p_j\right)\label{kper1}
\end{equation}
If we now compare this expression with Schur's matrix decomposition formula, which states
\begin{equation}
\operatorname{det}\left(\begin{array}{ll}
A & B \\
C & D
\end{array}\right)=\operatorname{det}(D)\operatorname{det}\left(A-C D^{-1} B\right)\label{schur1}
\end{equation}
we may guess that the $\kl{E+1}\cross\kl{E+1}$ matrix of all the momentum scalar products $\sigma$ involved takes the following form: 
\begin{equation}
\hat{G}=\left(\begin{array}{cc}
k^2 & k \cdot p_i \\
k \cdot p_j & G_{j i}
\end{array}\right)
\end{equation}
From this we may deduce for \eqref{kper1}, using \eqref{schur1} the following:
\begin{equation}
\frac{\operatorname{det}(\hat{G})}{\operatorname{det}(G)}=k^2-\sum_{i j=1}^E(k\cdot p)\left(G^{-1}\right)_{i j}\left(k \cdot p_j\right)
\end{equation}
and thus follows this expression:
\begin{equation}
\sqrt{\frac{\operatorname{det}(\hat{G})}{\det(G)}}=k_{\perp}\label{kper2}
\end{equation}
If we insert equations \eqref{dkpa}, \eqref{dkper1} and \eqref{kper2} into the old expression for the integration measure, we get the integration measure in our new coordinates $\sigma$. 
Thus, the integral now reads: 
\begin{align}
I&=\frac{1}{\Gamma\kl{\frac{D-E}{2}}i \pi^{\frac{E}{2}}} 
\int\frac{N(\sigma)}{\mathsf{D}_1(\sigma)^{a_1}\cdots \mathsf{D}_{n}(\sigma)^{a_n}} \frac{1}{2}\left(\frac{\det(\hat{G})}{\det(G)}\right)^{\frac{D-E-2}{2}}\det\kl{G\left(p_1,\dots,p_{E}\right)}^{\frac{1}{2}}d\sigma_0 \prod_{i=1}^E d^E\sigma_i\label{Int2}
\end{align}
we define
\begin{equation}
\mathcal{B}:=\operatorname{det} \hat{G}=\operatorname{det} G\left(k, p_1, \ldots, p_E\right)\label{BPLY}
\end{equation}
and 
\begin{equation}
\mathcal{G}:=\operatorname{det} G=\operatorname{det} G\left(p_1, \ldots, p_E\right)\label{GPLY}
\end{equation}
Thus, Eq. \eqref{Int2} reads: 
\begin{equation}
I=\frac{\mathcal{G}^{(E-D+1) / 2}}{\Gamma((D-E) / 2) i \pi^{E / 2}} \int \frac{N(\varsigma) \mathcal{B}(\varsigma)^{(D-E-2) / 2} d^{E+1} \varsigma}{\mathsf{D}_1(\varsigma)^{a_1} \cdots \mathsf{D}_{n}(\varsigma)^{a_{n}}}
\end{equation}
In order to arrive at the final expression with the Baikov representation, we have to change the variable to the Baikov representation's variables: 
\begin{equation}
    z_a=-q_a^2+m^2_a
\end{equation}
As seen above, the Baikov representation variables are nothing else than the inverse propagators from Eq. \eqref{BC2}. The inverse relation between them is expressed below: 
\begin{equation}
\sigma_a=\left(C^{-1}\right)_{a b}\left(z_b-f_b\right)
\end{equation}
The Jacobi is a commonly chosen to be:
\begin{equation}
    \mathcal{J}=2^{-E}\label{BJ}
\end{equation}
or in the final result, for the one-loop Baikov representation, it reads as follows: 
\begin{equation}
I=\frac{\mathcal{J} \mathcal{G}^{\frac{E-D+1}{2}}}{\Gamma\kl{\frac{D-E}{2}} i\pi^{\frac{E}{2}}}\int\frac{N(\mathbf{z}) \mathcal{B}(\mathbf{z})^{\frac{D-E-2}{2}} d^{E+1} z}{z_1^{a_1} \cdots z_{\mathrm{n}}^{a_{n}}}\label{Brep}
\end{equation}

\section{Baikov Representation of the Gravity Box and Cross-Box}

In the following section we will transform the Feynman integral of standard Feynman representation for the box and the cross-box contribution into the Baikov representation. During this procedure, we will essentially follow the quantities defined earlier in the derivation in section \ref{SECBR}. 
Since the external quantities of the box and the cross-box are identical, the quantities in the Baikov representation, which depend purely on the external quantities, are also identical for the box and the cross-box. 
For the Baikov Jacobian in Eq. \eqref{BJ} we get:
\begin{equation}
\mathcal{J}=2^{-3}
\end{equation}
For the Gram matrix of the external momenta $\mathcal{G}$, Eq. \eqref{GPLY}, we arrive when rewriting the external momenta in terms of the Mandelstam variables at the following expression:
\begin{equation}
\mathcal{G}=\frac{1}{4}t\left(m_1^4-2 m_1^2 \left(m_2^2+s\right)+\left(m_2^2-s\right)^2+s t\right)
\end{equation}
For the box the Baikov representation 
variables are as follows, 
\begin{equation}
\begin{aligned}
      z_1&=k^2\\
      z_2&=\kl{q_1+k}^2+m_1^2\\
      z_3&=\kl{q_4+k}^2+m_2^2\\
      z_4&=\kl{k+Q}^2     
\end{aligned} 
\end{equation}
The Baikov representation polynomial is then as follows: 
\begin{equation}
\begin{aligned}
\mathcal{B}_{B}=
&\frac{1}{16} \Big{(}m_1^4 \left(t^2+2t(z_1+z_4)+(z_1-z_4)^2\right)\\
&-2m_1^2(m_2^2 \left(t^2+2 t (z_1+z_4)+(z_1-z_4)^2\right)\\
&+s\left(2 z_4(t-z_1)+(t+z_1)^2+z_4^2\right)\\
&+t(z_2-z_3)(t+z_1-2 z_3+z_4))\\
&+m_2^4\left(t^2+2t(z_1+z_4)+(z_1-z_4)^2\right)\\
&-2 m_2^2(s\left(t^2+2 t (z_1+z_4)+(z_1-z_4)^2\right)\\
&-t(z_2-z_3)(t+z_1-2 z_2+z_4))+s^2 t^2+2 s^2 t z_1\\
&+s^2 z_1^2+s^2 z_4^2+2 s t^2 z_2+2 s t^2 z_3+2 s z_4 (s (t-z_1)\\
&+t (-2 z_1+z_2+z_3))+2 s t z_1 z_2+2 s t z_1 z_3-4 s t z_2 z_3\\
&+t^2 z_2^2-2 t^2 z_2 z_3+t^2 z_3^2\Big{)}  
\end{aligned}
\end{equation}
For the cross-box the Baikov representation 
variables are as follows, 
\begin{equation}
\begin{aligned}
      z_1&=k^2\\
      z_2&=\kl{q_2+k}^2+m_1^2\\
      z_3&=\kl{q_4-k}^2+m_2^2\\
      z_4&=\kl{k+Q}^2     
\end{aligned} 
\end{equation}
The Baikov representation polynomial is then as follows: 
\begin{equation}
\begin{aligned}
\mathcal{B}_{CB}=&\frac{1}{16}\Big{(}m_1^4 \left(t^2+2 t (z_1+z_4)+(z_1-z_4)^2\right)\\
&+2 m_1^2 (-\left(m_2^2 \left(t^2+2 t (z_1+z_4)+(z_1-z_4)^2\right)\right)\\
&-s\left(2 z_4 (t-z_1)+(t+z_1)^2+z_4^2\right)\\
&-t(t+z_1-2 z_3+z_4) (t+z_1-z_2-z_3+z_4))\\
&+m_2^4\left(t^2+2 t (z_1+z_4)+z_1-z_4)^2\right)\\
&-2m_2^2(s\left(t^2+2 t (z_1+z_4)+(z_1-z_4)^2\right)\\
&+t (t+z_1-2 z_2+z_4)(t+z_1-z_2-z_3+z_4))\\
&+s^2 t^2+2 s^2 t z_1+s^2 z_1^2+2 s t^3\\
&+4 s t^2 z_1-2 s t^2 z_2-2 s t^2 z_3\\
&+2 s t z_1^2+2 z_4 (s+t) (s (t-z_1)+t (t+z_1-z_2-z_3))\\
&-2 s t z_1 z_2-2 s t z_1 z_3+4 s t z_2 z_3\\
&+z_4^2 (s+t)^2+t^4+2 t^3 z_1-2 t^3 z_2-2 t^3 z_3\\
&+t^2 z_1^2-2 t^2 z_1 z_2-2 t^2 z_1 z_3\\
&+t^2 z_2^2+2 t^2 z_2 z_3+t^2 z_3^2\Big{)}
\end{aligned}
\end{equation}
The last part that is missing for the complete Baikov representation given by Eq. \eqref{Brep}, is the expression for the numerator. For the box it was
\begin{equation}
\begin{aligned}
N(\mathbf{z})_{B}=& \frac{i^4 i \kappa^4}{2(-2+d)^2} \\
& \left(( - 2 + d ) \left(\left(m_1^2+m_2^2-s\right)\left(m_1^2+m_2^2-s-z_3\right)\right.\right. \\
& \left.+\frac{1}{2}(z_1-z_2)\left(2\left(m_1^2+m_2^2-s\right)+z_1-z_3\right)\right) \\
& \left.-\left(2 m_1^2+z_1-z_2\right)\left(2 m_2^2+z_1-z_3\right)\right) \\
& \left(( - 2 + d ) \left(\left(m_1^2+m_2^2-s\right)\left(m_1^2+m_2^2-s-z_2-z_3\right)\right.\right. \\
& \left.+\left(-m_1^2-m_2^2+s+z_2-z_4\right)\left(-m_1^2-m_2^2+s+2 z_1+z_3-z_4\right)\right) \\
& \left.-2\left(2 m_1^2-z_2+z_4\right)\left(2 m_2^2-2 z_1-z_3+z_4\right)\right) \\
\end{aligned}
\end{equation}
and the cross-box:
\begin{equation}
\begin{aligned}
N_{CB}(\mathbf{z})=& \frac{i^4 i\kappa^4 }{2(-2+d)^2} \left(( - 2 + d ) \left(\left(m 1^2+m_2^2-u\right)\left(m 1^2+m_2^2-u-z_3\right)\right.\right. \\
& \left.+\frac{1}{2}(z_1-z_2)\left(2\left(m 1^2+m_2^2-u\right)+z_1-z_3\right)\right) \\
& \left.-\left(2 m_1^2+z_1-z_2\right)\left(2 m_2^2+z_1-z_3\right)\right) \\
& \left(( - 2 + d ) \left(\left(m 1^2+m_2^2-u\right)\left(m 1^2+m_2^2-u-z_2-z_3\right)\right.\right. \\
& \left.\left(+m_1^2+m_2^2+u+z_2-z_4\right)\left(+m_1^2+m_2^2+u+z_3-z_4\right)\right) \\
& \left.-2\left(2 m_1^2-z_2+z_4\right)\left(2 m_2^2-z_3+z_4\right)\right)
\end{aligned}
\end{equation}
The code used to obtain both the numerator of the Feynman integral and the the Baikov representation polynomial, can be found under \url{https://www.wolframcloud.com/obj/wfv651/Published/Line%20rule_1904.02667v1%20-%20new%20prop%202.nb} and \url{https://www.wolframcloud.com/obj/wfv651/Published/dotprules.nb}.
We could put it all together and write down the complete expression of the Feynman integral in Baikov representation, however, there is little to no didactic value in this. Therefore, we will simply end it here and encourage the motivated reader to write it down for himself.
\section{Integration by Parts Relations}
In this section, we will introduce the concept of integration by parts relation (IBPs), as well as their conceptual underpinnings, and sketch out their numerical implementation, it is based on the works of 
\citep{Gasparotto:2023cdl}
\citep{Grozin:2011mt}\citep{Frellesvig:2021vdl}\citep{Mattiazzi:2022zbo}\citep{Weinzierl:2022eaz}. Integration by parts identities enables us to express any Feynman integral as a linear combination of basis integrals, referred to as master integrals. These master integrals span a vector space, and if we consider the Feynman integrals as functions on a Manifold of the kinematic variables, then the master integrals give us a basis for a vector bundle in which the Feynman integral lies. We have already studied the family of Feynman integrals,
\begin{equation}
I=\int \prod_{j=1}^{l} \frac{d^D k_j}{i \pi^{D / 2}} \frac{1}{\mathsf{D}_1^{a_1} \ldots \mathsf{D}_n^{a_n}}
\end{equation}
IBPs relation give us the relations between the different members of a family, that is, the differences in the values of the indices $(a_1,...,a_n)$. They are based on the fact that, in a given dimensional regularisation, the integral of the total derivative vanishes:
\begin{equation}
0=\int \prod_{j=1}^{l} \frac{d^D k_i}{i \pi^{D / 2}} \frac{\partial}{\partial k_j^\mu}\left(\frac{\zeta^\mu}{\mathsf{D}_1^{a_1} \ldots \mathsf{D}_n^{a_n}}\right)\label{IBPR1}
\end{equation}
From the nonexistence of the boundary terms, we can then derive linear relations between different members of the family, that are exactly the IBPs. The vector $\zeta$, belongs to the set of $k^{\mu}_1,...,k_{l}^{\mu},p_1^{\mu},...,p_E^{\mu}$, or a linear combination formed from them.
Equation \eqref{IBPR1} can be derived as follows:
Let us first assume that $\zeta$ is a combination of the external momentum, and since the integrals are translation invariant within an equal dimensional regularization, it follows that:
\begin{equation}
    \int \frac{d^D k}{i \pi^{\frac{D}{2}}} f(k)=\int \frac{d^D k}{i \pi^{\frac{D}{2}}} f(k+\lambda\zeta)\label{tinv}
\end{equation}
Here the right side must be independent of $\lambda$, which implies, in particular, that the $\mathcal{O}(\lambda)$ term must vanish. If we now apply the total derivative to $\lambda$ to the RHS of Eq. \eqref{tinv} at $\lambda=0$, it gives: 
\begin{equation}
\left.\left[\frac{d}{d \lambda} f(k+\lambda q)\right]\right|_{\lambda=0}=\zeta^\mu \frac{\partial}{\partial k^\mu} f(k)=\frac{\partial}{\partial k^\mu}\left[\zeta^\mu \cdot f(k)\right]
\end{equation}
 thus, follows for $f(k)=\frac{1}{\mathsf{D}_1^{a_1} \ldots \mathsf{D}_n^{a_n}}$:
 \begin{equation}
0=\int \prod_{j=1}^{l} \frac{d^D k_i}{i \pi^{D / 2}} \frac{\partial}{\partial k_j^\mu}\left(\frac{\zeta^\mu}{\mathsf{D}_1^{a_1} \ldots \mathsf{D}_n^{a_n}}\right)\label{IBPR2}
 \end{equation}
If we perform the algebraic manipulations in \eqref{IBPR2} under the integral sign and express the results obtained in terms of Feynman integrals, we obtain linear relations between the integrals of the following kind $I_{\mathrm{ID},r,s}$, $I_{\mathrm{ID},r+1,s}$, $I_{\mathrm{ID},r,s+1}$ and $I_{\mathrm{ID},r+1,s+1}$. Here the coefficients are polynomials in $D$, masses $m_j$, and scaler products $s_{ij}$ among the external momentum. Simplifications among numerator and denominator could also produce integrals belonging to subsectors.\\

The integration-by-part identities can be generated systematically up to a certain explicit value of $(r,s)$ for a given sector and all its subsectors, see equation \eqref{sr}. The identities can be cast into a system where the Feynman integrals are considered as unknowns. Interestingly, the number of equations grows faster than the number of integrals. Thus, we can sort the integrals by their complexity. As mentioned earlier, this means that integrals with higher $r$ and $s$ are considered more complicated than integrals with lower values of $r$ and $s$. It is now possible to process the system via, for example, Gauss' elimination. This involves systematically scanning the system, equation by equation, expressing the most complex integrals in terms of other simpler integrals, and, subsequently, substituting these relations into the remaining equations. Eventually, all integrals can be expressed as combinations of a small set of remaining integrals, and is therefore known as master integrals. In our study, we will denote this set of master integrals as $(\mathcal{I}_1, \ldots, \mathcal{I}_{\nu})$. They form, as mentioned, the basis of our vector space on the manifold of the kinetic variables. As they form the basis elements of the vector space, their number is in most cases significantly lower than the number of Feynman integrals. This process is known as the Laporte algorithm. Various algorithms, such as \citep{Anastasiou:2004vj}, \citep{vonManteuffel:2012np},\citep{Lee:2013mka}, \citep{Smirnov:2019qkx}, can be employed for performing these computations. However, one of the main challenges encountered in practical implementations of these algorithms is the extensive amount of algebraic manipulation involved in solving the system of equations.
\section{Integration by parts identities in the Baikov representation:\label{IBPB}}
It is instructive for our intuition to consider IBPs identities in terms of the Baikov representation since it was originally proposed in \citep{Larsen:2015ped} and later improved in a number of papers \citep{GEORGOUDIS2017203}\citep{Georgoudis:2016wff}\citep{Bohm:2018bdy}\citep{Boehm:2020zig}. The main advantage of this strategy is that by benefiting from tools of computational algebraic geometry, it is possible to generate smaller subsets of IBPs identities. Consequently, the initial set of identities can be trimmed down, leading to a streamlined system-solving procedure. Once again, our starting point is the observation of the total derivative vanishing under the integral sign, which transforms Eq. \eqref{IBPR2} into: 
\begin{align}
0=\int_\gamma d\left(\frac{\mathcal{B}^{\frac{D-l-E-1}{2}}}{z_1^{a_1} \ldots z_n^{a_n}} \zeta\right)\label{B2}
\end{align}
where $\mathcal{B}(\partial \gamma)=0$, hence the Baikov representation polynomial vanishes on its boundaries.
$\zeta$ is an $(n-1)$ differential form $\zeta=\sum_{i=1}^n(-1)^{i+1} \zeta_i$, with $\zeta_i= \hat{\zeta} d z_1 \wedge \ldots \wedge \hat{d z_i} \wedge \ldots \wedge d z_n$ and each $\hat{\zeta}_i$ a polynomial in the variables $\mathbf{z}=\left(z_1, \ldots, z_n\right)$ whose coefficients may depend on the kinematic invariants.
\newpage

\graphicspath{{././images/}}
\chapter{Intersection Theory I, the Univariate case}
As we have already discussed \label{ISTU}, a given Feynman integral $I_{\nu_1,..,\nu_m}$ from a family of Feynman integrals is the linear combination of master integrals $I_{\nu_1},...,I_{N_{mastr}}$. The coefficients can then be determined by solving a system of linear equations originating from the symmetry relations and integration by part identities of the Feynman integral family. In principle, the system of equations can be easily solved. Although the equations can be solved in theory, the computational requirements are in practice considerable. Therefore, novel approaches are needed to efficiently determine the coefficients.  We will see that the master integrals span out a vector space and that a Feynman integral from a given family of Feynman integrals corresponds to a specific vector of this vector space. The decomposition of a Feynman integral into a linear combination of master integrals is thus nothing else than the decomposition of a vector into a linear combination of basis vectors. Finding the coefficients is particularly easy if the vector space is equipped with an inner product.\cite{Weinzierl:2022eaz}

This leads to the natural question of whether there is an inner product on the vector space of Feynman integrals. In the following section, we will see that the answer, roughly speaking, is yes. However, instead of working with the Feynman integrals themselves, we will work with their integrands.\cite{Weinzierl:2022eaz} %The difference between the two is that when working with Feynman integrals, both symmetry relations and integration by partial identities are considered, whereas the integrands, only integration by partial identities are considered. §As the reader may recall from Section $TBI/TBD$§. The number of master integrals obtained by considering both integrations by parts identities and symmetry relations is denoted by $I_{Master}$, while the number of integrations resulting solely from integration by partial identities is referred to as $I_{com}$. Since we are only concerned with the integrand of the Feynman integral in this section, $I_{com}$ is the quantity of interest.
\section{Mathematical foundations of intersection theory }\label{S2INTR1}
This section is largely based on the works of 
\cite{Gasparotto:2023cdl}
\cite{Frellesvig:2019kgj}\cite{Frellesvig:2021vem}\cite{Weinzierl:2022eaz} as well as other publications that are mentioned at the appropriate times. In this section we will introduce the reader to the concept of twisted co-homology, focusing first on the case of one-fold integrals,  i.e. the univariate case.  In \ref{Intersection1}, we will introduce the concept of the co-homology intersection number. Lastly, we will walk through an example that demonstrates how the various steps involved in determining the intersection number lead to the coefficient of the master integral associated with integration by parts relations. We will in \ref{Intersection_Theory_II} extend our analysis to the multivariate case, using the knowledge obtained from the univariate case. In the beginning, we have a general integral that has a form:
\begin{equation}
\int_{\gamma}P_1(y)^{\alpha_1}...P_m(y)^{\alpha_m}dy
\end{equation}
where $P_i(y),i\in\{1,..,m\}$ is a multivariate polynomial in $y$, $\alpha_i\in\mathbb{C}/\mathbb{Z}$ and $\gamma$ is the integration contour. That lies in $X=\mathbb{C}/\bigcup^{m}_{j=1}\mathcal{D}_j$. To get an intuition for what exactly $\mathcal{D}_j$ is, we have to recognise that, 
each polynomial equation $P_i(y)=0$ defines a hypersurface $\mathcal{D}_{j}=\{y\in\mathbb{C}:P_{j}=0\}$, such that $\mathcal{D}=\bigcup_{i=1}^m \mathcal{D}_i$ is the union of all hypersurfaces, for the mathematicians this is called a divisor. 

\begin{figure}[h]
\caption{Hyperplanes Example \cite{PhysRevLett.120.141602}}
\begin{center}
\includegraphics[width=0.66\textwidth]{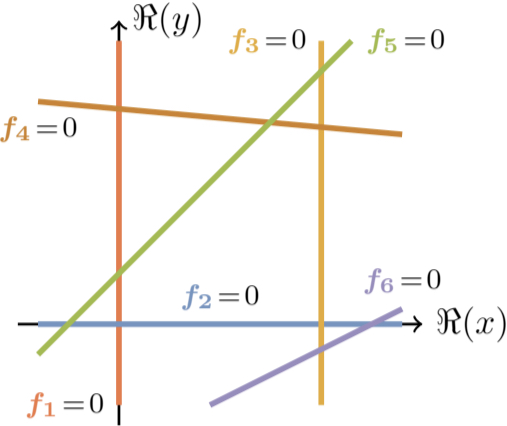}
\label{fig:hyperplanesEx}
\end{center}
\end{figure}

In order to get an intuition for what this means, let us consider an example, the complex projective 2 space $\mathbb{PC}^2$ with coordinates $(x,y)$, cut up by six hyperplanes defined through linear equations $\{f_i=0\} $\cite{PhysRevLett.120.141602}. We can easily visualise the real section of the space with the concrete choice of hyperplanes, for instance:
\begin{align*}
f_1&=x, \\
f_2&=y, \\
f_3&=1-x \\
f_4&=1-x / 4-y, \\
f_5&=1 / 4+x-y, \\
f_6&=5 / 4-x+2 y .
\end{align*}
The space of interest is now the original manifold with the hyperplanes removed:
\begin{equation}
    X=\mathbb{C P}^2 \backslash \bigcup_{i=1}^6\left\{f_i=0\right\}
\end{equation}
Graphically, this can be seen in figure \ref{fig:hyperplanesEx}.
%As mentioned before, physically the integral corresponds to our Feynman integral where each of the  $P_i(y)$ is a Baikov polynomial of the kinetic variable.

We will now introduce an example on which we will explore the most important ideas of intersection theory throughout the thesis, the Euler-Beta function:
\begin{align}
B(p, q) & =\int_0^1 d z z^{p-1}(1-z)^{q-1}, \quad \operatorname{re1}(p)>0, \text { and } \operatorname{re1}(q)>0 \\
& =\int_0^1 z^p(1-z)^q \frac{d z}{z(1-z)}\label{B11}
\end{align}
with
\begin{itemize}
    \item $P_1(z)=z$, and $\alpha_1=p-1$\\
    \item $P_2(z)=1-z$, and $\alpha_2=q-1$\\
    \item $\gamma=(0,1) \subset X=\mathbb{C} \backslash\{0,1\}$\\
\end{itemize}
From equation Eq. \eqref{B11}, it can be seen that $P_i(z)^{\alpha}$ can in general be decomposed into a multivalued function on $X$ and a part that belongs to the differential from $dz$, which form together a smooth $n$-form $\varphi$ on $X$. In our example this reads as follows $u(z)=z^p(1-z)^q$ and $\varphi(z)=\frac{dz}{z(1-z)}$. $u(z)$ corresponds to the Baikov polynomial of the Feynman integral, roughly speaking, it contains information about the number of external edges, loops, and the denominator that describes the process under investigation.
\begin{equation}
    u(z)=(\mathcal{B}(z))^{\frac{D-l-E-1}{2}}
\end{equation}
On the other hand, $\varphi$ is an $n$-form with poles at the zeros of the regulated Baikov which retains the information about the specific integral in consideration, such as powers to wish the denominators are raced, and what the numerator is:
\begin{equation}
\varphi(\mathbf{z})=\frac{\mathrm{N}(\mathbf{z})}{z_1^{a_1} \ldots z_n^{a_n}} d \mathbf{z}
\end{equation}
With $\mathrm{N}(\mathbf{z})$ being the numerator, which is a rational function, $\mathbf{z}=\{z_1,...,z_n\}$, $d\mathbf{z}=dz_1\wedge dz_2 \wedge..\wedge dz_n$ and $a_i$ are integer exponents
 $a_i\notin\mathbb{Z}$. This is to ensure the analytical continuation of the function.  
$\varphi$ is closed on $\mathbb{C}/\mathcal{D}$ \cite{Gasparotto:2023cdl} since it is a holomorphic $n$-form\label{closed}:
%\begin{equation}
   % \left(\frac{\partial}{\partial \bar{z}_j} \frac{f(\mathbf{z})}{z_1^{a_1} \ldots z_n^{a_n}}\right) d \bar{z}_j \wedge d z_n \wedge \cdots \wedge d z_1=0
%\end{equation}
%Since the derivative in the bracket ventures, and also
%\begin{equation}
%    \left(\frac{\partial}{\partial z_j} \frac{f(\mathbf{z})}{z_1^{n_1} \ldots z_m^{n_m}}\right) d z_j \wedge d z_n \wedge \cdots \wedge d z_1=0,
%\end{equation}
%since the wedge product contains $d z_j\wedge d z_j$.
In order to determine the integral of a multi-valued n-form $u\varphi$ on $\gamma$ in $\mathcal{D}$ , we have to fix a branch of $u$ on $\gamma$, this reads $\gamma\otimes u_{\gamma}$ where $u_{\gamma}$ is a fixed branch of $u$, that is integrated along $\gamma$. Then, the integral is 
\begin{equation}
    \int_{\gamma \otimes u_{\gamma}} u(z)\cdot\varphi(z):=\int_{\gamma}\left[\text { the fixed branch } u_{\gamma} \text { of } u \text { on } \gamma\right] \cdot \varphi \text {. }\label{A}
\end{equation}
As physicists, we abuse our notation and force the $\gamma$ into submission, so we simply write $u$ instead of $u_{\gamma}$. In the notation used in intersection theory, the integral Eq. \eqref{A} reads:
\begin{align}
\int_\gamma u(z) \varphi(z)&=\left\langle\varphi| \gamma \otimes u(z)\right]\\
&=\left\langle\varphi\vert\mathcal{C}\right]\label{ininter1}
\end{align}
which already contains the central notation of intersection theory.
The mathematicians among the readers may refer to $\bra{\varphi}$ as a cocycel and $\vert\mathcal{C}]$ as cycel \cite{HypergeometricI}.%, however, using these names only make sense in the context of a detailed discussion of co-homology and the homology group. Since we are in this paper more interested in the physical applications of intersection theory than in its mathematical underpinnings, we will only glanced over what exactly this means, and and only sketch out briefly the concept of co-homology group.

 For our example Eq. \eqref{B11}, with $u(t)=z^p(1-z)^q$ a branch of which is integrated integrated along a path $\gamma=(1,0)$ and the differential form $\varphi(z)=-\frac{dz}{z(1-z)}$. On which a particular branch of $u(z)$ is integrated along $\gamma$.
 In the new notation this looks as follows:
\begin{equation}
\int_0^1 z^p(1-z)^q \frac{d z}{z(1-z)}=\left\langle\frac{dz}{z(1-z)}\Bigg| (0,1) \otimes z^p(1-z)^q\right]
\end{equation}
\section{Aspects of The Twisted Co-Homology}
As the reader may recall from section \ref{IBPB} the integration by parts identity in the Baikov representation is \begin{align}
0=\int_\gamma d\left(\frac{\mathcal{B}^{\frac{D-l-E-1}{2}}}{z_1^{a_1} \ldots z_n^{a_n}} \zeta\right)
\end{align}
where $\mathcal{B}(\partial \gamma)=0, \zeta$ is an $(n-1)$ differential form $\zeta=\sum_{i=1}^n(-1)^{i+1} \zeta_i$, with $\zeta_i= \hat{\zeta} d z_1 \wedge \ldots \wedge \hat{d z_i} \wedge \ldots \wedge d z_n$ and each $\widehat{\zeta}_i$ a polynomial in the variables $\mathbf{z}=\left(z_1, \ldots, z_n\right)$ whose coefficients may depend on the kinematic invariants. If we identify the Baikov  polynomial $\mathcal{B}^{\frac{D-l-E-1}{2}}$, with the multivalued function $u(z)$ on $X$, and $\xi=\frac{\zeta}{z_1^{a_1}\ldots z_n^{a_n}}$, where different $\xi$ correspond to the different integrals with their respective integers, that are linked through IBPs.
Hence, we can simplify Eq. \eqref{B2} to 
\begin{align}
    \int_{\gamma}d(u(z)\xi)=0
\end{align}
Which reminds us of Stokes's theorem. If we exercise the differential we get: 
\begin{align}
0&=\int_\gamma d(u(z)\xi)\\ 
&=\int_\gamma(u(z)d\xi+d u(z)\wedge\xi)\\
& =\int_\gamma u(z)\left(d\xi(z)+\frac{d u(z)}{u(z)}\wedge\xi\right)
\label{GD1}
\end{align}
For our convenience, we introduce the following holomorphic one-form:
\begin{equation}
    \omega(z)=d\log u(z)=\frac{d u(z)}{u(z)}=\hat{\omega} dz \label{omegabas}
\end{equation}
where $\omega(z)$ is defined on $X$ 
the resulting operator reads 
\begin{align}
0&=\int_{\gamma}u(z)\kl{d(\xi(z))+\omega(z)\wedge\xi(z)}\label{Cov1}
\end{align}
It is indeed as the reader might suspect a connection i.e gauge derivative on $X$ and an element of the tangent vector space that is formed by the IBPs. Consequently, Eq. \eqref{GD1} can be written as:
\begin{align}
    0=\int_{\gamma} u(z) \nabla_\omega\xi\label{GD2}
\end{align}
Eq. \eqref{GD2} implies that the integral is invariant under a shift in the single-valued one form $\varphi$ by $\varphi+\nabla_{\omega}\xi$.
 \begin{align}
\int_{\gamma}u(z)\varphi=\int_{\gamma}u(z)(\varphi+\nabla_{\omega}\xi)
 \end{align}
Thus, $\varphi\to\varphi+\nabla_{\omega}\xi$ can intuitively be thought of as something like a gauge transformation on the tangent vector bundle of $X$. %at $z$ in its kinematic variables. $\xi$.
The form $\varphi$ and $\varphi+\nabla_{\omega}\xi$ describe Feynman integrals that are equivalent upon integration and are therefore equivalent through IBPs. This suggests that we can define an equivalence class of $\varphi$ and $\varphi+\nabla_{\omega}\xi$. We will do this in the following.
These ideas have first been formulated in the works of \cite{Gasparotto:2023cdl}\cite{Frellesvig:2019kgj}\cite{Frellesvig:2021vem}\cite{Weinzierl:2022eaz}.
\section{Homology and Co-Homology Group}\label{Group}
We will now briefly sketch a mathematically very sophisticated part of intersection theory. As the mathematical structure goes far beyond what can reasonably be discussed in this thesis we will outline the points relevant to us without proving or motivating them. This section is based on 
\cite{Gasparotto:2023cdl}
\cite{Mattiazzi:2022zbo}\cite{Frellesvig:2019kgj} and parts of \cite{HypergeometricI}. The group $\mathrm{H}^{n}$, which describes the group-theoretic action of the integration-by-parts relations and the equivalence class of $\varphi$ and $\varphi+d\xi$, is a quotient group of the twisted de Rham cohomology:
\begin{equation}
\mathrm{H}_\omega^n=\frac{\left\{\varphi \in \Omega^n(X) \mid \nabla_\omega \varphi=0\right\}}{\left\{\varphi \in \Omega^n(X) \mid \varphi=\nabla_\omega \xi\right\}},\label{ComGNabOm}
\end{equation}
Then the elements are $\langle\varphi|$ and capture a form that differs only by a shift through $\varphi\sim\varphi+\nabla_{\omega}$, so it follows that:
\begin{align}
\langle\varphi|\in \mathrm{H}^{k}:\varphi\sim\varphi+\nabla_{\omega}\xi
\end{align}
An integral is the pairing between an element of the twisted co-homology group and the twisted homology group
\begin{align}
\langle\bullet| \bullet]: \mathrm{H}^1\left(X, \nabla_\omega\right) \times \mathrm{H}_1(X, u) & \rightarrow \mathbb{C}\nonumber\\
(\langle\varphi|,| \gamma \otimes u(z)]) & \rightarrow\langle\varphi| \gamma \otimes u(z)]=\int_\gamma u(z) \varphi(z)\label{Hint}
\end{align}
or, more conveniently written:
\begin{equation}
\bra{\varphi}\mathcal{C}]=\int_\gamma u(z) \varphi(z)\label{Hint2}
\end{equation}
with, $\gamma\otimes u=\mathcal{C}$
while a dual integral is
\begin{align}
& {[\bullet\ket{\bullet}: \mathrm{H}_1\left(X, u^{-1}\right)\times\mathrm{H}^1\left(X,\nabla_{-\omega}\right)\rightarrow\mathbb{C}}\nonumber\\
&[\gamma^{\vee} \otimes u^{-1}(z)|,\ket{\varphi^{\vee}}) \rightarrow[\gamma^{\vee}\otimes u^{-1}(z)\ket{\varphi^{\vee}}=\int_{\gamma^{\vee}} u^{-1}(z)\varphi^{\vee}(z)\label{DHint}
\end{align}
or, more conveniently written:
\begin{equation}
[\mathcal{C}^{\vee}\ket{\varphi^{\vee}}=\int_{\gamma^{\vee}} u^{-1}(z)\varphi^{\vee}(z)\label{DHint2}
\end{equation}
Let $\left(\left\langle e_1\right|, \ldots,\left\langle e_\nu\right|\right)$ a basis for $\mathrm{H}^1\left(X, \nabla_\omega\right)\left(\left(\left|h_1\right\rangle, \ldots,\left|h_\nu\right\rangle\right)\right.$ a basis for $\left.\mathrm{H}^1\left(X, \nabla_{-\omega}\right)\right)$ and $\left(\left[\gamma_1 \otimes\right.\right.$ $u(z) \mid, \ldots,\left[\gamma_\nu \otimes u(z) \mid\right)$ a basis for $\mathrm{H}_1(X, u)\left(\left(\left[\gamma_1^{\vee} \otimes u^{-1}(z) \mid, \ldots,\left[\gamma_\nu^{\vee} \otimes u^{-1}(z) \mid\right)\right.\right.\right.$ a basis for $\left.\mathrm{H}_1\left(X, u^{-1}\right)\right)$
In the following, we will briefly discuss the dimension of the twisted co-homology group
if one rephrases the Feynman integral in terms of a pairing between cycles and cocycles the elements of the homology groups and twisted co-homology respectively, then we establish a connection between the number of master integrals and the dimensions of the co-homology and homology groups: 
\begin{equation}
\nu=\dim\kl{\mathrm{H}^{n}_{\pm\omega}}=\dim\kl{\mathrm{H}^{\pm\omega}_{n}}\label{nu}
\end{equation}
This involves a number of mathematical operations from algebraic geometry which we will skip in the thesis as it goes far beyond the mathematics that is required for obtaining the desired result. However, at the end of these mathematical operations, it was found 
for example in \cite{Lee:2013hzt} that we can compute the number of independent master integrals from the number of critical points of $\log|u|$, 
which is given by:
\begin{equation}
\hat{\omega}_i=\partial_{z_i}\logv{ u(\mathbf{z})}=0\quad i\in\{1,..,n\}\label{om1}
\end{equation}
This can be determined without explicit computation of the zeros. In our application the function $u(\mathbf{z})$ always take the form $u(\mathbf{z})=\prod_{j}\mathcal{B}_{j}^{\frac{D_j-l_j-E_j-1}{2}}(\mathbf{z})$, which gives us the following equation:
\begin{equation}
\hat{\omega}_i=\sum_j \frac{D_j-l_j-E_j-1}{2}\frac{\partial_{z_i} \mathcal{B}_j}{\mathcal{B}_j}, \quad i\in\{1,..,n\}
\end{equation}
\section{Co-Homology Intersection Number}\label{Intersection1}
We will now turn to the mathematical backbone of the thesis and give an introduction to the concept of co-homology intersection numbers, it is based on the original work of 
\cite{Gasparotto:2023cdl}\cite{Mastrolia:2018uzb} \cite{Weinzierl:2022eaz}. We introduce first a bilinear, non-degenerate pairing among the elements $\langle\varphi|\in \mathrm{H}^{1}(X,\nabla_{\omega})$ and their dual $|\varphi^\vee\rangle\in \mathrm{H}^{1}(X,\nabla_{\omega})$, where $u\to u^{-1}$ produces $\omega\to-\omega$ with $\varphi$ and $\varphi^\vee $ being holomorphic, and take in general the form 
\begin{equation}
\varphi=\hat{\varphi} d z_n \wedge \cdots \wedge d z_1, \quad \varphi^{\vee}=\hat{\varphi}^{\vee} d z_1 \wedge \cdots \wedge d z_n .\label{dfvphi1}
\end{equation}
 The pairing between twisted cocycles, and their dual counterparts plays a crucial role in deriving the master decomposition formula, and thus in translating IBP decomposition into the language of intersection theory. Similar to the pairing between cycles and cocycles defined in Eq. \eqref{Hint2}, and Eq. \eqref{Hint}, and following our intuition that into section number is roughly speaking, some kind of in a product,  we may guess
 \begin{align}
\langle\varphi|\varphi^\vee\rangle=\int_{X}\varphi(z)\wedge\varphi^{\vee}(z)\label{intX1}
\end{align}
It is referred to as the co-homology intersection number or short intersection number. The Integral Eq. \eqref{intX1} is not well-defined because the integrand can have divergent behaviour near the boundary of $X$. It consists of the set of points $\mathcal{P}_{\omega}=\{x_0,...,x_m+2\}$ excluded from $\mathbb{PC}$. Here $\mathbb{PC}$ is the projective plane, thus also includes the points at $\infty$ and $0$. Since $\varphi^{\vee}$ may have singularities at these particular points, and, the numerator of the integral Eq.\eqref{intX1} involves $dz\wedge dz = 0$, which roughly speaking leads to an indeterminacy of the form $0/0$. For solving, this problem, we require a form of regularisation of $\varphi$, which we can achieve with: $\mathcal{R}_{\omega}(\varphi)$.
\begin{align}
\langle\varphi|\varphi^\vee\rangle=\int_{X}\mathcal{R}(\varphi(z))\wedge\varphi^{\vee}(z)\label{intXr}
\end{align}
Assuming that there are no points at infinity, we can consider two small discs centred at a point $z_i$, $V_i$ and $U_i$, where the radius of $V_i$ is smaller than that of $U_i$, as illustrated in Figure \ref{figfigurebump}.
\begin{figure}[h]
\centering
\caption{holomorphic function $h_i(z, \bar{z})$ \cite{Gasparotto:2023cdl}}
\includegraphics[width=0.66\textwidth]{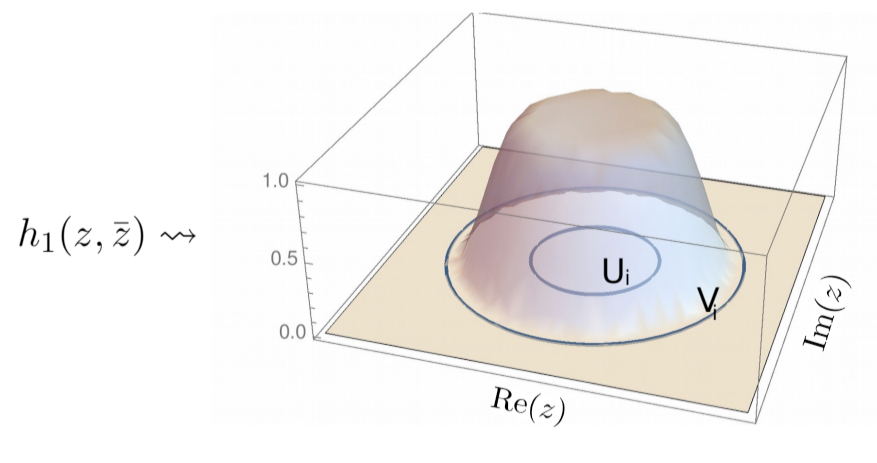}
\end{figure}\label{figfigurebump}
We also assume that the regions $U_i$ do not overlap. In order to ensure a smooth interpolation of these regions, we have to introduce a non-holomorphic function.
\begin{equation}
h_i(z, \bar{z})= \begin{cases}1 & \text { on } V_j \\ 0 \leq h_j \leq 1 & \text { on } U_j \backslash V_j \\ 0 & \text { outside } U_j\end{cases}\label{bump}     
\end{equation}
Where $h(z,\bar{z})$ is:
\begin{equation}
   h(z,\bar{z})=\hat{h}(z,\bar{z})dz\wedge d\bar{z}\label{hbase}  
\end{equation}
Since $\varphi$ and $\mathcal{R}_\omega(\varphi)$ are in the same co-homology class, they differ only by a covariant derivative:
\begin{equation}
\varphi-\mathcal{R}_{\omega}(\varphi)=\nabla_{\omega}\xi\label{dfreg}
\end{equation}
We can build a differential equation on $U_i/{z_i}$ that describes $\varphi$,
\begin{equation}
\nabla_{\omega}\psi_{j}=\varphi\label{dpsi}
\end{equation}
whose solution is $\psi$.
The differential equation implies by $\varphi\wedge\varphi=0$, that $\nabla_{\omega}\psi\wedge\phi=0$ which implies, that $\omega\wedge\varphi=0$\label{dpswpi}. As can be shown easily: 
\begin{align}
0 & =\varphi\wedge\varphi \\
& =\nabla_\omega \psi \wedge \varphi \\
& =d \psi \wedge \varphi+\omega \wedge \psi \wedge \varphi\label{expan} \\
& =d \psi \wedge \varphi-\psi \wedge \omega \wedge \varphi\label{antsym}\\
& =d \psi \wedge \varphi
\end{align}
which we will use later. Where we expanded the covariant derivative Eq. \eqref{Cov1} in line \ref{expan} and used the anti-symmetry of the wedge product in line \ref{antsym} and the fact that $\omega\wedge\varphi=0$, which comes from, the fact that $\omega=\hat{\omega} dz$ and thus $dz\wedge dz$. We set
\begin{equation}
\xi=\sum_{i=1}^{m}h_i\psi_i\label{dfxi}
\end{equation}
Using Eq. \eqref{dfreg} and Eq. \eqref{dfxi}, we find that
the regulator has the following form:
\begin{align}
\mathcal{R}_\omega\left(\varphi\right)  &=\varphi-\nabla_\omega \xi\\
&=\varphi-\sum_{i=1}^m \nabla_\omega\left(h_i \psi_{i}\right) \\
&=\varphi-\sum_{i=1}^m h_i \nabla_\omega \psi_{i}-\sum_{i=1}^m\left(\nabla_{\omega} h_i\right) \psi_{i}\\
&=\varphi-\sum_{i=1}^m h_i \nabla_\omega \psi_{i}-\sum_{i=1}^m\left(dh_i+\omega\wedge h_i\right) \psi_{i}\label{omdzhdz}\\
&=\varphi-\sum_{i=1}^m h_i\nabla_\omega \psi_{i}-\sum_{i=1}^m\left(d h_i\right) \psi_{i} \\
& =\varphi-\sum_{i=1}^m h_i \varphi-\sum_{i=1}^m\left(d h_i\right) \psi_{ i}\label{reg}
\end{align}
in line \eqref{omdzhdz} follows from  the definition of $\omega$ and $h_i$ in Eq. \eqref{omegabas} and Eq. \eqref{hbase} respectively, that that last term has $dz\wedge dz$ and thus is $0$.
We now need to distinguish between the three different regions where the intersection number is calculated on $V_i$, $h_i=0$ is, on $U_j$, $h_i=1$ is and $V_i/U_i$, $0 \leq h_i \leq 1$. First, for $V_i=0$ and $h_i=0$ follows that $\varphi\wedge\varphi^{\vee}=0$. Second, on $U_i$, $h_i=1$ and $h_i=0$ for $i\neq j$. Further we have on $U_i$ that the derivative of all functions $h_i$ vanishes $dh_i$ for all $i$, it follows that: 
\begin{equation}
\mathcal{R}_\omega\left(\varphi\right)=\varphi-h_i \varphi=\varphi-\varphi=0 .
\end{equation}
Thus we find that the intersection number Eq. \eqref{intXr} on $U_i$ is Zero.
\begin{equation}
\langle\varphi|\varphi^\vee\rangle=\int_{\bigcup_j U_j}\mathcal{R}(\varphi)\wedge\varphi^{\vee}(z)=0\label{intXr2} 
\end{equation}
This leaves us with only the region $V_i/U_i$. Let us now turn to the intersection number for the case $\mathcal{P}_{\omega}$. Playing the expression of the regulator of Eq. \eqref{reg} into Eq. \ref{intXr} we have 
\begin{align}
\left\langle\varphi \mid \varphi^{\vee}\right\rangle_\omega 
&=\int_{V_i/U_i}\mathcal{R}_\omega\left(\varphi\right)\wedge \varphi^{\vee}\\
&=\int_{V_i/U_i}\left[\varphi-\sum_{i=1}^m h_i \varphi-\sum_{i=1}^m\left(d h_i\right) \psi_{i}\right] \wedge \varphi^{\vee}\\
&=-\sum_{i=1}^m \int_{V_i/U_i}\left(d h_i\right) \psi_{i} \wedge \varphi^{\vee}
\end{align}
We can see that the first two terms evaluate to zero based on the definition of $\varphi$ as given in Eq. \eqref{dfvphi1}, which gives a $d z \wedge d z$. Only the last term yields a non-vanishing wedge product $d z \wedge d \bar{z}$. As $d h_i$ is non-zero only on ${V_i/U_i}$ we obtain:
\begin{align}
\left\langle\varphi\mid \varphi^{\vee}\right\rangle_\omega
&=-\sum_{i=1}^m \int_{V_i/U_i}\left(d h_j\right)\psi_{j}\wedge \varphi^{\vee}\\
&=-\sum_{j=1}^m \int_{V_i/U_i}\left[d\left(h_i \psi_{j} \varphi^{\vee}\right)
-h_i\left(d\psi_{i}\right) \wedge\varphi^{\vee}-h_i \psi_{i} \wedge d\varphi^{\vee}\right]\\
&=-\sum_{i=1}^m \int_{V_i/U_i} d\left(h_i \psi_{i} \varphi^{\vee}\right)
\end{align}
In the second line the last two terms vanish, $h_i\left(d\psi_{i}\right)\wedge\varphi^{\vee}$ vanishes because of the calculations done in  \ref{dpswpi}, where we found $d \psi \wedge \varphi=0$, while $h_i\psi_{i}\wedge d\varphi^{\vee}$ vanishes, because $\varphi$ is closed, as mentioned in section \ref{closed}.
We may now use Stokes theorem and obtain
\begin{align}
\left\langle\varphi \mid \varphi_R\right\rangle_\omega
&=-\sum_{i=1}^m \int_{\partial V_i/U_i}h_i\psi_{i}\varphi\\
&=-\sum_{i=1}^m \int_{\partial V_i} h_i\psi_{i} \varphi+\int_{\partial U_i}h_i\psi_{i}\varphi\\
&=-\sum_{i=1}^m \int_{\partial V_i} \psi_{i} \varphi\\
&=2\pi i\sum_{x_i\in\mathcal{P}_{\omega}}\Res_{z=x_i}\left(\psi_{i} \varphi^{\vee}\right)
\end{align}
where we used that $h_i=0$ on $\partial U_i$ and $h_i=1$ on $\partial V_i$. In the last line we understood that $\partial V_i$ is nothing else but the contour. Hence, the residue theorem applies and we got our expression for the intersection number. 
\begin{align}
\langle\bullet \mid \bullet\rangle: \mathrm{H}^1\left(X, \nabla_\omega\right) \times \mathrm{H}^1\left(X, \nabla_{-\omega}\right) & \rightarrow \mathbb{C}\nonumber \\
\left(\left\langle\varphi|,| \varphi^{\vee}\right\rangle\right) & \rightarrow\left\langle\varphi \mid \varphi^{\vee}\right\rangle=\sum_{x_i\in\mathcal{P}_{\omega}}\Res\left(\psi_i \varphi^{\vee}\right)\label{Res}
\end{align}
where $\psi_i$ is a local solution of $\nabla_\omega\psi_i=\varphi$ and $\mathcal{P}_\omega$ is the set of poles of $\omega$.
$\omega$ is, as the reader may remember, also our connection coefficient on our manifold $X$.
The intersection pairing is manifestly (bi) linear, i.e.
\begin{equation}
    \left\langle c_1 \varphi_1+c_2 \varphi_2 \mid \varphi^{\vee}\right\rangle=c_1\left\langle\varphi_1 \mid \varphi^{\vee}\right\rangle+c_2\left\langle\varphi_2 \mid \varphi^{\vee}\right\rangle
\end{equation} as well as
\begin{equation}
    \left.\left\langle\varphi \mid c_1 \varphi_1^{\vee}+c_2 \varphi_2^{\vee}\right\rangle=c_1\left\langle\varphi \mid \varphi_1^{\vee}\right\rangle+c_2\left\langle\varphi\mid\varphi_2^{\vee}\right\rangle\right)
\end{equation}
Furthermore, we stress that the intersection is a pairing between elements of $\mathrm{H}^1$ and its dual, so one can replace $\varphi$ by another representative in the same co-homology class, i.e. $\varphi \rightarrow \varphi+\nabla_\omega \xi$. This does not change the result, the same obviously is true for $\varphi^{\vee}$ when replaced by $\varphi^{\vee} \rightarrow \varphi^{\vee}+\nabla_{-\omega} \xi^{\vee}$.

In the following, we present a recursive powerseries method for solving the differential equations for $\nabla_{\omega}\psi=\phi$. After obtaining $\psi$, we can substitute it into the equation \eqref{Res} to obtain the intersection number. First, the local coordinates around each $x_i$ are chosen to be of the form $z = y - x_i$, and then a series expansion of the following form is performe:
\begin{align}
\omega&=\sum_{m=-1}^{\infty}\hat{\omega}_{m}y^{m}dy\label{os}\\
\varphi&=\sum_{m=\min}^{\infty}\hat{\varphi}y^{m}dy\label{phis}
\end{align}
The solution around each $x_i$ has then the form
\begin{equation}
\psi_i=\sum_{m=\min+1}^{\infty}\psi_{m}y^{m}\label{psis}
\end{equation}
Substituting the equations \eqref{os}, \eqref{phis} and \eqref{psis} into Eq.\eqref{dpsi}, $\nabla_{\omega}\psi_i=\varphi$, gives:
\begin{equation}
    \left(\sum_{m=\min _{\varphi}+1}^{\infty} m \psi_m y^{m-1}+\left(\frac{\hat{\omega}_{-1}}{y}+\hat{\omega}_0+\hat{\omega}_1 y+\ldots\right) \cdot \sum_{m=\min _{\varphi}+1}^{\infty} \psi_m y^m\right)=\sum_{m=\min _{\varphi}} \hat{\varphi}_m y^m
\end{equation}
Rearranging the products of two sums into a single sum gives
\begin{equation}
\sum_{m=\min_{\varphi}}^{\infty}\left((m+1) \psi_{m+1}+\sum_{q=-1}^{m-\min _{\varphi}-1} \hat{\omega}_q \psi_{m-q}\right) y^m=\sum_{m=\min _{\varphi}}^{\infty} \hat{\varphi}_m y^m\label{spd}
\end{equation}
Assuming that Eq. \eqref{spd} in $y$ holds in order by order, we obtain a linear system for the unknown coefficients of $\psi$ in Eq. \eqref{spd}.  Since the unknown coefficients obey a recursive relation, Eq. \eqref{spd} can be further simplified to
\begin{equation}
\psi_{m+1}=\frac{1}{m+1+\hat{\omega}_{-1}}\left(\hat{\varphi}_m-\sum_{q=0}^{m-\min _{\varphi}-1} \hat{\omega}_q \psi_{m-q}\right), \quad \psi_{\min _{\varphi}+1}=\frac{\hat{\varphi}_{\min _{\varphi}}}{\min _{\varphi}+1+\hat{\omega}_{-1}}\label{re1}
\end{equation}
where the coefficients on the RHS are known up to the $m$-th sept. While it is possible to calculate the coefficients in Eq.\eqref{psis} up to any order, our main interest is in the lower order terms. Since the aim of this calculation is to determine the residue  of $\psi\phi$, we are particularly interested in terms up to order $-1$. Thus, when considering the local expansion, the following applies: \begin{equation} 
\varphi=\sum_{m=\min_{\varphi^{\vee}}}^{\infty}\hat{\varphi}_{m}y^{m}dy
\end{equation}
Where we can see that it is sufficient to expand $\psi$ up to order $(-\min-1)$, which captures all meaningful contributions to the residue.
\section{Linear Relations and Master Decomposition Formula}\label{LiRe}
In this section, we are going to derive the master decomposition formula of intersection theory. For this we are going to follow the work of 
\cite{Gasparotto:2023cdl}\cite{Frellesvig:2019kgj} 
We recall from section \ref{S2INTR1}, Eq. \eqref{ininter1} the definition of the Feynman integral
\begin{equation}
    I=\int_{\gamma}u(z)\varphi(z)\label{F2}
\end{equation} 
over the path $\gamma$ in $X=\mathbb{CP}^{1}/\mathcal{P}_{\omega}$, where $\mathcal{P}_{\omega}$ is the set of poles of $\omega=\log\vert u\vert$, such that $u\kl{\partial\gamma}$. Then the integral family in Eq. \eqref{F2} can be decomposed into $\nu$ master integers $(\mathcal{I}_{1},. ...,\mathcal{I}_{\nu})$, where $\nu$ is the number of solutions of $\omega=0$, with
\begin{equation}
\mathcal{I}_i=\int_{\gamma}u(z)e_i(z), \; i\in\{0,...,\nu\}
\end{equation}
The decomposition of any given integral of the form Eq. \eqref{F2} into its basis of the master integral is then translated in the language of the co-homology group into the decomposition of $\bra{\varphi}\in\mathrm{H}^{1}(X,\nabla_{\omega})$ in terms of $\bra{e_i},...,\bra{e_{\nu}}\in\mathrm{H}^{1}(X,\nabla_{\omega})$. It follows intuitively that
\begin{equation}
I=\int_{\gamma} u \varphi=
\left\langle\varphi| u\otimes\gamma\right]=\sum_{i=1}^\nu c_i\left\langle e_i| \mathcal{\mathcal{C}}\right]=
\sum_{i=1}^\nu c_i \int_{\gamma} u e_i=\sum_{i=1}^\nu c_i \mathcal{I}_i\label{eqMIINTER}
\end{equation} 
This we will name the Feynman integral decomposition formula. 
The question now is how can we find the coefficients $c_i$ in terms of $\bra{\varphi}\in \mathrm{H}^{1}(X,\nabla_{\omega})$ in terms of $\bra{e_1},...,\bra{e_{\nu}}\in\mathrm{H}^{1}(X,\nabla_{\omega})$, and their respective duals $\ket{\varphi}\in\mathrm{H}^{1}(X,\nabla_{-\omega})$ in terms of $\kl{\ket{h_1},...,\ket{h_{\nu}}}\in\mathrm{H}^{1}(X,\nabla_{-\omega})$. For this, we write down a matrix that pairs the basis element and their respective duals, this gives the $\kl{\nu+1}\cross\kl{\nu+1}$ matrix.
\begin{equation}
    \mathbf{M}=\left(\begin{array}{cccc}\left\langle\varphi \mid \varphi^{\vee}\right\rangle & \left\langle\varphi \mid h_1\right\rangle & \ldots & \left\langle\varphi \mid h_\nu\right\rangle \\ \left\langle e_1 \mid \varphi^{\vee}\right\rangle & \left\langle e_1 \mid h_1\right\rangle & \ldots & \left\langle e_1 \mid h_\nu\right\rangle \\ \vdots & \vdots & \ddots & \vdots \\ \left\langle e_\nu \mid \varphi^{\vee}\right\rangle & \left\langle e_\nu \mid h_1\right\rangle & \ldots & \left\langle e_\nu \mid h_\nu\right\rangle\end{array}\right)=\left(\begin{array}{cc}\left\langle\varphi \mid \varphi^{\vee}\right\rangle & \mathbf{A}^{\top} \\ \mathbf{B} & \mathbf{C}\end{array}\right).
\end{equation}
As a consequence of the linearity of the intersection pairing, the matrix is degenerate, therefore we get:
\begin{equation}
0=\operatorname{det} \mathbf{M}=\operatorname{det}\left(\begin{array}{cc}\left\langle\varphi \mid \varphi^{\vee}\right\rangle & \mathbf{A}^{\top} \\ \mathbf{B} & \mathbf{C}\end{array}\right)
\end{equation}
In deriving the Baikov representation, we use the Schur matrix decomposition as before, and we get that:
\begin{equation}
0=\operatorname{det} \mathbf{M}=\operatorname{det} \mathbf{C} \cdot\det\left(\left\langle\varphi \mid \varphi^{\vee}\right\rangle-\mathbf{A}^{\top} \cdot \mathbf{C}^{-1} \cdot \mathbf{B}\right)
\end{equation}
As $\kl{\bra{e_i},...,\bra{e_{\nu}}}$ and $\kl{{\ket{h_1}},...,\ket{{h_{\nu}}}}$ are the basis elements of $H^{1}(X,\nabla_{\pm\omega})$, then it follows that $\det{C}\neq0$. Therefore, we get: 
\begin{equation}
\left\langle\varphi \mid \varphi^{\vee}\right\rangle-\mathbf{A}^{\top} \cdot \mathbf{C}^{-1} \cdot \mathbf{B}=0
\end{equation}
or explicitly written. 
\begin{equation}
\left\langle\varphi\mid\varphi^{\vee}\right\rangle=\sum_{i, j=1}^\nu\left\langle\varphi \mid h_j\right\rangle\left(\mathbf{C}^{-1}\right)_{j i}\left\langle e_i \mid \varphi^{\vee}\right\rangle\label{b1} 
\end{equation}
As Eq. \eqref{b1} holds for an arbitrary $\ket{\varphi^{\vee}}$, it can be dropped and we arrive at the master decomposition formula
\begin{equation}
\langle\varphi|=\sum_{i, j=1}^\nu\left\langle\varphi \mid h_j\right\rangle\left(\mathbf{C}^{-1}\right)_{j i}\left\langle e_i\right|\label{MDF}
\end{equation}
or, more compactly,
\begin{equation}
\langle\varphi|=\sum_{i=1}^\nu c_i\left\langle e_i\right| \quad\text{with:}\quad c_i=\sum_{j=1}^\nu\left\langle\varphi \mid h_j\right\rangle\left(\mathbf{C}^{-1}\right)_{j i}\quad\text{and}\quad\mathbf{C}_{i j}=\left\langle e_i \mid h_j\right\rangle \label{ci}
\end{equation}
Further, we have the dual master decomposition formula
\begin{equation}
\left|\varphi^{\vee}\right\rangle=\sum_{i j=1}^\nu\left|h_j\right\rangle\left(\mathbf{C}^{-1}\right)_{j i}\left\langle e_i \mid \varphi^{\vee}\right\rangle,
\end{equation}
or more compactly as
\begin{equation}
\left|\varphi^{\vee}\right\rangle=\sum_{j=1}^\nu c_j^{\vee}\left|h_j\right\rangle \quad \text{with:}\quad c_j^{\vee}=\sum_{i=1}^\nu\left(\mathbf{C}^{-1}\right)_{j i}\left\langle e_i \mid \varphi^{\vee}\right\rangle\quad\text{and}\quad \mathbf{C}_{i j}=\left\langle e_i \mid h_j\right\rangle
\end{equation}
Thus, the decomposition of $\bra{\varphi}$, is extracted in terms of the intersection number of $\braket{\varphi}{h_j} $ with the dual basis $\ket{h_i} $, normalised by the basis intersection number $\braket{e_i}{h_i}$. The intersection number is calculated via the residue of $\psi\varphi$, as given in Eq. \eqref{Res}, where $\psi$ is the solution to the differential equation $\nabla_{\omega}\psi=\varphi$, which is obtained by the recursive relation Eq. \eqref{re1}.
\section{Example I}
We will now go through an explicit example, let us return to the Euler beta function, 
\begin{equation}
    I_{1,1}=\int_{0}^{1} z^{p}(1-z)\frac{dz}{z(1-z)}\label{I11}
\end{equation}
where $I_{1,1}$ corresponds to the choice $(a_{1},a_{2})=(1,1)$ within the more general family of Feynman integrals:
\begin{equation}
    I_{a_1,a_2}=\int_{0}^{1}z^{p}(1-z)^{q}\frac{dz}{z^{a_1}(1-z)^{a_2}}, \quad (a_1,a_2)\in\mathbb{Z}^2\label{Ia1a2}
\end{equation}
The reader may realise that it is possible to relate the system of integration by parts relation to a single master integral $\mathcal{I}=I_{1,1}$. It follows that:
\begin{align}
    I_{a_1,a_2}=c_{a_1,a_2}I_{1,1}\\
    =c_{a_1,a_2}\mathcal{I}\label{Ia}
\end{align}
In this example, we focus on the case $(a_1,a_2)=(2,1)$ and solve for $c_{2,1}$ exploiting the twisted co-homology structure that underlies Eq. \eqref{I11} and Eq. \eqref{Ia1a2}. The formulation of the integral in the language of intersection theory is given by Eq. \eqref{Hint}
and Eq. \eqref{DHint}, the integral,
\begin{equation}
    I_{2,1}=\int_{0}^{1}z^{p}(1-z)^{q}\frac{dz}{z^2\kl{1-z}},\label{I21}
\end{equation}
translated into 
\begin{equation}
\Bbra{\frac{dz}{z^2(1-y)^2}}=\Bbra{\varphi}\in \mathrm{H}^{1}(X,\nabla_{\omega})
\end{equation} 
with $X=\mathbb{CP}/\{0,1,2\}$ and 
\begin{equation}
    \omega=d\logv{z^p(1-z)^q}=\kl{\frac{p}{z}-\frac{q}{1-z}}
\end{equation}
In particular, since there is only one independent principal integral, this means that $\mathrm{H}^{1}(X,\nabla_{\omega})$ is one-dimensional, which can be deduced from the fact that $\nu=1$. Following the differential form $\bra{\varphi}$ we choose the single basis element as, 
\begin{equation}
    \bra{e}=\bra{\frac{dz}{z(1-z)}}\in \mathrm{H}^{1}(X,\nabla_{\omega})
\end{equation}
Roughly speaking, $c_{1,2}$ in Eq. \eqref{Ia} can be understood as the projection of any given $\Bbra{\varphi}$ onto $\ket{h}$ the dual basis element. Thus, the coefficients can be extracted thanks to the existence of the scalar product defined between the elements of 
$\mathrm{H}^{1}$. Thus, according to equation \eqref{Ia}, we obtain the following expression for the coefficients
\begin{equation}
    c_{2,1}={\braket{\frac{dz}{z^2(1-z)}}{\frac{dz}{z(1-z)}}}\braket{\frac{dz}{z(1-z)}}{\frac{dz}{z(1-z)}}^{-1}\label{c12A}
\end{equation}
where the denominator is to be understood as the normalisation factor. In order to calculate the intersection number, we use the residue formula as derived in Section \ref{Intersection1}, Eq. \eqref{Res}. The reader may remember that for functions with pole $z_0$, the residue is the coefficient $a_{-1}$ of the Laurent series of the function at $z_0$. 
\begin{equation}
    f(z)=\sum_{n\in\mathbb{Z}}a_{n}(z-z_0)
\end{equation}
For this, we need to know the series expansion of $\psi\varphi$ and $\psi_i h$ up to order $-1$, but to do this, we first need to find the solution of the differential equation
\begin{equation}
   \nabla_{\omega}\psi=\frac{dz}{z^2(1-z)}
\end{equation}
as well as for the differential equation for the base element 
\begin{equation}
   \nabla_{\omega}\Xi=\frac{dz}{z(1-z)}
\end{equation}
around $\mathcal{P}=\{0,1,\infty\}$, which we will obtain via the recursive relation given by Eq. \eqref{re1}. So let us begin.
\paragraph{series solution around  $0$, for local coordinates $y=z$}
The series expansion of $\varphi=\frac{dz}{z^2(1-z)}$ around $0$ at local coordinates $y=z$ and $dy=dz$ has the form: 
\begin{align*}
    \frac{dy}{y^2}+\frac{dy}{y}+dy+ydy+\mathcal{O}\left(y^2\right)
\end{align*}
The coefficients are $\hat{\varphi}=1,\;\hat{\varphi}_{-1}=1,\; \hat{\varphi}_{0}=1,\;\hat{\varphi}_{1}=1$. The series expansion of $e_0=\frac{dz}{z(1-z)}$ around $0$ for the local coordinates $y=z$ and $dy=dz$ has the form:
\begin{equation*}
\frac{dy}{y}+dy+ydy+\mathcal{O}\left(y^2\right)  \end{equation*}
The coefficients are $\hat{e}_{-1}=1,\;\hat{e}_{0}=1,\;\hat{e}_{1}=1$. The series expansion of $\omega=\kl{\frac{p}{z}-\frac{q}{1-z}}dz$ around $0$ at local coordinates $y=z$ and $dy=dz$ has the form,
\begin{equation*}
\frac{pdy}{y}-qdy-q y dy+\mathcal{O}\left(y^2\right)
\end{equation*} 
The coefficients are
$\hat{\omega}_{-1}=p,\;\hat{\omega}_{0}=-q,\;\hat{\omega}_{1}=-q$. Thus follows by Eq. \eqref{re1},
and for $\min_\varphi=-2$,  
\begin{align*}
\psi_{-1}(y,0)&=\frac{1}{(1-p)y}
\end{align*}
For $m=-1$ follows by the recursive relation Eq. \eqref{re1}, that:
\begin{align*}
\psi_{0}(y,0)&=\frac{1}{-1+1+p}\kl{\hat{\varphi}_{-1}+\sum_{q=0}^{-1+2-1}\hat{\omega}_{q}\psi_{q-1}}\\
&=\frac{1}{-1+1+p}\kl{\hat{\varphi}_{-1}+\hat{\omega}_{0}\psi_{-1}}\\
&=\frac{1}{p}\kl{1-\frac{q}{1-p}}\\ 
&=\frac{1-q-p}{p(1-p)} 
\end{align*}
If we now combine both expressions together, we get the following series expansion for $\psi(y,0)$ we get:
\begin{equation*}
\psi(z,0)=\frac{1}{(1-p)y}+\frac{1-q-p}{p(1-p)} +\mathcal{O}(y)
\end{equation*}
The series expansion of $\psi(y,0) h_0$ around $0$ reads than: 
\begin{align}
\psi(y,0) h_0&=\kl{\frac{1}{(1-p)y}+\frac{1-q-p}{p(1-p)}}\kl{\frac{1}{z}+1}\nonumber\\
&=\frac{1}{(1-p)y^2}+\frac{1-2p-q}{p(1-p)y}+\frac{1-q-p}{p(1-p)}\label{PSER0}
\end{align} 
Thus, the residue is 
\begin{equation*}
\Res_{y_0=0}\kl{\psi(y,0) h_0}=\frac{1-2p-q}{(1-p)p}
\end{equation*}
We will now go through the same procedure as before, of obtaining a series expansion for $\Xi$. Thus follows by Eq.\eqref{re1}, and for $\min=-1$,  
\begin{equation*}
\Xi_{0}(y,0)=\frac{1}{p}
\end{equation*}
For $m=0$ follows:
\begin{align*}
\Xi_{1}(y,0)&=\frac{1}{1+p}\kl{1+\omega_0\psi_0}\\   
&=\frac{1}{1+p}\kl{1+\frac{q}{p}}y  
\end{align*}
If we now put both expressions together, we get the following series expansion for $\Xi(z,0)$:
\begin{align*}
\Xi(z,0)=\frac{1}{p}+\frac{1}{1+p}\kl{1+\frac{q}{p}}y+\mathcal{O}(y)
\end{align*}
The series expansion of $\Xi(y,0) h_0$ around $0$ reads: 
\begin{align*}
    \Xi(y,0) h_0&=\kl{\frac{1}{p}+\frac{1}{1+p}\kl{1+\frac{q}{p}}y}\kl{\frac{1}{y}+1}\\
    &=\frac{1}{py}+\frac{1}{p}+\frac{1}{1+p}\kl{1+\frac{q}{p}}+\frac{1}{1+p}\kl{1+\frac{q}{p}}y
\end{align*}
Thus, the residue is:
\begin{equation}
    \Res_{y_0=0}\kl{\Xi(y,0) h_0}=\frac{1}{p}\label{XSER0}
\end{equation}
\paragraph{series solution around $1$, for local coordinates $y=z-1$}
The series expansion of $\varphi=\frac{dz}{z^2(1-z)}$ around $1$ at local coordinates $y=z-1$ and $dy=dz$ has the form
\begin{align*}
   -\frac{dy}{y}+2dy-3ydy+\mathcal{O}\left(y^2\right)
\end{align*}
with the coefficients $\hat{\varphi}_{-1}=-1,\;\hat{\varphi}_{0}=2,\;\hat{\varphi}_{1}=3$
The series expansion of $e_1=\frac{dz}{z(1-z)}$ around $1$ at local coordinates $y=z-1$ and $dy=dz$ has the form
\begin{equation*}
    -\frac{dy}{y}+1dy-ydy+\mathcal{O}\left(y^2\right)
\end{equation*}
with the coefficients $\hat{e}_{-1}=1,\;\hat{e}_{0}=1,\;\hat{e}_{1}=-1$. The series expansion of $\omega=\kl{\frac{p}{z}-\frac{q}{1-z}}dz$ around $1$ at local coordinates $y=z-1$ and $dy=dz$ has the form,
\begin{align*}
  \frac{q}{y}dy+pdy-pydy+\mathcal{O}\left(y^2\right)
\end{align*}
with the coefficients $\hat{\omega}_{-1}=q,\;\hat{\omega}_0=p,\;\hat{\omega}_{1}=-p$. Thus follows by Eq.\eqref{re1},
and for $\min=-1$ follows:
\begin{align*}
\psi_{0}(y,1)&=-\frac{1}{q}
\end{align*}
As we are only interested in the terms up to order $0$, the full series expansion of $\psi(y,1)$ reads:
\begin{equation*}
\psi(y,1)=-\frac{1}{q}+\mathcal{O}(y)
\end{equation*}
The series expansion of $\psi(y,1)h_1$ around $1$ reads: 
\begin{align*}
\psi(y,1)h_1=\frac{1}{qy}+\frac{1}{q}
\end{align*}
Thus, the residue is 
\begin{equation}
     \Res_{y_1=1}\kl{\psi(y,1) h_1}=\frac{1}{q}\label{PSER1}
\end{equation}
We will now go through the same procedure as before, of obtaining a series expansion for $\Xi(y,1)e$. Thus follows by Eq. \eqref{re1}, and for $\min=-1$,  
\begin{equation*}
    \Xi_{0}(y,1)=\frac{1}{q}
\end{equation*}
For $m=0$ follows:
\begin{align*}
\Xi_{1}(y,1)&=\frac{1}{1+q}\kl{1+\omega_0\psi_0}\\   
&=\frac{1}{1+q}\kl{1-\frac{p}{q}}y  
\end{align*}
If we now put both expressions together, we get the following series expansion for $\Xi(y,1)$ we get
\begin{align*}
    \Xi(y,1)=\frac{1}{q}+\frac{1}{1+q}\kl{1-\frac{p}{q}}y+\mathcal{O}(y)
\end{align*}
The series expansion of $\Xi(y,1) h_1$ around $1$ reads: 
\begin{align*}
    \Xi(y,1) h_1&=\kl{\frac{1}{q}+\frac{1}{1+q}\kl{1+\frac{p}{q}}y}\kl{\frac{1}{y}+1}\\
    &=\frac{1}{qy}+\frac{1}{q}+\frac{1}{1+q}\kl{1+\frac{p}{q}}+\frac{1}{1+q}\kl{1+\frac{p}{q}}y
\end{align*}
Thus, the residue is:
\begin{equation}
    \Res_{y_0=0}\kl{\Xi(y,1) h_1}=\frac{1}{q}\label{XSER1}
\end{equation}
\paragraph{Series solution around at $x_\infty=\infty$, for local coordinates $y=\frac{1}{z}$}
The series expansion of $\varphi=\frac{dz}{z^2(1-z)}$ around $\infty$ at local coordinates $y=\frac{1}{z}$ and $dy=\frac{dz}{z^2}$, has the form
\begin{align*}
  \mathcal{O}\left(y^2\right)
\end{align*}
The series expansion of $e_1=\frac{dz}{z(1-z)}$ around $\infty$ at local coordinates $y=\frac{1}{z}$ and $dy=\frac{dz}{z^2}$ has the form
\begin{equation*}
    \mathcal{O}\left(y^2\right)
\end{equation*}
The series expansion of $\omega=\kl{\frac{p}{z}-\frac{q}{1-z}}dz$ around $\infty$ at local coordinates $y=\frac{1}{z}$ and  $dz=\frac{dy}{y^2}$ has the form
\begin{align*}
\kl{p+q}ydy+\mathcal{O}\left(y^2\right)
\end{align*}
with the coefficients $\hat{\omega}_{-1}=\kl{q+p}y$. Thus follows by Eq. \eqref{re1},
and for $\min=-2$ follows:
\begin{align*}
\psi_{-1}(y,\infty)&=\mathcal{O}\left(y^2\right)
\end{align*}
The full series expansion of $\psi(y,\infty)$ reads:
\begin{equation*}
\psi(y,\infty)=\mathcal{O}(y^2)
\end{equation*}
The series expansion of $\psi(y,\infty) h_{\infty}$ around $\infty$ reads then: 
\begin{align*}
\psi(y,1)h_{\infty}=\mathcal{O}\left(y^2\right)
\end{align*}
Therefore, there is no residue on the pole at $\infty$. Hence the pole does not contribute to the intersection number.
We will now go through the same procedure as before, of obtaining a serious expansion for $\Xi(y,\infty)$. Thus follows by Eq. \eqref{re1}, and for $\min=-1$,  
\begin{equation*}
    \Xi_{0}(y,\infty)=\mathcal{O}\left(y^2\right)
\end{equation*}
For $m=0$ follows:
\begin{align*}
\Xi_{1}(y,\infty)&=\mathcal{O}\left(y^2\right)
\end{align*}
If we now put both expressions together, we get the following series expansion for $\Xi(y,\infty)$ we get
\begin{align*}
\Xi(y,\infty)=\mathcal{O}\left(y^2\right)
\end{align*}
The series expansion of $\Xi(y,\infty) e_1$ around $\infty$ reads: 
\begin{align*}
    \Xi(y,\infty) h_{\infty}&=\mathcal{O}\left(y^2\right)
\end{align*}
Therefore, there is no residue at the pole at $\infty$. Consequently, the pole does not contribute to the intersection number. According to Eq. \eqref{Res} and Eq. \eqref{c12A}, since the pole does not contribute at $\infty$, we combine the expressions Eq. \eqref{PSER0} and Eq. \eqref{PSER1} to obtain the following expression.
\begin{equation*}
    \braket{\frac{dz}{z^2(1-z)}}{\frac{dz}{z\kl{1-z}}}=\frac{1-2p-q}{\kl{1-p}}+\frac{1}{q}
\end{equation*}
as well as Eq.\eqref{XSER0} and Eq.\eqref{XSER1}
\begin{equation*}
    \braket{\frac{dz}{z\kl{1-z}}}{\frac{z}{z\kl{1-z}}}=\frac{p+q}{pq}
\end{equation*}
Therefore follows by \ref{ci} for the intersection number:
\begin{equation*}
    c_{2,1}=\braket{\frac{dz}{z^2\kl{1-z}}}{\frac{dz}{z\kl{1-z}}}\braket{\frac{dz}{z\kl{1-z}}}{\frac{dz}{z\kl{1-z}}}^{-1}=\frac{1-p-q}{1-p}
\end{equation*}
\section{Example II}
Let us consider now the 4-loop vacuum diagram graph with the following denominators where:
\begin{center}
$$
\begin{array}{lll}D_1=k_1^2-1, & D_2=k_2^2-1, & D_3=k_3^2-1, \\ D_4=\left(k_1-k_2\right)^2-1, & D_5=\left(k_1-k_3\right)^2-1, & D_6=\left(k_2-k_3\right)^2-1, \\ D_7=\left(k_1-k_4\right)^2-1, & D_8=\left(k_2-k_4\right)^2-1, & D_9=\left(k_3-k_4\right)^2-1 ;
\end{array}
$$
\end{center}
the irreducible scaler product is chosen as:
\begin{equation}
    z=D_{10}=k_4^2
\end{equation}
The multi-valued function that corresponds to the Baikov Polynomial and the associated twist reads
\begin{equation}
    u(z)=\left(\frac{z}{2}-\frac{3 z^2}{16}\right)^{\frac{d-5}{2}}, \quad \omega(z)=\frac{(d-5)(3 z-4)}{z(3 z-8)} d z
\end{equation}
which was be obtained by following the algorithm laid out in section \ref{SECBR}.
The equation yields for $\omega=0$, Eq. \eqref{om1} has one solution, implying that there is $1$ master integral on the maximum cut:
\begin{equation}
\nu=1, \quad \mathcal{P}_\omega=\left\{0, \frac{8}{3}, \infty\right\}.
\end{equation}
Let us choose the master integral and the corresponding differential form as
\begin{equation}
\mathcal{J}=I_{1,1,1,1,1,1,1,1,1 ; 0} \rightsquigarrow\langle e|=\langle 1 d z|
\end{equation}
and 
\begin{equation}
I_{1,1,1,1,1,1,1,1,1 ;-1} \rightsquigarrow\langle\varphi|=\langle z d z|
\end{equation}
For the dual basis we choose:
\begin{equation}
|h\rangle=\left|h_1\right\rangle=|1 d z\rangle .
\end{equation}
Following Eq. \eqref{ci}, we will now derive expressions for $\braket{\varphi}{h_1}$ and $C_1=\braket{e}{h_1}$, to obtain an expression for 
\begin{equation}
 \langle\varphi|=\langle\varphi \mid h_1\rangle \mathbf{C}_1^{-1}\langle e|
\end{equation}
\paragraph{series solution around  $0$, for local coordinates $y=z$}
The series expansion of $\varphi=zdz$ around $0$ at local coordinates $y=z$ and $dy=dz$ has the form $\varphi=ydy$. The series expansion is: 
\begin{align*}
ydy+\mathcal{O}\left(y^2\right)
\end{align*}
The coefficients are $\hat{\varphi}_{-2}=0,\;\hat{\varphi}_{-1}=0,\; \hat{\varphi}_{0}=0,\;\hat{\varphi}_{1}=1$. The series expansion of $e_0=1dz$ around $0$ for the local coordinates $y=z$, has the form, $1$. The coefficient is then $\hat{e}_{0}=1$,
the series expansion of $\omega$ around $0$ at local coordinates $y=z$ and $dy=dz$ has the form,
\begin{equation*}
\frac{(d-5)dy}{2 y}-\frac{3 (d-5)dy}{16}-\frac{9}{128} (d-5) ydy-\frac{27 (d-5) y^2dy}{1024}+\mathcal{O}\left(y^3\right)
\end{equation*} 
The coefficients are
$\hat{\omega}_{-1}=\frac{d-5}{2},\;\hat{\omega}_{0}=-\frac{3 (d-5)}{16},\;\hat{\omega}_{1}=-\frac{9}{128} (d-5),\;\omega_{2}=-\frac{27 (d-5) y^2}{1024}$. From $\varphi=zdz$ follows that $\min=1$ for recursive relation \eqref{re1}. 
Hence follows that:
\begin{equation*}
    \psi_2=\frac{2}{d-1}
\end{equation*}
For $m=1$ follows by the recursive relation \eqref{re1}:
\begin{align*}
    \psi_3&=\frac{2}{6+d-5}\omega_0\psi_2\\
    &=\frac{2}{1+d}\frac{4}{3}(d-5)\frac{2}{d-1}\\
    &=\frac{16(d-5)}{3(d+1)(d-1)}
\end{align*}
As $e_0=1$, it follows by \eqref{Res} that the pole at $0$ has no residue.  
\begin{equation*}
     \Res_{y_0=0}\kl{\psi(y,0) h_0}=0
\end{equation*}
For $\Xi(z,0)e_0^{\vee}$ follows by a similar calculation that $\Res_{y_0}\kl{\Xi(z,0)e_0^{\vee}}=0$.
\paragraph{series solution around $\frac{8}{3}$, for local coordinates $y=z-\frac{8}{3}$} The series expansion of $\varphi=zdz$ around $\frac{8}{3}$ at local coordinates $y=z-\frac{8}{3}$ and $dy=dz$ has the form $\varphi=ydy$. The series expansion has the form: 
\begin{align*}
ydy+\mathcal{O}\kl{y^2}
\end{align*}
The coefficients are $\hat{\varphi}_{-2}=0,\;\hat{\varphi}_{-1}=0,\; \hat{\varphi}_{0}=0,\;\hat{\varphi}_{1}=1$. The series expansion of $e_{8/3}=1dz$ around $\frac{8}{3}$ for the local coordinates $y=z-\frac{8}{3}$ has the form $1dy$
The coefficients are $\hat{e}_{0}=1$. The series expansion of $\omega$ around $\frac{8}{3}$ at local coordinates $y=z-\frac{8}{3}$ and $dy=dz$ has the form
\begin{equation*}
\frac{(d-5)dy}{2 y}-\frac{3 (d-5)dy}{16}-\frac{9}{128} (d-5) ydy-\frac{27 (d-5) y^2dy}{1024}+\mathcal{O}\left(y^3\right)
\end{equation*} 
The coefficients are
$\hat{\omega}_{-1}=\frac{d-5}{2},\;\hat{\omega}_{0}=-\frac{3 (d-5)}{16},\;\hat{\omega}_{1}=-\frac{9}{128} (d-5)$. From $\varphi=zdz$ follows that $\min=1$ for recursive relation \eqref{re1}. 
Hence follows that:
\begin{equation*}
    \psi_2=\frac{2}{d-1}
\end{equation*}
For $m=1$ follows by the recursive relation \eqref{re1}:
\begin{align*}
    \psi_3&=\frac{2}{6+d-5}\omega_0\psi_2\\
    &=\frac{2}{1+d}\frac{4}{3}(d-5)\frac{2}{d-1}\\
    &=\frac{16(d-5)}{3(d+1)(d-1)}
\end{align*}
As $e_{8/3}=1dz$, follows by\eqref{Res} that the pole at $\frac{8}{3}$ has no residue.  
\begin{equation*}
     \Res_{y_{8/3}=8/3}\kl{\psi(y,8/3) h_{8/3}}=0
\end{equation*}
For $\Xi(z,8/3)h_{8/3}$ follows by a similar calculation that $\Res_{y_{8/3}}\kl{\Xi(z,8/3)e_{8/3}^{\vee}}=0$
\paragraph{series solution around $\infty$, for local coordinates $y=\frac{1}{z}$}
$\varphi=zdz$ at local coordinates $y=\frac{1}{z}$ and $dz=\frac{dy}{y^2}$, has the following form: 
\begin{align*}
\frac{-dy}{y^3}+\mathcal{O}\left(\frac{1}{y^2}\right)
\end{align*}
The series expansion of the coefficients are $\hat{\varphi}_{1}=-1$. The series expansion of $e_\infty dz=\frac{dz}{z^2}$ around $\infty$ for the local coordinates $y=\frac{1}{z}$ and $dz=\frac{dy}{y^2}$ has the form$e_{\infty}dy=\frac{dy}{y^2}$. The coefficients are $\hat{e}_{0}=1$. 

$\omega$ around $\infty$ at local coordinates $y=\frac{1}{z}$ and $dz=\frac{dy}{y^2}$ has the form
\begin{equation*}
    \omega=\frac{(d-5)(3-4 y)}{y(8 y-3)}dy
\end{equation*}
Thus reads the series expansion of $\omega$
\begin{equation*}
    \frac{(5-d)dy}{y}-\frac{4 (d-5)dy}{3}-\frac{32}{9} (d-5) y dy-\frac{256}{27} (d-5) y^2 dy-\frac{2048}{81} (d-5) y^3dy+O\left(y^4\right)
\end{equation*}
The coefficients are
$\hat{\omega}_{-1}=5-d,\;\hat{\omega}_{0}=-\frac{4 (d-5)}{3},\;\hat{\omega}_{1}=-\frac{32}{9}(d-5),\;\hat{\omega}_{2}=-\frac{256}{27} (d-5),\;\hat{\omega}_{3}=-\frac{2048}{81} (d-5)$. From $\varphi=\frac{dy}{y^3}$ follows that $\min=-3$ for recursive relation \eqref{re1}. 
Hence follows that:
\begin{align*}
\psi_{-2}&=\frac{-1}{-2-d+5} \\
\psi_{-2}&=\frac{-1}{3-d}
\end{align*}
For $m=-2$ follows by the recursive relation \eqref{re1}, 
\begin{align*}
\psi_{-1} & =\frac{1}{-1-\omega_{-1}} \omega_0 \psi_{-2} \\
\psi_{-1} & =\frac{1}{-1+5-d}\left(\frac{4(d-5)}{3(3-d)}\right)\\
& =\frac{4(d-5)}{3(d-3)(d-4)}
\end{align*}
For $m=-1$ follows by the recursive relation \eqref{re1},
\begin{align*}
\psi_0&=\frac{1}{\omega_{-1}}\left(\sum_{q=0}^1 w_q \psi_{-1-q}\right) \\
&=\frac{1}{d-5}\left(\omega_0 \psi_{-1}+\omega_1 \psi_{-2}\right) \\
&=\frac{1}{(d-5)}\left(-\frac{4}{3}(d-5)\frac{4(d-5)}{3(4-d)(3-d)}+\frac{32}{9}(d-5)\frac{1}{3-d}\right)\\
&=\frac{-16(d-5)}{9(4-d)(3-d)}-\frac{32}{9(3-d)} \\
&=\frac{-16(d-5)+32(4-d)}{9(4-d)(3-d)} \\
&=\frac{16(d-3)}{9(d-4)(d-3)} \\
&=\frac{16}{9(d-4)}
\end{align*}
For $m=0$ follows by the recursive relation \eqref{re1},
\begin{align*}
\psi_1&=\frac{1}{1+5-d} \sum_{q=0}^2 \omega_q \psi_{m-q} \\
& =\frac{1}{(6-d)}\left(\omega_0 \psi_0+\omega_1 \psi_{-1}+\omega_2 \psi_{-2}\right) \\
& =\frac{1}{(6-d)}\left(-\frac{4}{3}(d-5) \frac{16}{9(4-d)}-\frac{32}{9}(d-5)\frac{4(d-5)}{3(4-d)(3-d)}-\frac{256}{27}(d-5) \frac{1}{3-d}\right) \\
& =\frac{1}{(6-d)}\left(-\frac{64}{27} \frac{(d-5)}{(4-d)}-\frac{128(d-5)(d-5)}{27(4-d)(3-d)}-\frac{256(d-5)}{27(3-d)}\right) \\
& =\frac{1}{(6-d)}\frac{1}{27(4-d)(3-d)} \left(-64(d-5)(d-3)-128(d-5)(d-5)-256(d-5)(d-4)\right) \\
& =\frac{64(d-5)}{27(d-6)(d-4)} \\
\end{align*}
Thus follows, as a reader may already suspect, $\psi(y,\infty)$,
\begin{align*}
    \psi(y,\infty)&=\frac{1}{3-d} y^{-2}-\frac{4(d-5)}{3(4-d)(3-d)} y^{-1}+\frac{16}{9(4-d)} +\frac{64(5-d)y}{27(d-6)(d-4)}+\mathcal{O}(y^2)
\end{align*}
 and therefore the residue is
\begin{equation*}
     \Res_{y_0=0}\kl{\psi(y,0) h_\infty}=\frac{64(5-d) y}{27(6-d)(d-4)}
\end{equation*}
where $e_{\infty}=\frac{dy}{y^2}$.
Therefore follows by \ref{Res} for the intersection number:
\begin{align}
\left\langle\varphi \mid h_1\right\rangle&=\langle z d z \mid 1 d z\rangle \nonumber\\
&=\Res_{y_\infty=\infty}\kl{\psi(y,0) h_{\infty}}\nonumber\\
&=\frac{64(d-5)}{27(d-6)(d-4)}\label{64}
\end{align}
In the following we will perform a similar calculation for $\Xi(z,0)h_\infty$. From $\varphi=\frac{dy}{y^2}$ follows that $\min=-2$ for recursive relation \eqref{re1}. 
Hence follows that: 
\begin{align*}
\Xi_{-1} & =\frac{-1}{-2+1+\omega_{-1}}\\
& =\frac{-1}{-1+5-d}\\
& =\frac{-1}{4-d}
\end{align*}
For $m=-1$ follows by the recursive relation:
\begin{align*}
\Xi_0 & =\frac{1}{-1+1+\omega_{-1}}\left(\sum_{q=0}^0 \omega_q \psi_{m-q}\right) \\
& =\frac{1}{\omega_{-1}} \omega_0 \psi_{-1} \\
& =\frac{-1}{5-d}\frac{4}{3}(d-5) \frac{-1}{4-d} \\
& =\frac{4}{3(d-4)}
\end{align*}
For $m=0$ follows by the recursive relation:
\begin{align*}
\Xi_1&=\frac{1}{(1+5-d)} \sum_{q=0}^1 \omega_q \psi_{d-q} \\
&=\frac{1}{(6-d)}\left(\omega_0 \psi_0+\omega_1 \psi_{-1}\right) \\
&=\frac{1}{(6-d)}(-\frac{4}{3}(d-5) \frac{4}{3(d-4)}+\frac{32}{9}(d-5) \frac{1}{4-d} \\
&=\frac{16(d-5)}{9(d-6)(d-4)}
\end{align*}
So it follows that:
\begin{equation*}
\Xi\kl{y,\infty}=\frac{1}{(4-d)y}-\frac{4}{3(4-d)}+\frac{16(d-5)y}{9(d-6)(d-4)}
\end{equation*}
The residue is then 
\begin{equation*}
\Res_{y_\infty}\kl{\Xi(z,0)h_\infty}=\frac{16(d-5)}{9(d-6)(d-4)}
\end{equation*}
Therefore, follows by \ref{Res} for the intersection number:
\begin{align}
    C_1&=\langle e \mid h\rangle\nonumber\\
    &=\langle 1 d z \mid 1 d z\rangle\nonumber\\
    &=\Res_{y_\infty}\kl{\Xi(z,0)e_\infty^{\vee}}\nonumber\\
    &=\frac{16(d-5)}{9(d-6)(d-4)}\label{C_1}
\end{align}
Combining \eqref{64} and \eqref{C_1}
in equation \eqref{ci} gives:
\begin{align*}
\bra{\varphi}&=\braket{\varphi}{h}C_1^{-1}\bra{e}\\
&=\frac{4}{3}\bra{e}
\end{align*}
This implies that:
\begin{equation*}
    I_{1,1,1,1,1,1,1,1,1 ;-1}=\frac{4}{3} \mathcal{J}
\end{equation*}
To give an example for base independence, we will now go through a similar process for the dual basis $
\ket{h}=\ket{zdz}$. As $\varphi=zdz$, the intersection number is $\braket{zdz}{zdz}$, thus following equation\eqref{Res}, we need to go up to 3rd-order to find the residue for $\infty$.

The local coordinates are stil $y=\frac{1}{z}$ and  $dz=\frac{dy}{y^2}$ s.t $\varphi=\frac{-1}{y^3}dy$: 
\begin{align*}
\frac{-dz}{y^3}+\mathcal{O}\left(\frac{1}{y^2}\right)
\end{align*}
The coefficients are $\hat{\varphi}_{-2}=-1,\;\hat{\varphi}_{-1}=0,\; \hat{\varphi}_{0}=0,\;\hat{\varphi}_{1}=0$. The series expansion of $e_\infty dz=\frac{dz}{z^2}$ around $\infty$ for the local coordinates $y=\frac{1}{z}$ and  $dz=\frac{dy}{y^2}$ has the form $e_{\infty}dy=\frac{dy}{y^2}$. The coefficients are $\hat{e}_{-2}=0,\;\hat{e}_{-1}=-1,\;\hat{e}_{0}=0,\;\hat{e}_{1}=0$. 
The series expansion of $h_\infty dz=\frac{dy}{y^3}$ around $\infty$ for the local coordinates $y=\frac{1}{y}$ and  $dz=\frac{dy}{y^2}$ has the form, 
\begin{equation*}
    h_{\infty}dy=\frac{-dy}{y^3}+\mathcal{O}(y^{-1})
\end{equation*}
The coefficients are $\hat{e}_{-2}=-1,\;\hat{e}_{-1}=0,\;\hat{e}_{0}=-1,\;\hat{e}_{1}=0$.
Thus follows by the calculation above and our old familiar recursive relation \eqref{re1} for $\psi_2$:
\begin{align*}
\psi_2&=\frac{1}{3+\omega_{-1}} \sum_{q=0}^3\omega_q \psi_{m-q} \\
&=\frac{1}{(8-d)} \\
&\left(\omega_0 \psi_1+\omega_1 \psi_0+\omega_2 \psi_{-1}+\omega_3 \psi_{-2}\right) \\
&=\frac{1}{(8-d)}\Biggl(-\frac{4}{3}(d-5) \frac{64(5-d)}{27(6-d)(d-4)}-\frac{32}{9}(d-5) \frac{16}{9(d-4)}\\
&\;-\frac{256(d-5)}{27}\left(\frac{4(d-5)}{3(d-4)(d-3)}\right)+\frac{2048}{81}(d-5) \frac{1}{(d-3)}\Biggl) \\
&=\frac{256(d-5)}{81(d-6)(d-4)}
\end{align*}
Thus follows:
\begin{align*}
    &\psi(y,\infty)\\
    &=\frac{1}{3-d}y^{-2}-\frac{4(d-5)}{3(4-d)(3-d)}y^{-1}+\frac{16}{9(4-d)}\\
    &+\frac{64(5-d)y}{27(d-6)(d-4)}+\frac{256(d-5)y^2}{81(d-6)(d-4)}+\mathcal{O}(z^3)
\end{align*}
The residue is then 
\begin{equation*}
\Res_{y_\infty}\kl{\psi(z,0)e_\infty^{\vee}}=\frac{16(d-5)}{9(d-6)(d-4)}
\end{equation*}
Therefore, follows by \ref{Res} for the intersection number:
\begin{align}
    \left\langle\varphi \mid h_2\right\rangle&=\langle z d z \mid z d z\rangle\nonumber\\    &=\Res_{y_\infty}\kl{\psi(z,0)h_\infty^{\vee}}\nonumber\\
    &=\frac{256(d-5)}{81(d-6)(d-4)}\label{A256}
\end{align}
For the intersection number of $\braket{\varphi}{h_2}$, we can obviously use \eqref{64} and get the following:
\begin{align}
    C_2&=\left\langle e\mid h_2\right\rangle\nonumber\\
    &=\langle 1 d z \mid z d z\rangle\nonumber\\
    &=\Res_{y_\infty}\kl{\psi(z,0)h_\infty^{\vee}}\nonumber\\
   &=\frac{64(d-5)}{27(d-6)(d-4)}\label{C2}
\end{align}
Combining \eqref{A256} and \eqref{C2}
in equation \eqref{ci} gives:
\begin{align*}
\langle\varphi| & =\left\langle\varphi \mid h_2\right\rangle \mathbf{C}_2^{-1}\langle e| \\
& =\frac{4}{3}\langle e| .
\end{align*}
\newpage

\graphicspath{{./images/}}
\chapter{Intersection Theory II, the Multivariate Case}\label{Intersection_Theory_II}
In this chapter, we will discuss the multivariate intersection number as first introduced in \cite{Mizera:2019gea}, and its mathematical foundations and give an example of how to use the recursive relation. The algorithm has been successfully applied in the context of Feynman integrals as well as for hypergeometric functions in \cite{Frellesvig:2019kgj}\cite{Frellesvig:2019uqt}.
Furthermore, the section is built to large extent on work by 
\cite{Gasparotto:2023cdl}\cite{Mattiazzi:2022zbo}\cite{Weinzierl:2022eaz}. 
The recursive algorithm expresses the $n$-variable intersection number in terms of the next lower layer down the $(n-1)$-variant intersection number and so on, where the last term in the sequence is the univariate intersection. We consider integrals with $n$-integration variables $\{z_{i_{1}},...,z_{i_{n}}\}$, which can be thought of as iterative integrals, with a layout structure that follows from the chosen order $\{i_1,...,i_k\}$ of the integers $\{1,...,n\}$, somewhat similar to a volume integral. In order to calculate the multivariate intersection number for $n$-differential forms, it is necessary to calculate the dimension of the co-homology group for all $k$-differential forms from $k\in\{1,...,n\}$. It is obtained by counting the number of solutions $\nu_{\mathbf{k}}$ of the system of equations given by 
\begin{equation}
\hat{\omega}_j \equiv \partial_{z_j} \log u(\mathbf{z})=0, \quad j\in\{i_1,\dots,i_k\}\label{omegaj}
\end{equation}
where $\mathbf{k}\in\{i_1,...,i_k\}$ is a subset of $\{1,...,n\}$ with $k$ discrete elements. In this way, one obtains a set of dimensions $\{\nu_{\mathbf{1}},...,\nu_{\mathbf{n}}\}$ corresponding to the iterative integral in
$\{z_{i_1}\}$,in $\{z_{i_1},z_{i_2}\}$, 
$...$, in $ \{z_{i_1},...,z_{i_n}\}$ respectively.
Where we have used the vector notation,
$\mathbf{1}=\{i_1\}, \mathbf{2}=\{i_1,i_2\},...,\mathbf{n}=\{i_1,...,i_2\}$ to indicate the integration variables. 
We have to point out that while $\nu_{\mathbf{n}}$ is independent of the order of the integration variables, the dimensions of the subspace $\nu_{\mathbf{k}}$ may in fact depend on which specific subset $\mathbf{k}$ of $\{1,2,...,n\}$ is chosen and in which order. As a working principle, we can choose the order that minimises the size of $\nu_{\mathbf{k}}$ for all $\mathbf{k}$-forms $\mathbf{k}\in\{1,..,n\}$
\section{2-variable Intersection Number}
We start by examining an integral with two integration variables $\{z_1,z_2\}$ 
that is written as follows:
\begin{equation}
I=\int_{\gamma^{(\mathbf{2})}} \varphi^{(\mathbf{2})}(z_1, z_2\big) u\left(z_1, z_2\right)=\Big\langle\varphi^{(\mathbf{2})}\Big| \mathcal{C}^{(\mathbf{2})}\Big] \label{2vint}
\end{equation}
with $ \gamma^{(\mathbf{2})}\otimes u(z)= \mathcal{C}^{(\mathbf{2})}$ where $\mathbf{2}=\{1,2\}$, $\varphi$ is a differential $2$-form in the variables $z_1$ and $z_2$, and $\gamma^{(\mathbf{2})}$ is a two-dimensional integration domain embedded in some ambient manifold $X$ with complex dimension $2$. We assume that $X$ admits a fibration into one-dimensional spaces $z_1\in X_1$, and $z_2\in X_2$ with  $\varphi^{(\mathbf{2})}$. $\gamma^{(\mathbf{2})}$ can be decomposed into a similar manner. Analogously, we can consider the dual integral given by
\begin{equation}
\tilde{I}=\int_{\gamma^{(\mathbf{2})}} \varphi^{(\mathbf{2}) \vee}\Big(z_1, z_2\Big) u^{-1}\Big(z_1, z_2\Big)=\Big[\mathcal{C}^{(\mathbf{2}) \vee}\Big|\varphi^{(\mathbf{2}) \vee}\Big\rangle\label{d2vint}
\end{equation}
with all the variables defined analogously to the ones above. As before, we have
\begin{equation}
\omega=d \log u(\mathbf{z})=\sum_{i=1}^2 \hat{\omega}_i d z_i \label{omega2}
\end{equation}
and applying Eq. \eqref{om1} we can count the number of master integrals on the $X_1$ sub-manifold, which we denote as $\nu_1$ with $\mathbf{1}=\{1\}$. Our aim now is to find the $2$ variable intersection $\braket{\varphi^{(\mathbf{2})}}{\varphi^{\vee,\mathbf{2}}}$ in terms of the univariate intersection numbers on the manifold $X_1$ calculated via the univariate intersection number method discussed in \cite{Mastrolia:2018uzb}\cite{Frellesvig:2019kgj}. We start by decomposing the differential forms $\left\langle\varphi^{(\mathbf{2})}\right| \in \mathrm{H}^2\left(X, \nabla_\omega\right)$ and $\ket{\varphi^{(\mathbf{2}) \vee}}| \in \mathrm{H}^2\left(X, \nabla_{-\omega}\right)$ as
\begin{align}
\left\langle\varphi^{(\mathbf{2})}\right|&=\sum_{i=1}^{\nu_{(\mathbf{1})}}\left\langle e_i^{(\mathbf{1})}\right|\wedge\left\langle\varphi_{ i}^{(2)}\right|\label{phidicom}\\
\left|\varphi^{(\mathbf{2})\vee}\right\rangle&=\sum_{i=1}^{\nu_{(\mathbf{1})}}\left|h_i^{(\mathbf{1})}\right\rangle\wedge\left|\varphi_{i}^{(2)\vee}\right\rangle\label{vphidicom}
\end{align}
into arbitrary base forms $\left(\left\langle e_1^{\boldsymbol{1}}\right|, \ldots,\left\langle e_{\nu_{1}}^{\boldsymbol{1}}\right|\right)$ as the basis for $\mathrm{H}^1\left(X_1, \nabla_{\omega_1}\right)$ and $\left(\left|h_1^{\boldsymbol{1}}\right\rangle, \ldots,\left|h_{\nu_{1}}^{\boldsymbol{1}}\right\rangle\right)$ as a basis for $\mathrm{H}^1\left(X_1, \nabla_{-\omega_1}\right)$
In the above expression $\bra{\varphi^{(2)}_i}$ and $\ket{\varphi^{(2) \vee}_j}$ are one-foms in the variables $z_2$, and thus they are treated as coefficients of the basis expansions. They can be obtained by projecting using the multivariate version of the master decomposition formula
\begin{align}
\left\langle\varphi_{i}^{(2)}\right| & =\sum_{j}^{\nu_{(\mathbf{1})}}\left\langle\varphi^{(\mathbf{2})} \Big{\vert} h_j^{(\mathbf{1})}\right\rangle\left(\mathbf{C}_{(\mathbf{1})}^{-1}\right)_{j i}\label{Dphib2} \\
\left|\varphi_{i}^{(2)\vee}\right\rangle & =\sum_{j}^{\nu_{(\mathbf{1})}}\left(\mathbf{C}_{(\mathbf{1})}^{-1}\right)_{i j}\left\langle e_j^{(\mathbf{1})}\big{\vert} \varphi^{(\mathbf{2})\vee}\right\rangle\label{Dphib2v}
\end{align}
with $\left\langle\varphi_i^{(2)}\right|\in\mathrm{H}^1\left(X_2, \nabla_{\Omega^{(2)}}\right)$ and $\left|\varphi_j^{(2)\vee}\right\rangle \in \mathrm{H}^1\left(X_2, \nabla_{-\Omega^{(2)}}\right)$, with $\Omega^{(2)}$ is the connection coefficient on the total manifold $X$, and the metric matrix normalisation factor.
\begin{align}
\left(\mathbf{C}_{(\mathbf{1})}\right)_{i j} \equiv\left\langle e_i^{(\mathbf{1})} \Big{\vert} h_j^{(\mathbf{1})}\right\rangle 
\end{align}
which is also a univariate intersection matrix.
As $\left\langle\varphi^{(\mathbf{2})}\right|$ and $\left|\varphi^{(\mathbf{2}) \vee}\right\rangle$ are vector-valued they incorporate both the differential one form on $\left\langle\varphi_i^{(2)}\right|$ and $\left|\varphi_j^{(2) \vee}\right\rangle$ respectively, as well as the vector-valued basis elements $\left\langle e_i^{\boldsymbol{1}}\right|$ and $\left|h_1^{\boldsymbol{1}}\right\rangle$.
In order to determine the new connection coefficient $\Omega^{(2)}$, on our total manifold $X$ we proceed in the same manner as in the single-variable case, starting from the integral and defining the equivalence classes of the single-valued differential forms. We want to find an analogue of:
\begin{equation}
0=\int_{\gamma} d\left(\xi u\right)=\int_{\gamma}\left(d \xi+\omega\wedge \xi\right) u \equiv \int_{\gamma} \nabla_\omega \xi u
\end{equation}
Let us consider the original integral $I$ from Eq. \eqref{2vint} and apply the decomposition formula Eq. \eqref{phidicom}
\begin{align}
\Big\langle\varphi^{(\mathbf{2})}\Big| \mathcal{C}^{(\mathbf{2})}\Big] 
&=\int_{\gamma^{(\mathbf{2})}} \varphi^{(\mathbf{2})}\left(z_1, z_2\right) u\left(z_1, z_2\right)\\
&= \sum_{i=1}^{\nu_{(\mathbf{1})}}\left[\left\langle e_i^{(\mathbf{1})}\right|\wedge\left\langle\varphi_{ i}^{(2)}\right|\right]\Big| \gamma^{(\mathbf{2})}\otimes u(z_1,z_2) \Big]\\ 
& =\sum_{i=1}^{\nu_{(\mathbf{1})}} \int_{\gamma^{(2)}} \varphi_{i}^{(2)}\left(z_2\right) \int_{\gamma^{(\mathbf{1})}} e_i^{(\mathbf{1})}\left(z_1, z_2\right) u\left(z_1, z_2\right) \\
& =\sum_{i=1}^{\nu_{(\mathbf{1})}} \int_{\gamma^{(\mathbf{1})}} \varphi_{i}^{(2)}\left(z_2\right) u_i\left(z_2\right)
\end{align}
where we defined 
\begin{equation}
u_i\left(z_2\right)=\int_{\gamma ^{(\mathbf{1})}}e _i^{(\mathbf{1})}\left(z_1, z_2\right) u\left(z_1, z_2\right)
\end{equation}
There could be many forms $\varphi^{(2)}_i$ giving the same result. So let us consider a total derivative of $u_i$ times an arbitrary function $\xi_i(z_i)$ with correctly regulated poles.
\begin{equation}
0=\int_{\gamma^{(1)}}d_{z_2}
\left(u_i\left(z_2\right)\xi_i\left(z_2\right)\right)
=\int_{\gamma^{(1)}}
\left(u_i\left(z_2\right)d_{z_2}\xi_i\left(z_2\right) 
+d_{z_2} u_i\left(z_2\right)\wedge\xi_i\left(z_2\right)\right)\label{dXiu1}
\end{equation}
where $d_{z_2}$ denotes the differential acting only on $z_2$ i.e. $d_{z_2}=\partial_{z_2}dz_2$. From a detailed study of IBPs it also follows that the following differential equation is satisfied for $u_i(z_2)$ 
\begin{equation}
d_{z_2} u_i\left(z_2\right)=\Omega_{i j}^{(2)}\wedge u_j\left(z_2\right)\label{ODEdueqOM}
\end{equation}
where $\Omega^{(2)}$ is a $\nu_1\cross\nu_1$ matrix. We obtainen $\Omega^{(2)}$ directly by computing the $z_2$-differential of $u_i(z_2)$
\begin{align}
d_{z_2} u_i\left(z_2\right) & =d_{z_2} \int_{\gamma^{(\mathbf{1})}} e_i^{(\mathbf{1})}\left(z_1, z_2\right) u\left(z_1, z_2\right) \\
& =\int_{\gamma^{(\mathbf{1})}}\left(d_{z_2} e_i^{(\mathbf{1})}\left(z_1, z_2\right)+\frac{d_{z_2} u(z_1,u_2)}{u\left(z_1, z_2\right)}\wedge e_i^{(\mathbf{1})}\left(z_1, z_2\right)\right) u\left(z_1, z_2\right) \\
& =\int_{\gamma^{(\mathbf{1})}}\left(d_{z_2}+\omega_2 \wedge\right) e_i^{(\mathbf{1})}\left(z_1, z_2\right) u\left(z_1, z_2\right) \\
&
\left.=\left\langle\left(d_{z_2}+\omega_2 \wedge\right) e_i^{(\mathbf{1})}\right| \gamma^{(\mathbf{1})}\otimes u(z_1,z_2) \right]\\
&
\left.=\left\langle\nabla_{\omega_2} e_i^{(\mathbf{1})}\right| \mathcal{C}^{(\mathbf{1})}\right]
\end{align}
The last line can be simplified further by using the 
multivariate master decomposition formula Eq. \eqref{Dphib2} and inserting a well-chosen one,
\begin{equation}
\mathbb{I}_c=\sum_{k, j=1}^{\nu}\left|h_j\right\rangle(\mathbf{C})_{k j}^{-1}\left\langle e_j\right|
\end{equation}
such that:
\begin{align}
d_{z_2} u_i\left(z_2\right)=\sum_{k, j=1}^{\nu}\left\langle\nabla_{\omega_2} e_i^{(\mathbf{1})} \vert h_k^{(\mathbf{1})}\right\rangle\left(\mathbf{C}_{(\mathbf{1})}^{-1}\right)_{k j}\left\langle e_j^{(\mathbf{1})}\right|\mathcal{C}^{(\mathbf{1})\vee}\Big]    
\end{align}
where the last part can be rewritten as:
\begin{align}
&\left\langle e_j^{(\mathbf{1})}\right|\mathcal{C}^{(\mathbf{1})\vee}\Big]\\
=&\left\langle e_j^{(\mathbf{1})}\right|\gamma^{(\mathbf{1})}\otimes u_j(z_1,z_2)\Big]\\
=&\int_{\gamma^{(\mathbf{1})}}e_j^{(\mathbf{1})} u_j(z_1,z_2)\\
=& u_j(z_2)
\end{align}
Hence we get: 
\begin{equation}
d_{z_2} u_i\left(z_2\right)
=\sum_{k, j=1}^{\nu}\left\langle\nabla_{\omega_2} e_i^{(\mathbf{1})} \vert h_k^{(\mathbf{1})}\right\rangle\left(\mathbf{C}_{(\mathbf{1})}^{-1}\right)_{k j}\wedge u_j(z_2)    
\end{equation}
Using Eq. \eqref{ODEdueqOM} we can identify $\Omega^{(2)}$ as
\begin{equation}
\Omega^{(2)}_{kj}=
\sum_{j, k=1}^{\nu}\Big\langle\nabla_{\omega_{2}}e_i^{(\mathbf{1})}\vert h_k^{(\mathbf{1})}\Big\rangle\left(\mathbf{C}_{(\mathbf{1})}^{-1}\right)_{kj}
\end{equation}
Or explicitly written as
\begin{equation}
\Omega^{(2)}=\sum_{j, k=1}^{\nu}\frac{\left\langle\nabla_{\hat{\omega}_2}\left(e^{(\mathbf{1})}\right) \vert h^{(\mathbf{1})}\right\rangle}{\left\langle e^{(\mathbf{1})} \vert h^{(\mathbf{1})}\right\rangle}
\end{equation}
Therefore, $\Omega^{(2)}$ can be understood as the coefficient Connection of the $e^{(\mathbf{1})}$ basis along $\omega_2$ projected onto the $h^{(\mathbf{1})}$ basis normed by $\mathbf{C}^{(\mathbf{1})}=\left\langle e^{(\mathbf{1})} \vert h^{(\mathbf{1})}\right\rangle$.
Inserting Eq. \eqref{ODEdueqOM} into Eq. \eqref{dXiu1} we obtain 
\begin{align}
0 &=\int_{\gamma^{(2)}}\left(u_j\left(z_2\right) d \xi_j+\Omega_{j i}^{(2)} u_i\left(z_2\right) \wedge \xi_j\right) \\
&=\int_{\gamma^{(2)}} u_i\left(z_2\right)\left(\delta_{i j} d \xi_j+\left(\Omega^{(2) \top}\right)_{i j} \wedge \xi_j\right)\\
&=\int_{\gamma^{(2)}}
u(z_2)\left(\nabla_{\Omega^{(2)}} \xi\right)
\end{align}
Therefore, the connection reads as follows:
\begin{equation}
\left(\nabla_{\Omega^{(2)}}\right)_{i j}(\bullet)_j=\delta_{i j} d(\bullet)_j+\left(\Omega^{(2) \top}\right)_{i j} \wedge(\bullet)_j
\end{equation}
This makes intuitive sense if we recall that the matrix can be understood as a linear transformation of our coordinate system, where the $i$-th component changes relative to the $j$-th  component. Therefore, a matrix-valued connection can be understood as the derivative of the $i$-th component of a "vector" along $i$ and $j$ component. Hence, we have along the $i$ component the classical derivative plus $\Omega_{ii}$ to account for the rate of change $i$-th relative to the changes along in the $i$-th component of our basis on the manifold. While along the $j$-th component, we only have to take into account the rate of change of the $i$-th relative to the changes along the $j$-th component of our basis on the manifold.\\
From section \ref{Intersection1}, it is known that the univariate intersection number is given by:
\begin{equation}
\Big\langle e_i^{(\mathbf{1})}\Big\vert h_j^{(\mathbf{1})}\Big\rangle=\sum_{x_i\in \mathcal{P}_{\omega_1}} \operatorname{Res}_{z_1=x_i}\kl{\psi_{x_i} h_j^{(\mathbf{1})}}
\end{equation}
%where $\psi_{x_i}$ is the local solution of the differential equation:
\begin{equation}
\nabla_{\omega_1}\psi_{x_i}=e_i^{(\mathbf{1})}
\end{equation}
around every pole $x_i$ of $\omega_1$ denoted by the set $\mathcal{P}$. Here the Connection coefficient $\omega_1$  is just the $dz_1$ connection coefficient of $\omega$ and therefore,
\begin{equation}
\nabla_{\omega_1}\bullet=\partial_{z_1}\kl{\bullet}+\omega_1\wedge\kl{\bullet}
\end{equation}
is the covariant connection on $X_1$. 
As the connection  $\Omega^{(2)}$ is matrix-valued, the regularisation map has now the form.
\begin{equation}
\operatorname{reg}_{\Omega^{(2)}}: \varphi^{(2)} \rightarrow \operatorname{reg}_{\Omega^{(2)}}\left(\varphi^{(2)}\right)=\varphi^{(2)}-\sum_{x_i \in \mathcal{P}_{\Omega^{(2)}}} \nabla_{\Omega^{(2)}}\left(h_{x_i}\left(z_2, \bar{z}_2\right) \psi_{x_i}\right)
\end{equation}
where $\psi_{x_i}$ is a vector-valued solution of the differential equation
\begin{equation}
\nabla_{\Omega^{(2)}}\left(\psi_{x_i}\right)=\varphi^{(2)}
\end{equation}
around each point $x_i$ from the set of queries $\Omega^{(2)}$ from the set $\mathcal{P}^{(2)}_{\Omega}$. Following our intuition for multivariable integration, we obtain the intersection number by integrating from inside $X_1$ to the outside $X_2$. So the intersection number is given by the following expression: 
\begin{align}
\left\langle\varphi^{(\mathbf{2})} \vert \varphi^{(\mathbf{2}) \vee}\right\rangle & =\frac{1}{2 \pi i} \int_{X_2}\left(\operatorname{reg}_{\Omega^{(2)}} \varphi^{(2)}\right)_i \wedge \varphi_j^{(2) \vee} \cdot \mathbf{C}_{i j}^{(\mathbf{1})}\label{inrec21} \\
& =\sum_{x_i \in \mathcal{P}_{\Omega^{(2)}}} \operatorname{Res}\left(\psi_{x_i} \cdot \mathbf{C}^{(1)} \cdot \varphi^{(2) \vee}\right)\label{inreg22}
\end{align}
where Eq. \eqref{inreg22}  is derived from Eq.\eqref{inrec21} analogously as in section \ref{Intersection1}.
\section{$n$-Variable Intersection Number}\label{nVariableIntersectionNumber}
Following the discussion above, we generalise the 2-variable intersection to the $n$-variable case by considering an integral with $n$ integration variables $(z_1,z_2,...,z_n)$ written as follows:
\begin{equation}
I\left(z_1, z_2, \ldots, z_n\right)=\int_{\gamma^{(\mathbf{n})}} \varphi^{(\mathbf{n})}\left(z_1, z_2, \ldots, z_n\right) u\left(z_1, z_2, \ldots, z_n\right)=\left\langle\varphi^{(\mathbf{n})}\right| \mathcal{C}^{(\mathbf{n})}\Big]
\end{equation}
with the notation $\mathbf{n}=\{1,...,n\}$. The $\varphi^{(\mathbf{n})}$ is an $n$-variable differential form on a manifold $X$. In a similar manner, one can define a dual form $\varphi^{(\mathbf{n}) \vee}$. It is then assumed that the $n$-complex dimensional manifold with coordinates $(z_1,...,z_n)$ admits a fibration into an $(n-1)$-dimensional submanifold defined by $(z_1,... ,z_{n-1})$, denoted by $\mathbf{n}-\mathbf{1}$, which we refer to as the inner manifold, and a submanifold with $z_n$, which we refer to as the outer manifold. 
\begin{equation}
\omega=d \log u(\mathbf{z})=\sum_{i=1}^n \hat{\omega}_i d z_i\label{Cont}
\end{equation}
and employing Eq. \eqref{Cont}, we can count the number of master integrals on the inner manifold, which we define as $\nu_{\mathbf{n}-\mathbf{1}}$. The aim is then to express the $n$-variable intersection number  $\braket{\varphi^{(\mathbf{n})}}{\varphi^{(\mathbf{n}) \vee}}$ in terms of the intersection number in $(n-1)$-variables on the inner manifold, which are assumed to be known at this point%, following the recursive character of this algorithm.
The choice of variables and their orienting parametrisation of the inner and outer manifold is arbitrary: as before, we use $\mathbf{k} \equiv\left\{i_1, i_2, \ldots, i_k\right\}$ to denote the variables that are taking part in the specific calculation. Therefore the $\mathbf{n}$-forms $\left\langle\varphi^{(\mathbf{n})}\right| \in\mathrm{H}^n\left(X, \nabla_\omega\right)$ and $\left\langle\varphi^{(\mathbf{n}), \vee}\right|\in\mathrm{H}^n\left(X, \nabla_{-\omega}\right)$ can be decomposed, similar to \eqref{phidicom} and \eqref{vphidicom}, the decomposition formula now takes the following form:
\begin{align}
&\left\langle\varphi^{(\mathbf{n})}\right|=\sum_{i=1}^{\nu_{\mathbf{n}-{(\mathbf{1})}}}\left\langle e_i^{(\mathbf{n}-\mathbf{1})}\right| \wedge\left\langle\varphi_{i}^{(n)}\right|\label{MDFNV}\\
&\left|\varphi^{(\mathbf{n}) \vee}\right\rangle=\sum_{i=1}^{\nu_{\mathbf{n}-1}}\left|h_i^{(\mathbf{n}-\mathbf{1})}\right\rangle \wedge\left|\varphi_{i}^{(n) \vee}\right\rangle
\end{align}
with an arbitrary base 
$\left(\left\langle e_1^{\boldsymbol{n}-\boldsymbol{1}}\right|, \ldots,\left\langle e_{\nu_{\boldsymbol{n}-\boldsymbol{1}}}^{\boldsymbol{n}-\boldsymbol{1}}\right|\right)$
as the basis for
$\mathrm{H}^{n-1}\left(X_{n-1}, \nabla_{\omega|dz_n=0}\right)$
and\\$\left(\left|h_1^{\boldsymbol{n}-\boldsymbol{1}}\right\rangle, \ldots,\left|h_{\nu_{1}}^{\boldsymbol{n}-\boldsymbol{1}}\right\rangle\right)$
as a basis for
$\mathrm{H}^{n-1}\left(X_{n-1}, \nabla_{-{\omega_{n-1}}}\right)$.
$\nu_{\mathbf{n}-\mathbf{1}}$ 
is the number of master integrals on the inner manifold with an arbitrary base $\bra{e_i^{\mathbf{n}-\mathbf{1}}}$, $\ket{h_i^{\mathbf{n}-1}}$. In the above expression, $\bra{\varphi_i^{n}}$ and $\ket{\varphi^{n\vee}_i}$ are one-forms in the variables $z_n$ and are treated as coefficients of a base expansion. They can be obtained by projection, using  the multivariate version of the master decomposition formula:
\begin{align}
\left\langle\varphi_i^{(n)}\right|&=\sum_{j=1}^{\nu_{(\mathbf{n}-1)}}\left\langle\varphi^{(\mathbf{n})} \vert h_j^{(\mathbf{n}-\mathbf{1})}\right\rangle
\left(\mathbf{C}_{(\mathbf{n}-\mathbf{1})}^{-1}\right)_{j i}\label{mdicphi}\\
\left|\varphi_i^{(n)\vee}\right\rangle&=\sum_{j=1}^{\nu_{\mathbf{n}-1}}\left(\mathbf{C}_{(\mathbf{n}-\mathbf{1})}^{-1}\right)_{i j}\left\langle e_j^{(\mathbf{n}-\mathbf{1})} \vert\varphi^{(\mathbf{n}) \vee}\right\rangle\label{mdicvphi}
\end{align}
with
\begin{equation}
\left(\mathbf{C}_{(\mathbf{n}-\mathbf{1})}\right)_{i j}=\left\langle e_i^{(\mathbf{n}-\mathbf{1})} \vert h_j^{(\mathbf{n}-\mathbf{1})}\right\rangle.
\end{equation}
It is to be noted that the $(n-1)$-variable intersection number is assumed to be known at this stage of the calculation. The recursive formula for the intersection number is then:
\begin{equation}
\left\langle\varphi^{(\mathbf{n})} \vert \varphi^{(\mathbf{n}) \vee}\right\rangle=\sum_{x_i \in \mathcal{P}_{\Omega^{(n)}}} \operatorname{Res}_{z_n=x_i}\left(\psi_{x_i, i}^{(n)} \cdot \mathbf{C}_{i j}^{(\mathbf{n}-1)} \cdot \varphi_j^{(n) \vee}\right)\label{Resmulti}
\end{equation}
where the function $\psi_{x_i}$ are solutions of the differential equations
\begin{equation}
\nabla_{\Omega^{n}}\psi^{( \mathbf{n})}_{x_i}=\varphi^{(\mathbf{n})}
\end{equation}
and $\varphi$ are obtained through Eq. \eqref{mdicphi}. Here $\hat{\Omega}^{n}$ is a $\nu_{\mathbf{n}-\mathbf{1}}\cross\nu_{\mathbf{n}-\mathbf{1}}$ matrix, with entries are given by:
\begin{equation}
\hat{\Omega}_{i j}^{(n)}=\sum_{k=1}^{\nu_{(\mathbf{n}-\mathbf{1})}}\left\langle\nabla_{\omega_{n-1}} e_i^{(\mathbf{n}-\mathbf{1})} \vert h_k^{(\mathbf{n}-\mathbf{1})}\right\rangle\left(\mathbf{C}_{(\mathbf{n}-\mathbf{1})}^{-1}\right)_{k j}\label{omegaij}
\end{equation}
and finally, $\mathcal{P}_{\Omega^{n}}$ is a set of poles of $\hat{\Omega}$ defined as the union of the poles of the entries, including the possible at infinity. 

Let us briefly comment on \eqref{Resmulti}, we notice that the formula for the residue contains $\mathbf{C}_{(\mathbf{n}-\mathbf{1})}$, while $\left|\varphi^{(n) \vee}\right\rangle$ contains $\mathbf{C}_{(\mathbf{n}-1)}^{-1}$. Thus, the two $\mathbf{C}_{(\mathbf{n}-\mathbf{1})}$, $\mathbf{C}_{(\mathbf{n}-1)}^{-1}$ cancel. Consequently, the intersection number computed by the residue takes the following form
\begin{equation}
\left\langle\varphi^{(\mathbf{n})} \vert \varphi^{(\mathbf{n}) \vee}\right\rangle=\sum_{x_i \in \mathcal{P}_n} \operatorname{Res}_{z_n=x_i}\left(\psi_{x_i} \cdot \varphi_{\mathbf{C}}^{(n) v}\right)
\end{equation}
with 
\begin{equation}
\left|\varphi_{\mathbf{C}, i}^{(n) \vee}\right\rangle=\left\langle e_i^{(\mathbf{n}-1)} \vert \varphi^{(\mathbf{n}) \vee}\right\rangle
\end{equation}
Nevertheless, we can't avoid computing $\mathbf{C}_{(\mathbf{n}-\mathbf{1})}$ altogether, as it
appears in other quantities, e.g 
$\Omega^{(n)}$. 
\section{Intersection Number of the Generalised Beta Function}
In order to get a better understanding of the concepts, just laid out, we are going to calculate the intersection number of the multivariate beta function in great detail: 
\begin{equation}
B(p, q, r)=\int_{\gamma^{(\mathbf{2})}} z_1^p z_2^q\left(1-z_1-z_2\right)^r \frac{d z_1 \wedge d z_2}{z_1 z_2\left(1-z_1-z_2\right)}\label{B1}
\end{equation}
with
\begin{align}
&\gamma^{(\mathbf{2})}=z_1>0 \cap z_2>0 \cap z_1+z_2<1\in X\\
&X\in \mathbb{C}^2 \backslash\left(z_1=0 \cup z_2=0 \cup 1-z_1-z_2=0\right)\label{1X}
\end{align}
\eqref{B1} can be rewritten as
\begin{align}
B(p, q, r) & =\int_{\gamma^{(2)}}\left(z_2^q \int_{\gamma^{1}} z_1^p\left(1-z_1-z_2\right)^r \frac{d z_1}{z_1\left(1-z_1-z_2\right)}\right) \frac{d z_2}{z_2} \\
&=\frac{\Gamma(p) \Gamma(r)}{\Gamma(p+r)} \int_{\gamma^{(2)}} z_2^{q-1}\left(1-z_2\right)^{p+r-1} \frac{d z_2}{z_2} \\
& =\int_{\gamma^{(2)}} u\left(z_2\right) \frac{d z_2}{z_2},
\end{align}
with
$$
\begin{array}{ll}
\gamma^{1}=\left(0,1-z_2\right) \in X_1, & \text { with: } X_1=\mathbb{C} \backslash\left\{0,1-z_2\right\}=\mathbb{C P}^1 \backslash\left\{0,1-z_2, \infty\right\}, \\
\gamma^{(2)}=(0,1) \in X_2, & \text { with: } X_2=\mathbb{C} \backslash\{0,1\}=\mathbb{C P}^{1} \backslash\{0,1, \infty\} ;
\end{array}
$$
and 
\begin{equation}
\frac{\Gamma(p) \Gamma(r)}{\Gamma(p+r)}=\mathrm{B}\left(p, r\right)=\int_0^1 t^{p-1}(1-t)^{r-1} d t
\end{equation}
as well as  
\begin{equation}
u\left(z_2\right)=\frac{\Gamma(p) \Gamma(r)}{\Gamma(p+r)} z_2^{q-1}\left(1-z_2\right)^{p+r-1}
\end{equation}
This whole setup can be best understood in a graphic.
\begin{figure}[h]
\begin{center}
\includegraphics[width=0.66\textwidth]{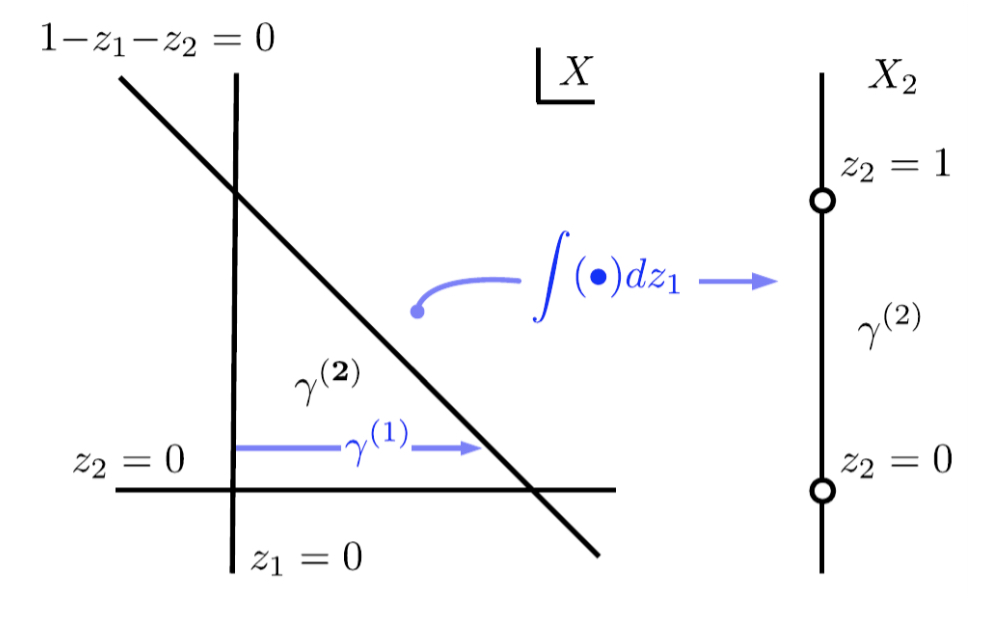}
\caption{Generalised Beta Function.\cite{Gasparotto:2023cdl}}
\label{fig:figure2}
\end{center}
\end{figure}
Furthermore, by \eqref{Cont} follows:
 \begin{equation}
\omega=\hat{\omega}_1 d z_1+\hat{\omega}_2 d z_2=\left(\frac{p}{z_1}-\frac{r}{1-z_1-z_2}\right) d z_1+\left(\frac{q}{z_2}-\frac{r}{1-z_1-z_2}\right) d z_2,\label{omegaE1}
 \end{equation}
and, as already mentioned in Eq. \eqref{1X} let:
\begin{equation}
X \in \mathbb{C}^2 \backslash\left(z_1=0 \cup z_2=0 \cup 1-z_1-z_2=0\right)
\end{equation}
Our aim is now to compute the intersection number
\begin{equation}
\left\langle\varphi^{(\mathbf{2})} \vert \varphi^{(\mathbf{2})\vee}\right\rangle=\left\langle\frac{d \mathbf{z}}{z_1 z_2\left(1-z_1-z_2\right)}\Bigg{\vert} \frac{d \mathbf{z}}{z_1^2 z_2\left(1-z_1-z_2\right)}\right\rangle
\end{equation}
$X$ can be decomposed in terms of  $X_1=\mathbb{C} \backslash\left\{0,1-z_2\right\}=\mathbb{C P}^1 \backslash\left\{0,1-z_2, \infty\right\}$ with the coordinate $z_1$ and $X_2=\mathbb{C} \backslash\{0,1\}=\mathbb{C P}^1 \backslash\{0,1, \infty\}$ with the coordinate $z_2$:
For the number of dimensions of $\mathrm{H}^1\left(X_1, \nabla_{ \pm\left.\omega\right|_{d z_2=0}}\right)$ we get via critical point analysis, 
\begin{equation}
\nu_{(\mathbf{1})}=\operatorname{dim}\left(\mathrm{H}^1\left(X_1, \nabla_{ \pm\left.\omega\right|_{d z_2=0}}\right)\right)=1 \text {;}
\end{equation}
as a basis, we choose as the reader might have already guessed:
\begin{equation}
\left\langle e^{(\boldsymbol{1})}\right|=\left\langle\frac{d z_1}{z_1\left(1-z_1-z_2\right)}\right|\in \mathrm{H}^1\left(X_1, \nabla_{\omega_1}\right), \quad\left| h^{(\boldsymbol{1})}\right\rangle=\left|\frac{d z_1}{z_1\left(1-z_1-z_2\right)}\right\rangle\in\mathrm{H}^1\left(X_1, \nabla_{-\omega_1}\right)
\end{equation}
From this, we will calculate the univariate intersection number, among these basis elements, for this, we are going to use the residue formula of the intersection number
, which in this case, takes the form:
\begin{equation}
\mathbf{C}^{(\mathbf{1})}=\braket{e^{(\boldsymbol{1})}}{h^{(\boldsymbol{1})}}=\sum_{x_i\in\mathcal{P}_{\omega}}\Res\left(\Xi h^{(\boldsymbol{1})}\right)\label{FE1}
\end{equation}
 where we obtain $\Xi$ by solving the following differential equation:
\begin{equation}
\nabla_{\omega}\Xi=\frac{dz}{z(1-z)}\label{ODEXi}
\end{equation}
around $\mathcal{P}=\{0,1,\infty\}$
via the recursive relation known from Eq. \eqref{re} Section  \ref{Intersection1}:
\begin{equation}
\psi_{m+1}=\frac{1}{m+1+\hat{\omega}_{-1}}\left(\hat{\varphi}_m-\sum_{q=0}^{m-\min _{\varphi}-1} \hat{\omega}_q \psi_{m-q}\right), \quad \psi_{\min _{\varphi}+1}=\frac{\hat{\varphi}_{\min _{\varphi}}}{\min _{\varphi}+1+\hat{\omega}_{-1}}\label{re}
\end{equation}
\subsection{Calculating $\mathbf{C}^{(\mathbf{1})}=\braket{e^{(\mathbf{1})}}{h^{(\mathbf{1})}}$, via Intersection number}
In this section, we are going to calculate, the intersection number of \eqref{FE1}, around the three poles $0$, $1-z_2$ and $\infty$
\paragraph{series solution around $0$}
The series expansion of $e=\frac{d z_1}{z_1(1-z_1-z_2)}$ around $0$ at local coordinates $y=z_1$ has the form: 
\begin{equation*}
\frac{d y}{y(1-y-z_2)}=\frac{dy}{y(1-z_2)}+\frac{dy}{(1-z_2)^2}+\frac{y dy}{(1-z_2)^3}+\frac{y^2_1 dy}{(1-z_2)^4}+O\left(y^3_1\right)
\end{equation*} 
The coefficients are
$\hat{e}_{-1}=\frac{1}{(1-z_2)},\;\hat{e}_{0}=\frac{1}{(1-z_2)^2},\;\hat{e}_{1}=\frac{1}{(1-z_2)^3},\;\hat{e}_{2}=\frac{y^2}{(1-z_2)^4}$. The series expansion of $\omega_1$,\eqref{omegaE1} around $0$ at local coordinats $y=z_1$ has the form,
\begin{equation*}
\left(\frac{p}{y}-\frac{r}{1-y-z_2}\right)dy=\frac{pdy}{y}+\frac{rdy}{z_2-1}-\frac{r y dy}{(z_2-1)^2}+\frac{r y^2 dy}{(z_2-1)^3}+O\left(y^3\right)
\end{equation*}
The coefficients are
$\hat{\omega}_{1,-1}=p,\;\hat{\omega}_{1,0}=\frac{r}{z_2-1},\;\hat{\omega}_{1,1}=\frac{r}{(z_2-1)^2},\;\hat{\omega}_{1,2}=\frac{r}{(z_2-1)^3}$
To obtain the solution for the differential equation Eq. \eqref{ODEXi} for $m=-1$ we use the Eq. \eqref{re} and get:
\begin{equation*}
\Xi_{0}(y,0)=\frac{1}{p(1-z_2)}
\end{equation*}
For $m=0$ follows:
\begin{align*}
\Xi_{1}(y,0)=\frac{p+r}{p(p+1)(z_2-1)^2}
\end{align*}
If we now put both expressions together, we get the following series expansion for $\Xi(y,0)$ we get:
\begin{align*}
\Xi(y,0)=\frac{1}{p(1-z_2)}+\frac{p+r}{p(p+1)(z_2-1)^2}y
\end{align*}
Thus the residue is:
\begin{equation}
\Res_{y_{0,1}=0}\kl{\Xi(y,0) h_0}=\frac{1}{p\kl{z_2-1}^2}\label{XSEM1R3}
\end{equation}
\paragraph{series solution around $1-z_2$,}
of $e=\frac{d z_1}{z_1(1-z_1-z_2)}$ at local coordinates $y=z_1+z_2-1$ has the form:
\begin{equation*}
\frac{dy}{y (z_2-y-
1)}=\frac{dy}{(z_2-1)y}+\frac{dy}{(z_2-1)^2}+\frac{ydy}{(z_2-1)^3}+\frac{y^2}{(z_2-1)^4}+O\left(y^3\right)
\end{equation*}
The coefficients are
$\hat{e}_{-1}=\frac{1}{(1-z_2)},\;\hat{e}_{0}=\frac{1}{(1-z_2)^2},\;\hat{e}_{1}=\frac{1}{(1-z_2)^3},\;\hat{e}_{2}=\frac{y^2}{(1-z_2)^4}$
The series expansion of $\omega_1$,\eqref{omegaE1} around $1-z_2$ at local coordinates $y=z_1+z_2-1$ has the form,
\begin{equation*}
\left(\frac{p}{y-z_2+1}+\frac{r}{y}\right)dy 
=\frac{rdy}{y}-\frac{pdy}{z_2-1}-\frac{py}{(z_2-1)^2}-\frac{py^2dy}{(z_2-1)^3}+O\left(y^3\right)
\end{equation*}
The coefficients are
$\hat{\omega}_{1,-1}=r,\;\hat{\omega}_{1,0}=\frac{p}{z_2-1},\;\hat{\omega}_{1,1}=\frac{p}{(z_2-1)^2},\;\hat{\omega}_{1,2}=\frac{r}{(z_2-1)^3}$.
To obtain the solution for the differential equation Eq. \eqref{ODEXi} for $m=-1$ we use the Eq. \eqref{re}
\begin{equation*}
\Xi_{0}(y,1-z_2)=\frac{1}{r(z_2-1)}
\end{equation*}
For $m=0$:
\begin{align*}
\Xi_{1}(y,1-z_2)=\frac{p+r}{r(r+1)(z_2-1)^2}
\end{align*}
If we now put both expressions together, we get the following series expansion for $\Xi(y,1-z_2)$ we get:
\begin{align*}
\Xi(y,1-z_2)=\frac{1}{r(z_2-1)}+\frac{p+r}{r(r+1)(z_2-1)^2}y
\end{align*}
Thus the residue is:
\begin{equation}
\Res_{y_{1,1-z_2}=0}\kl{\Xi(y,1-z_2) h_1}=\frac{1}{r\kl{z_2-1}^2}\label{XSEM1R2}
\end{equation}
\paragraph{series solution around $\infty$,}
of $e=\frac{d z_1}{z_1(1-z_1-z_2)}$ at local coordinates $y=\frac{1}{z_1}$ and $dz=-\frac{dy}{y^2}$ has the form:
\begin{equation*}
\frac{dy}{y(z_2-1)+1}=
dy +O\left(y^3\right)
\end{equation*}
The coefficient is
$\hat{e}_{2}=1$,%$$\;\hat{e}_{0}=\frac{1}{(1-z_2)^2},\;\hat{e}_{1}=\frac{1}{(1-z_2)^3},\;\hat{e}_{2}=\frac{y^2}{(1-z_2)^4}$
The series expansion of $\omega_1$, \eqref{omegaE1} around $\infty$ at local coordinates $y=\frac{1}{z_1}$ and $dz_1=-\frac{dy}{y^2}$ has the form,
\begin{equation*}
-\frac{dy(py (z_2-1)+p+r)}{y (y (z_2-1)+1)}=-\frac{dy (p+r)}{y}-dy y \left(r z_2^2-2 r z_2+r\right)+dy (r z_2-r)+O\left(y^3\right)
\end{equation*}
the coefficients are
$\hat{\omega}_{1,-1}=(p+r), \hat{\omega}_{1,0}= (r z_2-r)
,\hat{\omega}_{1,1}=(z_2-1),\hat{\omega}_{1,2}=-r(z_2-1)$.
We can already see for the series expansion around $\infty$, that it does not contain any residues, therefore it does not contribute to the intersection number.
\paragraph{Calculating the intersection number}
By equation Eq.\eqref{FE1} the intersection number is the sum over all the residues, so if we combine Eq. \eqref{XSEM1R3} and Eq. \eqref{XSEM1R2} we get the following expression:
\begin{align}
\mathbf{C}^{(\mathbf{1})}\nonumber
&=\braket{e^{(\boldsymbol{1})}}{h^{(\boldsymbol{1})}}\nonumber\\
&=\Res_{y_{1,0}=0}\kl{\Xi(y,0) h_0}\nonumber
+\Res_{y_{1}=1-z_2}\kl{\Xi(y,1-z_2) h_1}\nonumber\\
&=\frac{1}{p\kl{z_2-1}^2}
+\frac{1}{r\kl{z_2-1}^2}\nonumber\\
&=\frac{p+r}{pr\kl{z_2-1}^2}\label{F1c}
\end{align}
\subsection{Calculating $\Omega$, via Intersection number}
In the next part of the calculation, we are going to determine $\Omega$, as defined by Eq. \eqref{omegaij}. Which, as mentioned, can be understood, as the covariant derivative of $e^{(\boldsymbol{1})}$ along $\omega_2$ projected onto $h^{(\boldsymbol{1})}$ normed by $\mathbf{C}_{(\mathbf{1})}=\braket{e^{(\boldsymbol{1})}}{h^{(\boldsymbol{1})}}$:
\begin{equation}
\Omega^{(2)}=\left\langle\nabla_{\omega_2}e^{(\boldsymbol{1})}\Big{|}h^{(\boldsymbol{1})}\right\rangle \cdot \mathbf{C}_{(\mathbf{1})}^{-1} \label{OMEGA2D}
\end{equation}
For this, we are going to calculate the intersection number of $\left\langle\nabla_{\hat{\omega}_2}e^{(\boldsymbol{1})} \vert h^{(\boldsymbol{1})}\right\rangle$ and combine it with the result obtained in the last calculation, see \eqref{F1c}. As a reminder $\nabla_{\omega_2}$,is as stated in \eqref{naom1}:
\begin{equation}
\nabla_{\hat{\omega}_2}(\bullet)=\partial_{z_2}(\bullet)+\hat{\omega}_2 \wedge(\bullet)\label{naom1}
\end{equation}
where by \eqref{omegaE1}, $\omega_2$ is given by:
\begin{equation}
\omega_2=\frac{p}{z_2}-\frac{r}{-z_1-z_2+1}
\end{equation}
and $e^{(\boldsymbol{1})}$ is given by: 
\begin{equation}
   e^{(\boldsymbol{1})}=\frac{d z_1}{z_1\left(1-z_1-z_2\right)}
\end{equation}
Thus the whole expression is:
\begin{equation}
\nabla_{\omega_{z_2}}e^{(\boldsymbol{1})}=\frac{1}{z_1\kl{-z_1-z_2+1}^2}-\kl{\frac{p}{z_2}-\frac{r}{-z_1-z_2+1}}\wedge\frac{1}{z_1\kl{-z_1-z_2+1}
}\label{nabome1}
\end{equation}
To find the intersection number of \eqref{OMEGA2D} as defined by the residue formula
\begin{equation}
\left\langle\nabla_{\hat{\omega}_2}e^{(\boldsymbol{1})}\Big{|}h^{(\boldsymbol{1})}\right\rangle=\Res_{z=x_j}\kl{\zeta_i h_i}\label{Resze}
\end{equation}
For this, we will find the solution to the differential equation,
\begin{equation}
\nabla_{\omega_1}\zeta(z_1)=\frac{d z_1}{z_1(1-z_1-z_2)}\label{ODEze}
\end{equation}
via the recursive relation:
\begin{equation}
\psi_{m+1}=\frac{1}{m+1+\hat{\omega}_{-1}}\left(\hat{\varphi}_m-\sum_{q=0}^{m-\min _{\varphi}-1} \hat{\omega}_q \psi_{m-q}\right), \quad \psi_{\min _{\varphi}+1}=\frac{\hat{\varphi}_{\min _{\varphi}}}{\min _{\varphi}+1+\hat{\omega}_{-1}}
\end{equation}
and put the obtained solution $\zeta$ back into the residue formula for the intersection number Eq. \eqref{Resze}.
We will do this for each of the poles $\mathcal{P}=\{0,1-z_2,\infty\}$, and combine the resulting expression with the expression for $\mathbf{C}^{(\mathbf{1})}$, \eqref{F1c}, derived earlier.
\paragraph{The series solution around $0$,}
of $e=\frac{d z_1}{z_1(1-z_1-z_2)}$ around $0$ at local coordinats $y=z_1$  has the form: 
\begin{equation*}
\frac{dy}{y(1-y-z_2)}=
\frac{dy}{y(1-z_2)}+\frac{dy}{(1-z_2)^2}+\frac{y dy}{(1-z_2)^3}+\frac{y^2 dy}{(1-z_2)^4}+O\left(y^3\right)
\end{equation*} 
the coefficients are
$\hat{e}_{-1}=\frac{1}{(1-z_2)},\;\hat{e}_{0}=\frac{1}{(1-z_2)^2},\;\hat{e}_{1}=\frac{1}{(1-z_2)^3},\;\hat{e}_{2}=\frac{y^2}{(1-z_2)^4}$.
The series expansion of $\omega_1$,\eqref{omegaE1} around $0$ at local coordinates $y=z_1$  has the form,
\begin{equation*}
\left(\frac{p}{y}-\frac{r}{1-y-z_2}\right)d y=\frac{p dy}{y}+\frac{r dy}{z_2-1}-\frac{r y dy}{(z_2-1)^2}+\frac{r y^2}{(z_2-1)^3}+O\left(y^3\right)
\end{equation*}
the coefficients are
$\hat{\omega}_{1,-1}=p,\;\hat{\omega}_{1,0}=\frac{r}{z_2-1},\;\hat{\omega}_{1,1}=\frac{r}{(z_2-1)^2},\;\hat{\omega}_{1,2}=\frac{r}{(z_2-1)^3}$. The series expansion of $\omega_2$,\eqref{omegaE1} around $0$ at local coordinates $y=z_1$, has the form,
\begin{equation*}
\left(\frac{q}{z_2}-\frac{r}{1-y-y}\right) d z_2=
\frac{q dz_2}{z_2}+\frac{r dz_2}{z_2-1}-\frac{r y dz_2}{(z_2-1)^2}
+\frac{r y^2 dz_2}{(z_2-1)^3}+O\left(y^3\right)
\end{equation*}
the coefficients are
$\hat{\omega}_{2,0}=\frac{q}{z_2}+\frac{r}{z_2-1},\;\hat{\omega}_{2,1}=-\frac{r}{(z_2-1)^2},\;\hat{\omega}_{2,2}=\frac{r}{(z_2-1)^3}$
the series expansion of $\nabla_{\omega_2}e^{(\boldsymbol{1})}$, Eq. \eqref{nabome1} around $0$ at local coordinates $y=z_1$  has thus the form:
\begin{align*}
\nabla_{\omega_2}e^{(\boldsymbol{1})}(y,0)&=\frac{-qz_2+q-r z_2+z_2}{y (z_2-1)^2 z_2}dy\wedge dz_2+\frac{q z_2-q+2 r z_2-2 z_2}{(z_2-1)^3 z_2}dy\wedge dz_2\\
&+\frac{y (-qz_2+q-3 r z_2+3 z_2)}{(z_2-1)^4 z_2}dy\wedge dz_2+
\frac{y^2 (q z_2-q+4 r z_2-4 z_2)}{(z_2-1)^5 z_2}dy\wedge dz_2+
O\left(y^3\right)
\end{align*}
the coefficients are
$$
\begin{array}{cc}
\hat{\nabla_{\omega_2}e^{(\boldsymbol{1})}}_{-1}=\frac{-qz_2+q-rz_2+z_2}{(z_2-1)^2 z_2},&\hat{\nabla_{\omega_2}e^{(\boldsymbol{1})}}_{0}=\frac{q z_2-q+2 r z_2-2 z_2}{(z_2-1)^3 z_2} \\
\hat{\nabla_{\omega_2}e^{(\boldsymbol{1})}}_{1}=\frac{q (-z_2)+q-3 r z_2+3 z_2}{(z_2-1)^4 z_2},&\hat{\nabla_{\omega_2}e^{(\boldsymbol{1})}}_{2}=\frac{q z_2-q+4 r z_2-4 z_2}{(z_2-1)^5 z_2}
\end{array}
$$
As before we obtain the solution for the differential equation Eq. \eqref{ODEze}
via the recursive relation Eq. \eqref{re}
for $min=-1$, $m+1=0$ follows:
\begin{align*}
\zeta(y,0)_{0}=\frac{-qz_2+q-r z_2+z_2}{p (z_2-1)^2 z_2}
\end{align*}
for $m+1=1$ follows:
\begin{align*}
\zeta(y,0)_{1}=\frac{p z_2 (q+2 r-2)-q (p+r)+r z_2 (q+r-1)}{p (p+1) (z_2-1)^3 z_2}
\end{align*}
for $m+1=2$ follows:
\begin{align*}
\zeta(y,0)_2=\frac{q (p+r) (p+r+1)-z_2 \left(p^2 (q+3 r-3)+p \left(2 q r+q+3 r^2-3\right)+r (r+1) (q+r-1)\right)}{p (p+1) (p+2) (z_2-1)^4 z_2}
\end{align*}
If we now put these expressions together, we get for $\zeta(y,0)$ the following series expansion:
\begin{align*}
\zeta(y,0)=
&\frac{-qz_2+q-r z_2+z_2}{p (z_2-1)^2 z_2}+\frac{p z_2 (q+2 r-2)-q (p+r)+r z_2 (q+r-1)}{p (p+1) (z_2-1)^3 z_2}y\\
&\frac{q (p+r) (p+r+1)-z_2 \left(p^2 (q+3 r-3)+p \left(2 q r+q+3 r^2-3\right)+r (r+1) (q+r-1)\right)}{p (p+1) (p+2) (z_2-1)^4 z_2}y^2
\end{align*}
Thus the residue is:
\begin{equation}
\Res_{y=0}\kl{\zeta(y,0) h_0}=\frac{z_2 (q+r-1)-q}{p\left(z_2-1\right)^3 z_2}\label{Rze0}
\end{equation}
\paragraph{Series solution around $1-z_2$,} of $e=\frac{d z_1}{z_1(1-z_1-z_2)}$ around $1-z_2$ at local coordinats $y=z_1+z_2-1$  has the form:
\begin{equation}
-\frac{dy}{y(z_2-y-z_2+1)}=
\frac{dy}{y(z_2-1)}+\frac{dy}{(z_2-1)^2}+\frac{y dy}{(z_2-1)^3}+\frac{y^2 dy}{(z_2-1)^4}+O\left(y^3\right)\label{e11}
\end{equation}
The coefficients are
$\hat{e}_{-1}=\frac{1}{z_2-1},\;\hat{e}_{0}=\frac{1}{(z_2-1)^2},\;\hat{e}_{1}=\frac{1}{(z_2-1)^3},\;\hat{e}_{2}=\frac{1}{(z_2-1)^4}$. The series expansion of $\omega_1$,\eqref{omegaE1} around $1-z_2$ at local coordinates $y=z_1+z_2-1$  has the form,
\begin{equation*}
\left(\frac{r}{y}+\frac{p}{y-z_2+1}\right) d y=\frac{r dy}{y}-\frac{p dy}{z_2-1}-\frac{p y dy}{(z_2-1)^2}
-\frac{p y^2 dy}{(z_2-1)^3}+O\left(y^3\right)
\end{equation*}
the coefficients are
$\hat{\omega}_{1,-1}=r,\;\hat{\omega}_{1,0}=-\frac{p}{z_2-1},\;\hat{\omega}_{1,1}=-\frac{p}{(z_2-1)^2},\;\hat{\omega}_{1,2}=\frac{p}{(z_2-1)^3}$
The series expansion of $\omega_2$,\eqref{omegaE1} around $1-z_2$ at local coordinates $y=z_1+z_2-1$ has the form,
\begin{equation*}
\left(\frac{q}{z_2}-\frac{r}{y}\right) d z_2=\frac{q dz_2}{z_2}+\frac{r dz_2}{y}+O\left(y^3\right)
\end{equation*}
the coefficients are
$\hat{\omega}_{2,-1}=r,\;\hat{\omega}_{2,0}=\frac{q}{z_2}$
The series expansion of $\nabla_{\omega_2}e^{(\boldsymbol{1})}$, Eq. \eqref{nabome1} around $1-z_2$ at local coordinates $y=z_1+z_2-1$ has the form:
\begin{align*}
\nabla_{\omega_2}e^{(\boldsymbol{1})}(y,1-z_2)&=\frac{r-1}{y^2 (z_2-1)}+\frac{q z_2-q+r z_2-z_2}{y (z_2-1)^2 z_2}+\frac{q z_2-q+r z_2-z_2}{(z_2-1)^3 z_2}\\
&+\frac{y(q z_2-q+rz_2-z_2)}{(z_2-1)^4 z_2}+
\frac{y^2 (q z_2-q+r z_2-z_2)}{(z_2-1)^5}+O\left(y^3\right)
\end{align*}
the coefficients are:
$$
\begin{array}{cc}
\hat{\nabla_{\omega_2}e^{(\boldsymbol{1})}}_{-2}=\frac{r-1}{z_2-1},
&\hat{\nabla_{\omega_2}e^{(\boldsymbol{1})}}_{-1}=\frac{q z_2-q+r z_2-z_2}{(z_2-1)^2 z_2}\\
\hat{\nabla_{\omega_2}e^{(\boldsymbol{1})}}_{0}=\frac{qz_2-q+r z_2-z_2}{(z_2-1)^3 z_2},
&\hat{\nabla_{\omega_2}e^{(\boldsymbol{1})}}_{1}=\frac{q z_2-q+r z_2-z_2}{(z_2-1)^3
z_2}\\
\hat{\nabla_{\omega_2}e^{(\boldsymbol{1})}}_{2}=\frac{q z_2-q+r z_2-z_2}{(z_2-1)^5 z_2}&
\end{array}
$$
As before we obtain the solution for the differential equation \eqref{ODEze} for $min=-2$ via the recursive relation Eq. \eqref{re} for $m+1=-1$ follows:
\begin{align*}
\zeta(y,1-z_2)_{-1}=\frac{1}{z_2-1}
\end{align*}
for $m+1=0$ follows:
\begin{align*}
\zeta(y,1-z_2)_{0}=\frac{z_2 (p+q+r-1)-q}{r (z_2-1)^2 z_2}
\end{align*}
for $m+1=1$ follows:
\begin{align*}
\zeta(y,1-z_2)_{1}=\frac{(p+r) (z_2 (p+q+r-1)-q)}{r (r+1) (z_2-1)^3 z_2}
\end{align*}
for $m+1=2$ follows:
\begin{align*}
\zeta(y,1-z_2)_2=\frac{(p+r) (p+r+1)(z_2 (p+q+r-1)-q)}{r (r+1) (r+2) (z_2-1)^4 z_2}
\end{align*}
If we now put these expressions together, we get the following series expansion for $\zeta(y,1-z_2)$ we get:
\begin{align*}
\zeta(y,1-z_2)&=
\frac{1}{(z_2-1)y}+\frac{z_2 (p+q+r-1)-q}{r (z_2-1)^2 z_2}+\frac{(p+r) (z_2 (p+q+r-1)-q)}{r (r+1) (z_2-1)^3 z_2}y\\
&+\frac{(p+r) (p+r+1) (z_2 (p+q+r-1)-q)}{r (r+1) (r+2) (z_2-1)^4 z_2}y^2
\end{align*}
Thus the residue is:
\begin{align}
\Res_{y=1-z_2}\kl{\zeta(y,1-z_2) h_1}
&=\frac{1}{(z_2-1)}\frac{1}{(z_2-1)^2}+\frac{1}{(z_2-1)}\frac{z_2 (p+q+r-1)-q}{r (z_2-1)^2 z_2}\nonumber\\
&=\frac{z_2 (p+q+2 r-1)-q}{r \left(z_2-1\right)^3 z_2}\label{Rze1}
\end{align}
where we used the series expansion of $e_1$ as given in \eqref{e11}. 
\paragraph{series solution around $\infty$,}of $e=\frac{d z_1}{z_1(1-z_1-z_2)}$ at local coordinates $y=\frac{1}{z_1}$ and $dz_1=-\frac{dy}{y^2}$ has the form:
\begin{equation*}
\frac{dy}{1+y (z_2-1)}=dy-y(z_2-1)dy+y^2 \left(z_2^2-2z_2+1\right)dy+O\left(y^3\right)
\end{equation*}
The coefficients are
$\hat{e}_{0}=-1\;\hat{e}_{1}=(z_2-1),\;\hat{e}_{2}=\left(z_2^2-2z_2+1\right)$
The series expansion of $\omega_1$,\eqref{omegaE1} around $\infty$ at local coordinates $y=-\frac{1}{z_1}$ and $dz_1=-\frac{dy}{y^2}$ has the form,
\begin{align*}
-\frac{dy(p y (z_2-1)+p+r)}{y(y(z_2-1)+1)}=&\frac{(p+r)dy}{y}-r(z_2-1)dy
+r y \left(z_2^2-2z_2+1\right)dy+O\left(y^3\right)
\end{align*}
The coefficients are
$\hat{\omega}_{1,-1}=(p+r),\;\hat{\omega}_{1,0}=-r (z_2-1),\hat{\omega}_{1,1}=r \left(z_2^2-2z_2+1\right)$
The series expansion of $\omega_2$,\eqref{omegaE1} around $\infty$ at local coordinates $y=\frac{1}{z_1}$ and $dz_1=-\frac{dy}{y^2}$ has the form,
\begin{equation*}
\left(\frac{q}{z_2}+\frac{r y}{y(z_2-1)+1}\right)dz_2
=\frac{q dz_2}{z_2}+ydz_2 r-2ry^2(z_2-1)dz_2+O\left(y^3\right)
\end{equation*}
The coefficients are
$\hat{\omega}_{2,0}=\frac{dz_2 q}{z_2},\;\hat{\omega}_{2,1}=rdz_2,;\hat{\omega}_{2,2}=- r (z_2-1)dz_2$
The series expansion of $\nabla_{\omega_2}e^{(\boldsymbol{1})}$, Eq. \eqref{nabome1} around $\infty$ at local coordinates $y=\frac{1}{z_1}$ and $dz_1=-\frac{dy}{y^2}$ has thus the form:
\begin{align*}
\nabla_{\omega_2}e^{(\boldsymbol{1})}(y,\infty)&=\frac{q y^2}{z_2}+O\left(y^3\right)
\end{align*}
the coefficients is $\hat{\nabla_{\omega_2}e^{(\boldsymbol{1})}}_{0}=-\frac{q}{z_2}$
For the series expansion around $\infty$, we can already see that it does not contain any residues and therefore does not contribute to 
the intersection number Eq. \eqref{Resze}. 
\paragraph{Calculating the intersection number}
By equation Eq. \eqref{Resze} the intersection number is the sum over all the residues, so if we combine Eq. \eqref{XSEM1R3} and Eq. \eqref{XSEM1R2} we get for Eq. \eqref{FE1} the following expression. 
\begin{align}
\left\langle\nabla_{\hat{\omega}_2}e^{(\boldsymbol{1})}\Big{|}h^{(\boldsymbol{1})}\right\rangle
&=\Res_{y_{1,0}=0}\kl{\zeta(y,0) h_0}
+\Res_{y_{1}=1-z_2}\kl{\zeta(y,1-z_2) h_1}\\
&=\frac{z_2 (p+q+2 r-1)-q}{r \left(z_2-1\right)^3 z_2}+\frac{z_2 (q+r-1)-q}{p\left(z_2-1\right)^3 z_2}\\
&=\frac{(p+r) \left(z_2 (p+q+r-1)-q\right)}{p r \left(z_2-1\right)^3 z_2}\label{INOec}
\end{align}
Following the definition of $\Omega^{(2)}$, Eq. \eqref{OMEGA2D} we combine \eqref{INOec} and arrive at the following expression: 
\begin{align}
\Omega^{(2)}=\left\langle\nabla_{\hat{\omega}_2}e^{(\boldsymbol{1})}\Big{|}h^{(\boldsymbol{1})}\right\rangle \cdot \mathbf{C}_{(\mathbf{1})}^{-1}=\frac{-p-r+1}{1-z_2}+\frac{q}{z_2}\label{OmeC}
\end{align}
\subsection{Calculate $\left\langle\varphi^{(\mathbf{2})} \vert \varphi^{(\mathbf{2}) \vee}\right\rangle$, via the Intersection Number}
In order to calculate the intersection number of 
\begin{equation}
\left\langle\varphi^{(\mathbf{2})} \vert \varphi^{(\mathbf{2}) \vee}\right\rangle=\left\langle\frac{d \mathbf{z}}{z_1 z_2\left(1-z_1-z_2\right)}\Bigg{\vert} \frac{d \mathbf{z}}{z_1^2 z_2\left(1-z_1-z_2\right)}\right\rangle
\end{equation}
that are element of $\bra{\varphi^{(\mathbf{2})}}\in H^2(X,\nabla_{\omega})$ and $\bra{\varphi^{(\mathbf{2}) \vee}}\in H^2(X,\nabla_{\omega})$ respectively. 
We use the associated residue formula for the vector-valued intersection number Eq.\eqref{inreg22}
%\begin{equation}
%\left\bra{\varphi^{(\mathbf{2})}}\varphi^{(\mathbf{2}) \vee}\right\rangle=\sum_{x_i \in \mathcal{P}_{\Omega^{(2)}}}\Res\left(\psi_{x_i}\cdot \mathbf{C}^{(\mathbf{1})}\cdot \varphi^{(2) \vee}\right)\label{Intpb2pbv2}
%\end{equation}
\begin{equation}
\left\langle{\varphi^{(\mathbf{2})}} \Big| \varphi^{(\mathbf{2}) \vee}\right\rangle=\sum_{x_i \in \mathcal{P}_{\Omega^{(2)}}}\Res\left(\psi_{x_i}\cdot \mathbf{C}^{(\mathbf{1})}\cdot \varphi^{(2) \vee}\right)\label{Intpb2pbv2}
\end{equation}
We have already obtained $\mathbf{C}^{(\mathbf{1})}$ in Eq.\eqref{F1c}. So our next goal is to obtain $\varphi^{(2)}\in \mathrm{H}^{1}(X_2,\nabla_{\Omega_{2}})$ and $\psi_{x_i}$, which a reader probably already understands at this point, the solution to the following differential equation
\begin{equation}
\nabla_{\Omega_2}\kl{\psi_{x_i}}=\varphi^{(2)}\label{ODEpsi}
\end{equation}
We can obtain $\varphi^{(2)}$ by simply rearranging Eq. \eqref{Dphib2} and Eq. \eqref{Dphib2v}, into the form:
\begin{align}
\bra{\varphi^{(2)}}&=\braket{\varphi^{(\mathbf{2})}}{h^{(\boldsymbol{1})}}\mathbf{C}^{-1}_{(\mathbf{1})}\label{Dphi2}\\
\ket{\varphi^{(2) \vee}}&=\mathbf{C}^{-1}_{(\mathbf{1})}\braket{e^{(\boldsymbol{1})}}{\varphi^{(\mathbf{2}) \vee}}\label{Dphi2v}
\end{align}
So, in order to obtain the intersection number of $\braket{\varphi^{(\mathbf{2})}}{h^{(\boldsymbol{1})}}$, we will use as usual, the residue formula for the intersection number, which in this case takes the form:
\begin{equation}
\braket{\varphi^{(\mathbf{2})}}{h^{(\boldsymbol{1})}}=\sum_{x_i\in\mathcal{P}_{\Omega^(2)}}\Res\kl{\alpha\varphi^{(\mathbf{2})}}\label{Resapl}
\end{equation}
where $\alpha$ is the solution to the following differential equations:
\begin{equation}
\nabla_{\omega_1}\alpha=\varphi^{(\mathbf{2})}\label{ODEalpha}
\end{equation}
Furthermore, to obtain the intersection number of $\braket{e^{(\boldsymbol{1})}}{\varphi^{(\mathbf{2}) \vee}}$. We will use the residue formula for the intersection number, which in this case takes the form:
\begin{equation}
\braket{e^{(\boldsymbol{1})}}{\varphi^{(\mathbf{2}) \vee}}=\sum_{x_i\in\mathcal{P}_{\Omega^2}}\Res\kl{\beta\varphi^{(\mathbf{2}) \vee}}\label{Resbet}
\end{equation}
where $\beta$ is the solution to the following differential equations:
\begin{equation}
\nabla_{\omega_1}\beta=e^{(\boldsymbol{1})}\label{ODEbeta}
\end{equation}
So let's begin.
\subsubsection{Calculating $\left\langle\varphi^{(2)}\right|=\left\langle\varphi^{(\mathbf{2})} \vert h^{(\boldsymbol{1})}\right\rangle \cdot \mathbf{C}_{(\mathbf{1})}^{-1}$, via the Intersection Number}

\paragraph{The series solution around $0$,} of $\varphi^{(\mathbf{2})}=\frac{d\mathbf{z}}{z_1z_2(1-z_1-z_2)}$ at local coordinats $y=z_1$  has the form: 
\begin{equation*}
\frac{dy dz_2}{z_1z_2(1-z_1-z_2)}=-\frac{dydz_2}{y (z_2-1) z_2}+\frac{dydz_2}{(z_2-1)^2 z_
2}-\frac{y dy dz_2}{(z_2-1)^3 z_2}+\frac{y^2 dydz_2}{(z_2-1)^4 z_2}
+O\left(y^3\right)
\end{equation*} 
the coefficients are
$
\hat{\varphi}^{(\mathbf{2})}_{-1}=-\frac{dz_2}{(z_2-1) z_2},\;
\hat{\varphi}^{(\mathbf{2})}_{0}=\frac{dz_2}{(z_2-1)^2 z_2},\;
\hat{\varphi}^{(\mathbf{2})}_{1}=-\frac{dz_2}{(z_2-1)^3 z_2},\;
\hat{\varphi}^{(\mathbf{2})}_{2}=\frac{dz_2}{(z_2-1)^4 z_2}$.
The series expansion of $e=\frac{d z_1}{z_1(1-z_1-z_2)}$ around $0$ at local coordinates $y=z_1$  has the form: 
\begin{equation}
\frac{d y}{y(1-y-z_2)}=
\frac{dy}{y(1-z_2)}+\frac{dy}{(1-z_2)^2}+\frac{y dy}{(1-z_2)^3}+\frac{y^2 dy}{(1-z_2)^4}+O\left(y^3\right)
\end{equation} 
the coefficients are
$\hat{e}_{-1}=\frac{1}{(1-z_2)},\;\hat{e}_{0}=\frac{1}{(1-z_2)^2},\;\hat{e}_{1}=\frac{1}{(1-z_2)^3},\;\hat{e}_{2}=\frac{1}{(1-z_2)^4}$.
The series expansion of $\omega_1$,\eqref{omegaE1} around $0$ at local coordinates $y=z_1$  has the form,
\begin{equation}
\left(\frac{p}{y}-\frac{r}{1-y-z_2}\right) d y=\frac{p dy}{y}+\frac{r dy}{z_2-1}-\frac{r y dy}{(z_2-1)^2}+\frac{r y^2 dy}{(z_2-1)^3}+O\left(y^3\right)
\end{equation}
the coefficients are
$\hat{\omega}_{1,-1}=p,\;\hat{\omega}_{1,0}=\frac{r}{z_2-1},\;\hat{\omega}_{1,1}=\frac{r}{(z_2-1)^2},\;\hat{\omega}_{1,2}=\frac{r}{(z_2-1)^3}$
As before we obtain the solution for the differential equation Eq. \eqref{ODEalpha} via the recursive relation Eq. \eqref{re}
for $min=-1$ and thus $m+1=0$ follows: 
\begin{equation}
\alpha(y,0)_0=-\frac{1}{p (z_2-1)z_2}
\end{equation}
for $m+1=1$ follows
\begin{equation}
\alpha(y,0)_1=\frac{p+r}{p (p+1) (z_2-1)^2 z_2}
\end{equation}
for $m+1=2$ follows
\begin{equation*}
\alpha(y,0)_2=-\frac{(p+r) (p+r+1)}{p (p+1) (p+2) (z_2-1)^3 z_2}
\end{equation*}
If we now put these expressions together, we get for $\alpha(y,0)$ the following series expansion:
\begin{align*}
\alpha(y,0)=-\frac{1}{p (z_2-1)z_2}+\frac{p+r}{p (p+1) (z_2-1)^2 z_2}y-\frac{(p+r) (p+r+1)}{p (p+1) (p+2) (z_2-1)^3 z_2}y^2+O\left(y^3\right)
\end{align*}
Thus the residue is:
\begin{equation}
\Res_{y=0}\kl{\alpha(y,0) h_0}=\frac{1}{p\left(z_2-1\right)^2 z_2}\label{Ral0}
\end{equation}

\paragraph{The series solution around $1-z_2$,}
of $\varphi^{(\mathbf{2})}=\frac{d\mathbf{z}}{z_1z_2(1-z_1-z_2)}$ at local coordinats $y=z_1$  has the form: 
\begin{equation*}
\frac{dydz_2}{y z_2 (y+z_2-1)}=\frac{dydz_2}{y(z_2-1) z_2}+\frac{dydz_2}{(z_2-1)^2 z_2}
+\frac{ydydz_2}{(z_2-1)^3 z_2}+\frac{y^2dydz_2}{(z_2-1)^4 z_2}
+O\left(y^3\right)
\end{equation*} 
the coefficients are
$\hat{\varphi}^{(\mathbf{2})}_{-1}=\frac{1}{(z_2-1)z_2},\;\hat{\varphi}^{(\mathbf{2})}_{0}=\frac{1}{(z_2-1)^2 z_2},\;\hat{\varphi}^{(\mathbf{2})}_{1}=\frac{1}{(z_2-1)^3 z_2},\;\hat{\varphi}^{(\mathbf{2})}_{2}=\frac{1}{(z_2-1)^4 z_2}$.
The series expansion of $e=\frac{dz_1}{z_1(1-z_1-z_2)}$ around $1-z_2$ at local coordinates $y=1-z_2-z_1$  has the form: 
\begin{equation*}
\frac{d y}{y(1-y-z_2)}=\frac{dy}{y(z_2-1)}+\frac{dy}{(z_2-1)^2}+\frac{y dy}{(z_2-1)^3}
\frac{y^2 dy}{(z_2-1)^4}+O\left(y^3\right)
\end{equation*} 
the coefficients are
$\hat{e}_{-1}=\frac{1}{z_2-1},\;\hat{e}_{0}=\frac{1}{(1-z_2)^2},\;\hat{e}_{1}=\frac{1}{(1-z_2)^3},\;\hat{e}_{2}=\frac{1}{(1-z_2)^4}$.
The series expansion of $\omega_1$,\eqref{omegaE1} around $1-z_2$ at local coordinates $y=1-z_2-z_1$  has the form,
\begin{equation*}
\left(\frac{r}{y}-\frac{p}{1-y-z_2}\right) d y=\frac{r}{y}dy-\frac{p}{z_2-1}dy-\frac{p y}{(z_2-1)^2}dy-\frac{p y^2}{(z_2-1)^3}dy+O\left(y^3\right)
\end{equation*}
the coefficients are
$\hat{\omega}_{1,-1}=r,\;\hat{\omega}_{1,0}=-\frac{p}{z_2-1},\;\hat{\omega}_{1,1}=-\frac{p}{(z_2-1)^2},\;\hat{\omega}_{1,2}=-\frac{p}{(z_2-1)^3}$
As before we obtain the solution for the differential equation Eq. \eqref{ODEalpha}
via the recursive relation Eq. \eqref{re}
for $min=-1$ and thus $m+1=0$ follows: 
\begin{equation*}
\alpha(y,1-z_2)_0=\frac{1}{r(z_2-1)z_2}
\end{equation*}
for $m+1=1$ follows
\begin{equation*}
\alpha(y,1-z_2)_1=\frac{p+r}{r(r+1)(z_2-1)^2z_2}
\end{equation*}
for $m+1=2$ follows
\begin{equation*}
\alpha(y,1-z_2)_2=\frac{(p+r)(p+r+1)}{r(r+1)(r+2)(z_2-1)^3z_2}
\end{equation*}
If we now put these expressions together, we get for $\alpha(y,1-z_2)$ the following series expansion:
\begin{align*}
\alpha(y,1-z_2)=\frac{1}{r(z2-1)z_2}+\frac{p+r}{r(r+1)(z_2-1)^2z_2}y-\frac{(p+r)(p+r+1)}{r (r+1)(r+2)(z_2-1)^3 z_2}y^2+O\left(y^3\right)
\end{align*}
Thus the residue is:
\begin{equation*}
\Res_{y=1-z_2}\kl{\alpha(y,1-z_2) h_1}=\frac{1}{r\left(z_2-1\right)^2 z_2}\label{Ral1}
\end{equation*}
\paragraph{series solution around $\infty$,}
of $\varphi^{(\mathbf{2})}=\frac{d z_1}{z_1z_2(1-z_1-z_2)}$ at local coordinats $y=\frac{1}{z_1}$ and $dz_1=-\frac{dy}{y^2}$ has the form:
\begin{equation*}
\frac{ydy}{y(z_2-1)z_2+z_2}=-\frac{y dy}{z_2}
+O\left(y^3\right)
\end{equation*}
The coefficients are
$\hat{e}_{1}=\frac{1}{z_1},\hat{e}_2=\frac{(-1 + z_2)}{z_2}$
The series expansion of $e=\frac{d z_1}{z_1(1-z_1-z_2)}$ around $\infty$ at local coordinats $y=\frac{1}{z_1}$ and $dz_1
=-\frac{dy}{y^2}$ has the form:
\begin{equation*}
\frac{ydy}{y(z_2-1)+1}
=ydy-\frac{dy y^2 (z_2-1)}{z_2}+O\left(y^3\right)
\end{equation*}
The coefficients are
$\hat{e}_{2}=-1$,%$$\;\hat{e}_{0}=\frac{1}{(1-z_2)^2},\;\hat{e}_{1}=\frac{1}{(1-z_2)^3},\;\hat{e}_{2}=\frac{y^2}{(1-z_2)^4}$
The series expansion of $\omega_1$,\eqref{omegaE1} around $\infty$ at local coordinates $y=\frac{1}{z_1}$ and $dz_1=-\frac{dy}{y^2}$ has the form,
\begin{equation*}
\left(p+\frac{r}{y(z_2-1)+1}\right)=- ydy+\mathcal{O}(y^2)
\end{equation*}
We can already see for the series expansion around $\infty$, that it does not contain any residues, and therefore does not contribute to the intersection number.
\paragraph{Calculating the intersection number}
By equation Eq. \eqref{Resapl} the intersection number is the sum over all the residues, so if we combine Eq. \eqref{Ral0} and Eq. \eqref{Ral1} we get for Eq. \eqref{Resapl} the following expression:
\begin{align}
&\left\langle\varphi^{(\mathbf{2})} \vert h^{(\boldsymbol{1})}\right\rangle\nonumber\\
&=\Res_{y=0}\kl{\alpha(y,0) h_0}\nonumber
+\Res_{y=1-z_2}\kl{\alpha(y,1-z_2) h_1}\nonumber\\
&=\frac{1}{p\left(z_2-1\right)^2 z_2}+\frac{1}{r\left(z_2-1\right)^2 z_2}\nonumber\\
&=\frac{p+r}{pr\left(z_2-1\right)^2 z_2}\label{p2hc}
\end{align}
Following Eq. \eqref{Dphi2} we combine Eq. \eqref{p2hc} and Eq. \eqref{F1c} and arrive at the following expression: 
\begin{align}
\left\langle\varphi^{(2)}\right|=\left\langle\varphi^{(\mathbf{2})} \vert h^{(\boldsymbol{1})}\right\rangle \cdot \mathbf{C}_{(\mathbf{1})}^{-1}=\left\langle\frac{d z_2}{z_2}\right|\label{phi2c}
\end{align}
\subsubsection{Calculate $\left|\varphi^{(2) \vee}\right\rangle=\mathbf{C}_{(\mathbf{1})}^{-1} \cdot\left\langle e^{(\mathbf{1})} \vert \varphi^{(\mathbf{2}) \vee}\right\rangle$ , via the Intersection Number}
\paragraph{The series solution around $0$, for local coordinats $y=z_1$}
The series expansion of $\varphi^{(\mathbf{2})}=\frac{d\mathbf{z}}{z_1^2z_2(1-z_1-z_2)}$ around $0$ at local coordinates $y=z_1$ and $dy=dz_1$ has the form: 
\begin{align*}  
\frac{dy dz_2}{y^2z_2(1-y-z_2)}=&
-\frac{dy dz_2}{y^2(z_2-1) z_2}
+\frac{dy dz_2}{y(z_2-1)^2 z_2}
-\frac{dy dz_2}{(z_2-1)^3z_2}\\
&+\frac{y dy dz_2}{(z_2-1)^4 z_2}
-\frac{y^2 dy dz_2}{(z_2-1)^5 z_2}+O\left(y^3\right)\\
\end{align*}
the coefficients are
$\hat{\varphi}^{(\mathbf{2})}_{-2}
=-\frac{1}{(z_2-1)z_2},
\;\hat{\varphi}^{(\mathbf{2})}_{-1}=\frac{1}{(z_2-1)^2 z_2},\;\hat{\varphi}^{(\mathbf{2})}_{0}=-\frac{1}{(z_2-1)^3 z_2},\;\hat{\varphi}^{(\mathbf{2})}_{1}=\frac{1}{(z_2-1)^4 z_2}\;\hat{\varphi}^{(\mathbf{2})}_{2}=-\frac{1}{(z_2-1)^5 z_2}$.
The series expansion of $e=\frac{d z_1}{z_1(1-z_1-z_2)}$ around $0$ at local coordinats $y=z_1$, $dy=dz_1$  has the form: 
\begin{equation*}
\frac{d y}{y(1-y-z_2)}=
\frac{dy}{y(1-z_2)}+\frac{dy}{(1-z_2)^2}+\frac{y dy}{(1-z_2)^3}+\frac{y^2 dy}{(1-z_2)^4}+O\left(y^3\right)
\end{equation*} 
the coefficients are
$\hat{e}_{-1}=\frac{1}{(1-z_2)},\;\hat{e}_{0}=\frac{1}{(1-z_2)^2},\;\hat{e}_{1}=\frac{1}{(1-z_2)^3},\;\hat{e}_{2}=\frac{1}{(1-z_2)^4}$.
The series expansion of $\omega_1$,\eqref{omegaE1} around $0$ at local coordinates $y=z_1$ and $dy=dz_1$ has the form,
\begin{equation*}
\left(\frac{p}{y}-\frac{r}{1-y-z_2}\right) d y=-\frac{p dy}{y}+\frac{r dy}{z_2-1}-\frac{r y dy}{(z_2-1)^2}-\frac{r y^2 dy}{(z_2-1)^3}+O\left(y^3\right)
\end{equation*}
the coefficients are
$\hat{\omega}_{1,-1}=-p,\;\hat{\omega}_{1,0}=\frac{r}{z_2-1},\;\hat{\omega}_{1,1}=-\frac{r y}{(z_2-1)^2},\;\hat{\omega}_{2}=-\frac{r}{(z_2-1)^3}$
As before we obtain the solution for the differential equation Eq. \eqref{ODEbeta}
via the recursive relation Eq. \eqref{re}
for $min=-1$ and thus $m+1=0$ follows: 
\begin{equation*}
\beta(y,1-z_2)_0=\frac{1}{p(1-z_2)}
\end{equation*}
for $m+1=1$ follows
\begin{equation*}
\beta(y,1-z_2)_1=\frac{p+r}{p(p+1)(z_2-1)^2}
\end{equation*}
for $m+1=2$ follows
\begin{equation*}
\beta(y,1-z_2)_2=-\frac{(p+r)(p+r+1)}{p(p+1)(p+2)(z_2-1)^3}
\end{equation*}
If we now put these expressions together, we get for $\beta(y,0)$ the following series expansion:
\begin{align*}
\beta(y,0)=\frac{1}{p(1-z_2)}+\frac{p+r}{p(p+1)(z_2-1)^2}y-\frac{(p+r)(p+r+1)}{p(p+1)(p+2)(z_2-1)^3}y^2+O\left(y^3\right)
\end{align*}
Thus the residue is:
\begin{align}
\Res_{y=0}\kl{\beta(y,0) \varphi^{(\mathbf{2}),
\vee}_0}&=\frac{-2 p-r-1}{p (p+1) (z_2-1)^3z_2}
\label{Rbe0}
\end{align}
\paragraph{The series solution around $1-z_2$, for local coordinats $y=1-z_2-z_1$}
The series expansion of $\varphi^{(\mathbf{2})}=\frac{d\mathbf{z}}{z_1^2z_2(1-z_1-z_2)}$ around $1-z_2$ at local coordinats $y=1-z_2-z_1$  has the form: 
\begin{align*}  
\frac{1}{y z_2(-y-z_2+1)^2}=&
\frac{dydz_2}{y(z_2-1)^2z_2}-\frac{2dydz_2}{(z_2-1)^3z_2}\\
+&\frac{3ydydz_2}{(z_2-1)^4 z_2}
-\frac{4y^2dydz_2}{(z_2-1)^5 
z_2}+O\left(y^3\right)
\end{align*}
the coefficients are
$\hat{\varphi}^{(\mathbf{2})}_{-1}=\frac{1}{(z_2-1)^2 z_2},\;\hat{\varphi}^{(\mathbf{2})}_{0}=-\frac{2}{(z_2-1)^3 z_2}\;\hat{\varphi}^{(\mathbf{2})}_{1}=\frac{3}{(z_2-1)^4 z_2}\;\hat{\varphi}^{(\mathbf{2})}_{2}=-\frac{4}{(z_2-1)^5 z_2}$.
The series expansion of $e=\frac{dz_1}{z_1(1-z_1-z_2)}$ around $1-z_2$ at local coordinates $y=1-z_2-z_1$  has the form: 
\begin{equation*}
\frac{d y}{y(1-y-z_2)}=
-\frac{dy}{y(z_2-1)}
+\frac{dy}{(z_2-1)^2}
-\frac{ydy}{(z_2-1)^3}
+\frac{y^2dy}{(z_2-1)^4}
+O\left(y^3\right)
\end{equation*} 
the coefficients are
$\hat{e}_{-1}=-\frac{1}{z_2-1},\;\hat{e}_{0}=\frac{1}{(1-z_2)^2},\;\hat{e}_{1}=-\frac{1}{(1-z_2)^3},\;\hat{e}_{2}=\frac{1}{(1-z_2)^4}$.
The series expansion of $\omega_1$,\eqref{omegaE1} around $1-z_2$ at local coordinates $y=1-z_2-z_1$  has the form,
\begin{equation}
\left(\frac{p}{-y-z_2+1}-\frac{r}{y}\right) d y=
-\frac{r}{y}dy-\frac{p}{z_2-1}dy
+\frac{p y}{(z_1-1)^2}dy
+\frac{p y^2}{(z_2-1)^3}dy+O\left(y^3\right)
\end{equation}
the coefficients are
$\hat{\omega}_{1,-1}=-r,\;\hat{\omega}_{1,0}=-\frac{p}{z_2-1},\;\hat{\omega}_{1,1}=\frac{p}{(z_2-1)^2},\;\hat{\omega}_{1,2}=\frac{p}{(z_2-1)^3}$
As before we obtain the solution for the differential equation Eq. \eqref{ODEbeta}
via the recursive relation Eq. \eqref{re}
for $min=-1$ and thus $m+1=0$ follows: 
\begin{equation}
\beta(y,1-z_2)_0=\frac{1}{r(z_2-1)}
\end{equation}
for $m+1=1$ follows
\begin{equation}
\beta(y,1-z_2)_1=\frac{p+r}{r (r+1)(z_2-1)^2}
\end{equation}
for $m+1=2$ follows
\begin{equation}
\beta(y,1-z_2)_2=\frac{(p+r) (p+r+1)}{r(r+1)(r+2)(z_2-1)^3}
\end{equation}
If we now put these expressions together, we get for $\beta(y,1-z_2)$ the following series expansion:
\begin{align}
\beta(y,1-z_2)=\frac{1}{r(z_2-1)}+\frac{p+r}{r(r+1)(z_2-1)^2}y
+\frac{(p+r)(p+r+1)}{r(r+1)(r+2)(z_2-1)^3}y^2+O\left(y^3\right)
\end{align}
Thus the residue is:
\begin{align}  
\Res_{y=1-z_2}\kl{\beta(y,1-z_2)\varphi^{(\mathbf{2}) \vee}}
&=-\frac{1}{r\left(z_2-1\right){}^3 z_2}\label{Rbe1}
\end{align}
\paragraph{series solution around $\infty$, for local coordinats $y=\frac{1}{z_1}$ and $dz_1=-\frac{dy}{y^2}$,}
The series expansion of $\varphi^{(\mathbf{2})}=\frac{d z_1}{z_1z_2(1-z_1-z_2)}$ around $\infty$ has the form:
\begin{equation}
\frac{ydy}{y z_2(z_2-1) +z_2}=\frac{ydy}{z_2}-\frac{dy_1 y^2 (z_2-1)}{z_2}+O\left(y^3\right)
\end{equation}
The coefficients are
$\hat{\varphi}_{1}=\frac{1}{z_2},\hat{\varphi}_2=\frac{(z_2-1)}{z_2}$
The series expansion of $e=\frac{d z_1}{z_1(1-z_1-z_2)}$ around $\infty$ at local coordinats $y=\frac{1}{z_1}$ and $dz_1=-\frac{dy}{y^2}$ has the form:
\begin{equation*}
\frac{dy}{y(z_2-1)z_2+z_2}=dy-ydy(z_2-1)+\left(z_2^2-2 z_2+1\right)y^2 dy+\mathcal{O}(y^3)
\end{equation*}
The coefficients are
$\hat{e}_{2}=1$,%$$\;\hat{e}_{0}=\frac{1}{(1-z_2)^2},\;\hat{e}_{1}=\frac{1}{(1-z_2)^3},\;\hat{e}_{2}=\frac{y^2}{(1-z_2)^4}$
The series expansion of $\omega_1$,\eqref{omegaE1} around $\infty$ at local coordinates $y=\frac{1}{z_1}$ and $y^2dy=dz_1$ has the form,
\begin{equation*}
y^3 \left(p+\frac{r}{y (z_2-1)+1}\right)dy=O\left(y^3\right)
\end{equation*}
We can already see for the series expansion around $\infty$, that it does not contain any residues, and therefore does not contribute to the intersection number. 
\paragraph{Calculating the intersection number}
By equation \eqref{inreg22} the intersection number is the sum over all the residues, so if we combine Eq. \eqref{Rbe0} and Eq. \eqref{Rbe1} we get for Eq. \eqref{Resbet} the following expression:
\begin{align}
&\left\langle\varphi^{(\mathbf{2})} \vert h^{(\boldsymbol{1})}\right\rangle\nonumber\\
&=\Res_{y=0}\kl{\beta(y,0) h_0}
+\Res_{y=1-z_2}\kl{\beta(y,1-z_2) h_1}\nonumber\\
&=\frac{-2 p-r-1}{p (p+1) (z_2-1)^3z_2}-\frac{1}{r\left(z_2-1\right){}^3 z_2}\nonumber\\
&=-\frac{(p+r) (p+r+1)}{p(p+1)r(z_2-1)^3 z_2}\label{ber}
\end{align}
Following the definition of Eq. \eqref{Dphi2v}, we combine Eq. \eqref{ber} and Eq. \eqref{F1c} and arrive at the following expression: 
\begin{align}
\left|\varphi^{(2) \vee}\right\rangle=\mathbf{C}_{(\mathbf{1})}^{-1} \cdot\left\langle e^{(\mathbf{1}
)} \vert \varphi^{(\mathbf{2}) \vee}\right\rangle=\left|\frac{(1+p+r) d z_2}{(1+p)\left(1-z_2\right) z_2}\right\rangle
\end{align}
\subsection{Calculate $\left\langle\varphi^{(2)}\right|$ , via the Intersection Number}
Following equation \eqref{inreg22}, our goal is now to obtain the solution for the differential equation Eq. \eqref{ODEpsi} via the recursive relation Eq. \eqref{re}, which, if we plug it back into Eq. \eqref{inreg22} will give us together with the results we already obtained Eq. \eqref{F1c}and Eq \eqref{phi2c}. The vector-valued intersection number.
So let's continue.
\paragraph{The series solution around $0$,}$\varphi^{(2)}=\frac{dz_2}{z_2}$ at local coordinats $y=z_2$ has the form: 
\begin{equation*}
\frac{dy}{y}=\frac{dy}{y}
+O\left(y^3\right)
\end{equation*} 
the coefficients are
$
\hat{\varphi}^{(2)}_{-1}=1$.
The series expansion of $\varphi^{(2) \vee}=\frac{(p+r+1)dz_2}{(p+1)(1-z2) z2}$ around $0$ at local coordinates $y=z_1$ has the form: 
\begin{equation*}
\frac{(p+r+1)dy}{(p+1)(1-y)y}=
+\frac{(p+r+1)dy}{(p+1)y}
+\frac{(p+r+1)dy}{p+1}
+\frac{y(p+r+1)dy}{p+1}
+\frac{y^2(p+r+1)dy}{p+1}
+O\left(y^3\right)
\end{equation*} 
the coefficients are
$\hat{\varphi}_{-1}^{(2)\vee}=\frac{(p+r+1)}{p+1},\;\hat{\varphi}_{0}^{(2)\vee}=\frac{(p+r+1)}{p+1},\;\hat{\varphi}^{(2)\vee}_{1}=\frac{(p+r+1)}{p+1},\;\hat{\varphi}^{(2)\vee}_{2}=\frac{(p+r+1)}{p+1}$.
The series expansion of $\Omega$,\eqref{OmeC} around $0$ at local coordinates $y=z_2$  has the form,
\begin{equation*}
\left(\frac{-p-r+1}{1-y}+\frac{q}{y}\right)d y=\frac{qdy}{y}-(p+r-1)dy-y(p+r-1)dy
-y^2(p+r-1)dy+O\left(y^3\right)
\end{equation*}
the coefficients are
$\hat{\Omega}_{0}=q,\;\hat{\Omega}_{1}=(1-p-r),\;\hat{\Omega}_{2}=(1-p-r),\;\hat{\Omega}_{2}=(1-p-r)$
The series expansion of $\mathbf{C}^{(\mathbf{1})}$,Eq. \eqref{F1c} around $0$ at local coordinates $y=z_2$  has the form,
\begin{equation*}
\frac{p+r}{pr(1-y)^2}=
\frac{p+r}{pr}
+\frac{2y(p+r)}{pr}
+\frac{3y^2(p+r)}{pr}
+O\left(y^3\right)
\end{equation*}
the coefficients are
$\hat{\mathbf{C}}^{(\mathbf{1})}_{0}=\frac{p+r}{pr},\;\hat{\mathbf{C}}^{(\mathbf{1})}_{1}=\frac{2(p+r)}{pr},\;\hat{\mathbf{C}}^{(\mathbf{1})}_{2}=\frac{3 (p+r)}{pr}$
As before we obtain the solution for the differential equation Eq. \eqref{ODEpsi} via the recursive relation Eq. \eqref{re}
for $min=-1$ and thus $m+1=0$ follows: 
\begin{equation*}
\psi(y,0)_0=\frac{1}{q}+O\left(y\right)
\end{equation*}
For the pole around $0$ an expansion in $0$-order, is sufficient. Following the residue formula for the intersection number Eq. \eqref{Intpb2pbv2}, we arrive at the following expression:
\begin{equation}
\Res_{y=0}\kl{\psi(y,0)\mathbf{C}^{(\mathbf{1})}\varphi^{(2) \vee}_0}=\frac{(p+r)(p+r+1)}{(p+1)prq}\label{Rpsi0}
\end{equation}
\paragraph{The series solution around $1-z_2$,}
of $\varphi^{(2)}=\frac{dz_2}{z_2}$ at local coordinates $y=z_2-1$ and $dy=dz_2$ has the form: 
\begin{equation*}
\frac{dy}{y+1}=1-ydy+y^2dy+O\left(y^3\right)
\end{equation*} 
the coefficients are
$\hat{\varphi}^{(\mathbf{2})}_{0}=1,\;\hat{\varphi}^{(\mathbf{2})}_{1}=-1 ,\;\hat{\varphi}^{(\mathbf{2})}_{2}=1$.
The series expansion of $\varphi^{(2) \vee}=\frac{dz_2 (p+r+1)}{(p+1)(1-z_2)z_2}$ around $1$ at local coordinates $y=z_2-1$ and $dy=dz_2$ has the form: 
\begin{equation*}
-\frac{(p+r+1)dy}{(p+1)y(y+1)}=
-\frac{p+r+1}{(p+1)y}
+\frac{p+r+1}{p+1}
-\frac{y(p+r+1)}{p+1}
+\frac{y^2(p+r+1)}{p+1}+O\left(y^3\right)
\end{equation*} 
the coefficients are
$\varphi^{(2) \vee}_{-1}=-\frac{p+r+1}{p+1},\;\varphi^{(2) \vee}_{0}=\frac{p+r+1}{p+1},\;\varphi^{(2) \vee}_{1}=-\frac{p+r+1}{p+1},\;\varphi^{(2) \vee}_{2}=\frac{p+r+1}{p+1}$.
The series expansion of $\Omega^2$,\eqref{OmeC} around $1$ at local coordinates $y=z_2-1$ has the form,
\begin{equation*}
\kl{\frac{q}{y+1}-\frac{-p-r+1}{y2}}d y=\frac{(p+r-1)dy}{y}+qdy-q yd_2+q y^2dy+O\left(y^3\right)
\end{equation*}
the coefficients are
$\hat{\Omega}_{-1}=(p+r-1),\;\hat{\omega}_{0}=q,\;\hat{\Omega}_{1}=-q,\;\hat{\Omega}_{2}=q$
As before we obtain the solution for the differential equation Eq. \eqref{ODEpsi}
via the recursive relation Eq. \eqref{re}
for $min=0$ and thus $m+1=0$ follows: 
\begin{equation*}
\psi(y,1)_0=0
\end{equation*}
for $m+1=1$ follows
\begin{equation*}
\psi(y,1)_1=\frac{1}{p+r}
\end{equation*}
for $m+1=2$ follows
\begin{equation*}
\psi(y,1)_2=-\frac{p+q+r}{(p+r) (p+r+1)}
\end{equation*}
If we now put these expressions together, we get for $\psi(y,1)$ the following series expansion:
\begin{align*}
\psi(y,1)=\frac{y}{p+r}-\frac{y^2(p+q+r)}{(p+r) (p+r+1)}+O\left(y^3\right)
\end{align*}
Thus the residue is:
\begin{equation}
\Res_{y=1}\kl{\psi(y,1)\mathbf{C}^{(\mathbf{1})}\varphi^{(2) \vee}_0}=\frac{2 p+q+2 r+1}{pr(1+p)}\label{Rpsi1}
\end{equation}
\paragraph{The series solution around $\infty$,}
of $\varphi^{(2)}=\frac{dz_2}{z_2}$ at local coordinates $y=\frac{1}{z_2}$ and $dz_2=-\frac{dy}{y^2}$ has the form: 
\begin{equation*}
-\frac{dy}{y}=-\frac{dy}{y}+O\left(y^3\right)
\end{equation*} 
the coefficients are
$\hat{\varphi}^{(\mathbf{2})}_{0}=1$. The series expansion of $\varphi^{(2) \vee}=\frac{dz_2(p+r+1)}{(p+1)(1-z_2)z_2}$ around $\infty$ at local coordinates $y=\frac{1}{z_2}$ and $dz_2=\frac{dy}{y}$ has the form: 
\begin{equation*}
-\frac{(p+r+1)dy}{(p+1)(y-1)}=
+\frac{(p+r+1)dy}{p+1}
+\frac{y(p+r+1)dy}{p+1}
+\frac{y^2(p+r+1)dy}{p+1}{p+1}+O\left(y^3\right)
\end{equation*} 
the coefficients are
$\varphi^{(2) \vee}_{0}=\frac{(p+r+1)}{p+1},\;\varphi^{(2) \vee}_{1}=\frac{p+r+1}{p+1},\;\varphi^{(2) \vee}_{2}=\frac{p+r+1}{p+1}$.
The series expansion of $\Omega^{(\mathbf{2})}$,\eqref{OmeC} around $\infty$ at local coordinates $z_2=\frac{1}{y}$ and $dz_2=\frac{dy}{y^2}$ has the form,
\begin{equation*}
-\frac{dy(p-q y+q+r-1)}{(y-1) y}
=\frac{dy(p+q+r-1)}{y}+dy(p+r-1)+dy y (p+r-1)+dy y^2(p+r-1)+O\left(y^3\right)
\end{equation*}
the coefficients are
$\hat{\Omega}_{-1}=(p+q+r-1),\;\hat{\Omega}_{0}=(p+r-1),\;\hat{\Omega}_{1}=(-1 + p + r),\;\hat{\Omega}_{2}=(-1+p+r)$
As before we obtain the solution for the differential equation Eq. \eqref{ODEpsi}
via the recursive relation Eq. \eqref{re}
for $min=0$ and thus $m+1=0$ follows: 
\begin{equation*}
\psi(y,\infty)_0=\frac{1}{p+q+r-1}
\end{equation*}
For the pole around $\infty$ an expansion in 0-order, is sufficient. Following the residue formula for the intersection number Eq. \eqref{Intpb2pbv2}, we arrive at the following expression:
\begin{align*}
\psi(y,\infty)=\frac{y}{p+r}-\frac{y^2(p+q+r)}{(p+r) (p+r+1)}+O\left(y^3\right)
\end{align*}
We find that the Phol arund $\infty$ has no resindue, 
\paragraph{Calculating the intersection number}
By equation \eqref{inreg22} the intersection number is the sum over all the residues, so if we combine Eq. \eqref{Rpsi0} and Eq. \eqref{Rpsi1} we get for Eq. \eqref{inreg22} the following expression:
\begin{align*}
&\left\langle\frac{d \mathbf{z}}{z_1 z_2\left(1-z_1-z_2\right)} \Bigg{|} \frac{d \mathbf{z}}{z_1^2 z_2\left(1-z_1-z_2\right)}\right\rangle\\
&=\Res_{y=0}\kl{\psi(y,0)\mathbf{C}^{(\mathbf{1})}\varphi^{(2) \vee}_0}
+\Res_{y=1}\kl{\psi(y,1)\mathbf{C}^{(\mathbf{1})}\varphi^{(2) \vee}_0}\\
&=\frac{(p+r)(p+r+1)}{(p+1)prq}+\frac{2 p+q+2 r+1}{pr(1+p)}\\
&=\frac{(p+q+r)(p+q+r+1)}{(p+1)prq}
\end{align*}
\newpage

%\graphicspath{{./images/}}
\chapter{The Cut Box and Cross-Box and Code}\label{GravBox}
\section{Feynman Integrals Reduction via Intersection Numbers}
We have now reached the point where we can discuss a full reduction, including the sub-topologies of the Feynman integral, via intersection theory. This section is mainly based on the work of \cite{Gasparotto:2023cdl} and \cite{Mattiazzi:2022zbo}. For the discussion ahead, it is beneficial to recall some definitions, namely the Feynman integral in Baikov representation: 
\begin{equation}
I_{a_1, \ldots, a_n}=\int_\gamma u(\mathbf{z}) \varphi(\mathbf{z}),
\end{equation}
Such that, as stated before :
\begin{equation}
u(\mathbf{z})=(\mathcal{B}(\mathbf{z}))^{\frac{D-l-E-1}{2}}, \quad \varphi(\mathbf{z})=\frac{\mathrm{N}(\mathbf{z})}{z_1^{a_1} \ldots z_n^{a_n}} d \mathbf{z}
\end{equation}
and
\begin{equation}
    \gamma \in X=\mathbb{C}^n \backslash(\mathcal{B}=0), \quad \text { s.t. } \quad \mathcal{B}(\partial \gamma)=0,\quad\left(a_1, \ldots, a_n\right) \in \mathbb{Z}^n
\end{equation}
Since $a_i$ can take positive values, it is possible for some of the Baikov variables to be part of the numerator of $\varphi$, while only the Baikov variables for which $a_i$ is negative and thus part of the denominator of $\varphi$ are called $\kl{a_1,...,a_{n_{den}}}$. This poses a problem since $\varphi$ is not well defined on $X$, and the poles of $\varphi(\mathbf{z})$ are not regulated by $u(\mathbf{z})$. Therefore, the techniques developed in section \ref{Intersection Theory II} cannot be applied directly. A possible solution is a slightly modified version of the original $u(\mathbf{z})$, namely.
\begin{equation}
u_{\mathrm{reg}}(z)=\prod_{i=1}^{n_{\mathrm{den}}} z_i^{\rho_i} u(\mathbf{z})\label{ureg}
\end{equation}
where $\rho\notin\mathbb{Z}$, for $i\in[1,n_{den}]$, are termed regulators. In this case,
\begin{equation}
X=\mathbb{C}^{n}/(\mathcal{B}=0,\bigcup^{n_{den}}_{i=1}z_i=0)
\end{equation}
The regulated Baikov  polynomial, for the Box is than: 
\begin{equation}
u_{\mathrm{reg},\mathrm{B}}(z)=\prod_{i=1}^{n_{\mathrm{den}}} z_i^{\rho_i} \mathcal{B}^{\frac{D-5}{2}}_{\mathrm{B}}(\mathbf{z})\label{uregB}
\end{equation}
and the cut vision of the box takes form:
\begin{equation}
\begin{aligned}
u_{\mathrm{reg},\mathrm{B},\mathrm{cut}}(z)
=&4^{5-d}z_2^{\rho_1}z_3^{\rho_2}\left(t \left(m_1^4 t+m_2^4 t+t(s+z_2)^2-2m_2^2\right.\right. \\
& (s t-(t-2z_2)(z_2-z_3))-2m_1^2\left(m_2^2 t+s t+(t-2z_3)(z_2-z_3)\right) \\
&+\left.\left.2 s tz_3-2(2 s+t)z_2z_3+tz_3^2\right)\right)^{\frac{(D-5)}{2}}
\end{aligned}
\end{equation}
For the Cross-box they are:
\begin{equation}
u_{\mathrm{reg},\mathrm{CB}}(z)=\prod_{i=1}^{n_{\mathrm{den}}} z_i^{\rho_i} \mathcal{B}^{\frac{D-5}{2}}_{\mathrm{BC}}(\mathbf{z})\label{uregCB}
\end{equation}
and the cut vision of the cross-box takes form:
\begin{equation}
\begin{aligned}
u_{\mathrm{reg},\mathrm{B},\mathrm{cut}}(z)
=&4^{(5-d)} z_2^{\rho_1}z_3^{\rho_{2}}\left(t \left(m_1^{4} t+m_2^{4} t+t(s+t-z_2)^{2}-2 m_2^{2}\right.\right. \\
&(s t+(t-2 z_2)(t-z_2-z_3))-2 m_1^{2}\left(m_2^{2} t+s t+(t-2 z_3) \right. \\
&\left.\left.(t-z_2-z_3))-2 t(s+t) z_3+2(2 s+t) z_2 z_3+t z_3^{2}\right)\right)^{\frac{(D-5)}{2}}
\end{aligned}
\end{equation}
It is now possible, with the help of \eqref{ureg}, to perform the reduction of the Feynman integral to its master integrals. The regulators $\rho_i$ are set to $0$ when the reduction coefficients are determined. However, the equation \eqref{ureg} has the disadvantage that now more non-physical parameters are involved in the reduction. To minimise the effort for algebraic manipulations, it is often useful to set all regulators equal, $\rho_i=\rho$ for $i\in\{1,...,n_{den}\}$. We consider a slightly modified version of \eqref{ureg}, in which we apply the multivariate intersection to the individual sectors of Feynman integrals, as in \ref{FIE}. The sectors of the full set of possible poles are then denoted by $\Sigma\subset\{1,...,n_{den}\}$ and $u(\mathbf{z})$ by $u_{\Sigma}(\mathbf{z})$.
%In this case, we could not consider differential forms with poles at $z_i$ with $i\in\Sigma$. 
%The auxiliary object $u_{\Sigma}$ is useful, since:
\begin{equation}
\omega_{\Sigma}=d\log\left\vert u_{\Sigma}\right\vert\label{omsig}
\end{equation}
which gives the number of master integrals in a sector $\Sigma$ for all possible subsectors. 
This results in a possible strategy for identifying a suspected master integral by, 
\begin{itemize}
    \item[1] start from the smallest possible sectors
    \item[2] consider the corresponding $d\log(u_{\Sigma})$
    \item[3] count the number of critical points of $\nu_{\Sigma}$
    \item[4] update the list of master integrals, without overcounting the master integrals, common to the subsectors.
\end{itemize}
This procedure does not take into account symmetry relations, such as those implemented in the public code so that the number of independent master integrals can in principle be further reduced. 

Further, can the strategy be used to determine the base elements for each internal layer required for the multivariate intersection number algorithm described in section 
\ref{nVariableIntersectionNumber}
as we will do the following.
\section{Dimension of $\mathrm{H}^n$}\label{Dimension2}
From section \ref{Group}, Eq. \eqref{nu} we recall that a number of solutions of $\omega$,
\begin{equation}
\nu_{(\mathbf{n})}=\# \text { solutions of: } \omega_1=0, \ldots, \omega_n=0\label{nsolw}
\end{equation}
gives us the dimension of $\mathrm{H}^n$ i.e 
\begin{equation}\nu_{(\mathbf{n}),\Sigma}=\operatorname{dim} \mathrm{H}^n_{\Sigma}\label{nunum1}
\end{equation}
and thus for our basis.
As mentioned earlier, we need to introduce a variable ordering such that we have a suitable base for each layer of the Fibration, as defined in chapter \ref{Intersection_Theory_II}. The variable ordering should be chosen such that the size of the base on each layer of the Fibration is kept minimal. This can be achieved by using \eqref{nsolw}, as mentioned earlier, by solving for each combination of integration variables and then taking the order that gives the lowest count. In order to calculate the dimension $\nu$ of each fibration, a Mathematica code was used, which has been explained in  Apendex \ref{dimensioncode}
In addition to having the right number of elements also the basis also needs to be in the correct subsector of the master integral vector space as explained in \ref{FIE}.
Following the Feynman integrals decomposition formula of Eq. \eqref{eqMIINTER} stated earlier in section \ref{LiRe}, we get for the different sectors: 
\begin{equation}
  I=\int_{\gamma} u_{\Sigma} \varphi=\left\langle\varphi| u_{\Sigma}\otimes\gamma\right]=\sum_{i=1}^{\nu_{\Sigma}} c_i\left\langle e_i| \mathcal{\mathcal{C}}\right]=
  \sum_{i=1}^{\nu_{\Sigma}} c_i \int_{\gamma} u_{\Sigma} e_i=\sum_{i=1}^{\nu_{\Sigma}} c_i \mathcal{I}_i\label{eqMIINTER2}
\end{equation}
with
\begin{equation}
c_i=\sum_{j=1}^\nu\left\langle\varphi \mid h_j\right\rangle\left(\mathbf{C}^{-1}\right)_{j i} \quad\text{and}\quad\mathbf{C}_{i j}=\left\langle e_i \mid h_j\right\rangle\label{eqCof2}
\end{equation}

\subsection{Box}\label{Basis_fibration}
In order to find the variable ordering for each layer of the fibration that gives the minimal basis of the box, we solve \eqref{omsig} for the set of integration variables and their possible subsets, which is nothing but the dimensions of the twisted co-homology groups at each layer of the fibration. The code used can be found under \url{https://www.wolframcloud.com/obj/wfv651/Published/uregEd_Box3grav%20(1)%20(1).nb}
\begin{center}
\begin{tabular}{|c|c|c|c|c|c|}
\hline solve & $\nu_{\Sigma}$& solve & $\nu_{\Sigma}$& solve & $\nu_{\Sigma}$ \\
\hline$z_1$ & 2 & $z_1, z_3$ & 4 &$z_1, z_2, z_3$ & 8 \\
\hline$z_2$ & 2 & $z_1, z_4  $ & 4 &$z_1, z_2, z_4$ & 8 \\
\hline$z_3$ & 2 & $z_2, z_3$   & 4 &$z_1, z_3, z_4$ & 8 \\
\hline$z_4$ & 2 & $z_2, z_4$ & 4 &$z_2, z_3, z_4$ & 8 \\
\hline$z_1, z_2$ & 4 & $z_3, z_4$ & 4 & $z_1, z_2, z_3, z_4$ & 7 \\
\hline
\end{tabular}
\end{center}
we choose the variable ordering from the inside out $z_1,z_2,z_3,z_4$, The minimum dimension of the twisted co-homology group at  the individual layers of the fibration is then:
\begin{equation}
\nu_{\Sigma,1}=2\quad\nu_{\Sigma,1,2,}=4\quad\nu_{\Sigma,1,2,3}=8\quad\nu_{\Sigma,1,2,3,4}=7
\end{equation}
In any other order, we choose, we would either get the same or a larger base in each layer of the fibration. Further, it follows from the equation \eqref{omsig} and \eqref{nsolw} that:  For $\nu_{\Sigma,1}$ there are $2$ master integrals with poles at $z_1$. 
For $\nu_{\Sigma,1,2}$ there are $4$ master integrals with poles at either $z_1$ or $z_2$, or both.  For $\nu_{1,2,3}$ there are $8$ master integrals with poles either at $z_1$, $z_2$ or $z_3$ or in any of the possible pairs or the complete triplet. For $\nu_{1,2,3,4}$ there are $7$ master integrals with polls either at $z_1$, $z_2$, $z_3$ or $z_4$ or in any possible subsets. Having determined the dimension of the twisted co-homology group at each level and the base order, we now need to find suitable base elements that have the desired size. This is done by varying which variables get regulated when applying eq. \eqref{omsig} at the corresponding layer. The base size of each fibrillation layer is calculated with a Mathematica code that can be found in the appendix \ref{base_size_code}, including a short description. The code used can be found at. \url{https://www.wolframcloud.com/obj/wfv651/Published/uregEd_Box3grav%20(1)%20(1).nb}
\paragraph{Laya 1 set $z_1$:}
\begin{center}
\begin{tabular}{|c|c|c|c|c|c|c|c|}
\hline Variable & Basis size & Variable & Basis size &  Variable & Basis size & Variable & Basis size \\
\hline $0$ & 1 & $z_4$ & 1 & $z_2, z_3$ & 1 & $z_1, z_2, z_4$ & 2 \\
\hline $z_1$ & 2 & $z_1, z_2$ & 2 & $z_2, z_4$ & 1& $z_1, z_3, z_4$ & 2 \\
\hline$z_2$ & 1 & $z_1, z_3$ & 2 & $z_3, z_4$ & 1& $z_2, z_3, z_4$ & 1 \\
\hline$z_3$ & 1 & $z_1, z_4$ & 2 & $z_1, z_2, z_3$ & 2 & $z_1, z_2, z_3, z_4$ & 2 \\
\hline
\end{tabular}
\end{center}
 The associated basis differential forms are:
\begin{equation}
\left\langle e_1^{(4)}\right|=\left\langle d \mathbf{z}\right|\quad\left\langle e_2^{(\mathbf{4})}\right|=\left\langle\frac{d \mathbf{z}}{z_1}\right|
\end{equation}
%\begin{align}
%\left\langle e_3^{(4)}\right|&=\left\langle\frac{d \mathbf{z}}{z_1 z_4}\right|\\
%\left\langle e_4^{(4)}\right|&=\left\langle\frac{d \mathbf{z}}{z_2 z_3}\right|\\
%\left\langle e_5^{(4)}\right|&=\left\langle\frac{d \mathbf{z}}{z_1 z_2 z_4}\right|\\ \left\langle e_6^{(4)}\right|&=\left\langle\frac{d \mathbf{z}}{z_1 z_3 z_4}\right|\\
%\left\langle e_7^{(4)}\right|&=\left\langle\frac{d \mathbf{z}}{z_1 z_2 z_3 z_4}\right|
%\end{align}
\paragraph{Laya 2 set $z_1,z_2$:}
\begin{center}
\begin{tabular}{|c|c|c|c|}
\hline Variable & Basis size & Variable & Basis size \\
\hline $0$ & 1 & $z_4$ & 1 \\
\hline$z_1$ & 2 & $z_1, z_2$ & 4 \\
\hline$z_2$ & 2 & $z_1, z_3$ & 2 \\
\hline$z_3$ & 1 & $z_1, z_4$ & 2\\
\hline
\end{tabular}
\end{center}
\begin{center}
\begin{tabular}{|c|c|c|c|}
\hline Variable & Basis size & Variable & Basis size \\
\hline$z_2, z_3$ & 2 & $z_1, z_2, z_4$ & 4 \\  
\hline$z_2, z_4$ & 2 & $z_1, z_3, z_4$ & 2 \\
\hline$z_3, z_4$ & 1 & $z_2, z_3, z_4$ & 2 \\
\hline$z_1, z_2, z_3$ & 4 & $z_1, z_2, z_3, z_4$ & 4 \\
\hline
\end{tabular}
\end{center}
 The associated basis differential forms are:
 \begin{equation}
\left\langle e_1^{(4)}\right|=\left\langle d \mathbf{z}\right|
\quad
\left\langle e_2^{(\mathbf{4})}\right|=\left\langle\frac{d \mathbf{z}}{z_1}\right|
\quad
\left\langle e_2^{(\mathbf{4})}\right|=\left\langle\frac{d \mathbf{z}}{z_1}\right|\quad \left\langle e_4^{(4)}\right|=\left\langle\frac{d \mathbf{z}}{z_1 z_2}\right|
\end{equation}
%\begin{align}
%\left\langle e_1^{(4)}\right|&=\left\langle d \mathbf{z}\right|\\
%\left\langle e_2^{(\mathbf{4})}\right|&=\left\langle\frac{d \mathbf{z}}{z_1}\right|\\
%\left\langle e_3^{(4)}\right|&=\left\langle\frac{d \mathbf{z}}{z_2}\right|\\
%\left\langle 
%e_4^{(4)}\right|&=\left\langle\frac{d \mathbf{z}}{z_1 z_2}\right|
%\left\langle e_4^{(4)}\right|&=\left\langle\frac{d \mathbf{z}}{z_2 z_3}\right|\\
%\left\langle e_5^{(4)}\right|&=\left\langle\frac{d \mathbf{z}}{z_1 z_2 z_4}\right|\\ \left\langle e_6^{(4)}\right|&=\left\langle\frac{d \mathbf{z}}{z_1 z_3 z_4}\right|\\
%\left\langle e_7^{(4)}\right|&=\left\langle\frac{d \mathbf{z}}{z_1 z_2 z_3 z_4}\right|
%\end{align}\\\\
\paragraph{Laya 3 set $z_1,z_2,z_3$:}
\begin{center}
\begin{tabular}{|c|c|c|c|}
\hline Variable & Basis size & Variable & Basis size\\
\hline $0$ & 1 & $z_4$ & 1 \\
\hline$z_1$ & 2 & $z_1, z_2$ & 4 \\
\hline$z_2$ & 2 & $z_1, z_3$ & 4 \\
\hline$z_3$ & 2 & $z_1, z_4$ & 2\\
\hline
\end{tabular}
\end{center}
\begin{center}
\begin{tabular}{|c|c|c|c|}
\hline Variable & Basis size & Variable & Basis size \\
\hline$z_2, z_3$ & 4 & $z_1, z_2, z_4$ & 4 \\
\hline$z_2, z_4$ & 2 & $z_1, z_3, z_4$ & 4 \\
\hline$z_3, z_4$ & 2 & $z_2, z_3, z_4$ & 4 \\
\hline$z_1, z_2, z_3$ & 8 & $z_1, z_2, z_3, z_4$ & 8 \\
\hline
\end{tabular}
\end{center}
The associated basis differential forms are:
\begin{center}
$$
\begin{array}{cccc}
 \left\langle e_1^{(4)}\right|=\left\langle d \mathbf{z}\right|
&  
\left\langle e_2^{(\mathbf{4})}\right|=\left\langle\frac{d \mathbf{z}}{z_1}\right|
&
\left\langle e_3^{(4)}\right|=\left\langle\frac{d \mathbf{z}}{z_2}\right| 
&
\left\langle e_4^{(4)}\right|=\left\langle\frac{d \mathbf{z}}{z_3}\right|\\\\
\left\langle e_5^{(4)}\right|=\left\langle\frac{d \mathbf{z}}{z_1 z_2}\right|
&
\left\langle e_6^{(4)}\right|=\left\langle\frac{d\mathbf{z}}{z_1 z_3 }\right|
&
\left\langle e_7^{(4)}\right|=\left\langle\frac{d \mathbf{z}}{z_2 z_3}\right|
&
\left\langle e_8^{(4)}\right|=\left\langle\frac{d \mathbf{z}}{z_1 z_2 z_3}\right|
\end{array}
$$
\end{center}
%\begin{align}
%\left\langle e_1^{(4)}\right|&=\left\langle d \mathbf{z}\right|\\
%\left\langle e_2^{(\mathbf{4})}\right|&=\left\langle\frac{d \mathbf{z}}{z_1}\right|\\
%\left\langle e_3^{(4)}\right|&=\left\langle\frac{d \mathbf{z}}{z_2}\right|\\
%\left\langle e_4^{(4)}\right|&=\left\langle\frac{d \mathbf{z}}{z_3}\right|\\
%\left\langle e_5^{(4)}\right|&=\left\langle\frac{d \mathbf{z}}{z_1 z_2}\right|\\ \left\langle e_6^{(4)}\right|&=\left\langle\frac{d \mathbf{z}}{z_1 z_3 }\right|\\
%\left\langle e_7^{(4)}\right|&=\left\langle\frac{d \mathbf{z}}{z_2 z_3}\right|\\
%\left\langle e_8^{(4)}\right|&=\left\langle\frac{d \mathbf{z}}{z_1 z_2 z_3}\right|
%\end{align}
\paragraph{layer 4 set $z_1,z_2,z_3,z_4$}
\begin{center}
\begin{tabular}{|c|c|c|c|}
\hline Variable & Basis size & Variable & Basis size \\
\hline $0$ & 0 & $z_4$ & 0 \\
\hline$z_1$ & 0 & $z_1, z_2$ & 1 \\
\hline$z_2$ & 1 & $z_1, z_3$ & 1 \\
\hline$z_3$ & 1 & $z_1, z_4$ & 1 \\
\hline
\end{tabular}
\end{center}
\begin{center}
\begin{tabular}{|c|c|c|c|}
\hline Variable & Basis size & Variable & Basis size \\
\hline$z_2, z_3$ & 3 & $z_1, z_2, z_4$ & 3 \\
\hline$z_2, z_4$ & 1 & $z_1, z_3, z_4$ & 3 \\
\hline$z_3, z_4$ & 1 & $z_2, z_3, z_4$ & 3 \\
\hline$z_1, z_2, z_3$ & 3 & $z_1, z_2, z_3, z_4$ & 7 \\
\hline
\end{tabular}
\end{center}
The associated differential forms are:
\begin{center}
$$
\begin{array}{cccc}
\left\langle e_1^{(4)}\right|=\left\langle\frac{d \mathbf{z}}{z_2}\right|&
\left\langle e_2^{(\mathbf{4})}\right|=\left\langle\frac{d \mathbf{z}}{z_3}\right|&
\left\langle e_3^{(4)}\right|=\left\langle\frac{d \mathbf{z}}{z_1 z_4}\right|&
\left\langle e_4^{(4)}\right|=\left\langle\frac{d \mathbf{z}}{z_2 z_3}\right|\\\\
\left\langle e_5^{(4)}\right|=\left\langle\frac{d \mathbf{z}}{z_1 z_2 z_4}\right|&
\left\langle e_6^{(4)}\right|=\left\langle\frac{d \mathbf{z}}{z_1 z_3 z_4}\right|&
\left\langle e_7^{(4)}\right|=\left\langle\frac{d \mathbf{z}}{z_1 z_2 z_3 z_4}\right|
\end{array}
$$
\end{center}
\subsection{CrosBox}
The following tables show the dimension of $\nu_{\Sigma}$, e.g. the dimensions of the twisted co-homology groups at to each layer of the fibration for the CrosBox. The code that was used can be found under \url{https://www.wolframcloud.com/obj/wfv651/Published/uregEdCrosBoxGrav%20(2).nb}
\begin{center}
\begin{tabular}{|c|c|c|c|c|c|}
\hline solve & $\nu$ &  solve & $\nu$ &  solve & $\nu$ \\
\hline
$z_1$ & 2 & $z_1, z_3$ & 4 & $z_1, z_2, z_3$ & 8\\
\hline
$z_2$ & 2 & $z_1, z_4$ & 4 & $z_1, z_2, z_4$ & 8 \\
\hline$z_3$ & 2 & $z_2, z_3$ & 4 & $z_1, z_3, z_4$ & 8 \\
\hline$z_4$ & 2 & $z_2, z_4$ & 4 & $z_2, z_3, z_4$ & 8 \\
\hline $z_1, z_2$ & 4 & $z_3, z_4$ & 4 & $z_1, z_2, z_3, z_4$ & 7 \\
\hline
\end{tabular}
\end{center}
we choose the variable ordering from the inside out $z_1,z_2,z_3,z_4$, The minimum dimension of the twisted co-homology group at the individual layers of the fibration is then:
\begin{equation}
\nu_{\Sigma,1}=2\quad\nu_{\Sigma,1,2,}=4\quad\nu_{\Sigma,1,2,3}=8\quad\nu_{\Sigma,1,2,3,4}=7\label{Xnu}
\end{equation}
From the equation \eqref{nsolw} it follows that for $\nu_{\Sigma,1}$ there are $2$ master integrals with poles at $z_1$, for $\nu_{\Sigma,1,2}$ there are $4$ master integrals with poles at either $z_1$ or $z_2$ or both, for $\nu_{1,2,3}$ there are, $8$ master integrals with poles either at $z_1$, $z_2$ or $z_3$ or in any possible couples or the complete triple, and for $\nu_{1,2,3,4}$ there are $7$ master integrals with poles either at $z_1$, $z_2$, $z_3$ or $z_4$ or in any possible supsets.
The expist code that was used can be found under  \url{https://www.wolframcloud.com/obj/wfv651/Published/uregEdCrosBoxGrav%20(2).nb}
\paragraph{layer 1 set $z_1$}
\begin{center}
\begin{tabular}{|c|c|c|c|}
\hline Variable & Basis size & Variable & Basis size\\
\hline $0$ & 1 & $z_4$ & 1 \\
\hline$z_1$ & 2 & $z_1, z_2$ & 2 \\
\hline$z_2$ & 1 & $z_1, z_3$ & 2 \\
\hline$z_3$ & 1 & $z_1, z_4$ & 2 \\
\hline
\end{tabular}
\end{center}
\begin{center}
\begin{tabular}{|c|c|c|c|}
\hline Variable & Basis size & Variable & Basis size \\
\hline$z_2, z_3$ & 1 & $z_1, z_2, z_4$ & 2 \\
\hline$z_2, z_4$ & 1 & $z_1, z_3, z_4$ & 2 \\
\hline$z_3, z_4$ & 1 & $z_2, z_3, z_4$ & 1 \\
\hline$z_1, z_2, z_3$ & 2 & $z_1, z_2, z_3, z_4$ & 2 \\
\hline
\end{tabular}
\end{center}
The associated basis differential forms are:
\begin{equation}
\left\langle e_1^{(4)}\right|=\left\langle d \mathbf{z}\right|\quad
\left\langle e_2^{(\mathbf{4})}\right|=\left\langle\frac{d \mathbf{z}}{z_1
}\right|
\end{equation}
%\begin{align}
%\left\langle e_3^{(4)}\right|&=\left\langle\frac{d \mathbf{z}}{z_1 z_4}\right|\\
%\left\langle e_4^{(4)}\right|&=\left\langle\frac{d \mathbf{z}}{z_2 z_3}\right|\\
%\left\langle e_5^{(4)}\right|&=\left\langle\frac{d \mathbf{z}}{z_1 z_2 z_4}\right|\\ \left\langle e_6^{(4)}\right|&=\left\langle\frac{d \mathbf{z}}{z_1 z_3 z_4]}\right|\\
%\left\langle e_7^{(4)}\right|&=\left\langle\frac{d \mathbf{z}}{z_1 z_2 z_3 z_4}\right|
%\end{align}

\paragraph{layer 2 set $z_1,z_2$}
\begin{center}
\begin{tabular}{|c|c|c|c|}
\hline Variable & Basis size & Variable & Basis size \\
\hline $0$  & 1 & $z_4$ & 1  \\
\hline$z_1$ & 2 &$z_1, z_2$ & 4 \\
\hline$z_2$ & 2 &$z_1, z_3$ & 2  \\
\hline$z_3$ & 1 &$z_1, z_4$ & 2 \\
\hline
\end{tabular}
\end{center}
\begin{center}
\begin{tabular}{|c|c|c|c|}
\hline Variable & Basis size & Variable & Basis size \\
\hline$z_2, z_3$ & 2 & $z_1, z_2, z_4$ & 4\\
\hline$z_2, z_4$ & 2 & $z_1, z_3, z_4$ & 2 \\
\hline$z_3, z_4$ & 1& $z_2, z_3, z_4$ & 2 \\
\hline$z_1, z_2, z_3$ & 4 & $z_1, z_2, z_3, z_4$ & 4 \\
\hline
\end{tabular}
\end{center}

The associated basis differential forms are:
 \begin{equation}
\left\langle e_1^{(4)}\right|=\left\langle d \mathbf{z}\right|
\quad
\left\langle e_2^{(\mathbf{4})}\right|=\left\langle\frac{d \mathbf{z}}{z_1}\right|
\quad
\left\langle e_2^{(\mathbf{4})}\right|=\left\langle\frac{d \mathbf{z}}{z_1}\right|\quad \left\langle e_4^{(\mathbf{4})}\right|=\left\langle\frac{d \mathbf{z}}{z_1 z_2}\right|
\end{equation}
%\begin{align}
%\left\langle e_5^{(4)}\right|&=\left\langle\frac{d \mathbf{z}}{z_1 z_2 z_4}\right|\\ \left\langle e_6^{(4)}\right|&=\left\langle\frac{d \mathbf{z}}{z_1 z_3 z_4]}\right|\\
%\left\langle e_7^{(4)}\right|&=\left\langle\frac{d \mathbf{z}}{z_1 z_2 z_3 z_4}\right|
%\end{align}
\paragraph{layer 3 set $z_1,z_2,z_3$}
\begin{center}
\begin{tabular}{|c|c|c|c|}
\hline Variable & Basis size & Variable & Basis size \\
\hline $0$ & 1 & $z_4$ & 1 \\
\hline$z_1$ & 2 & $z_1, z_2$ & 4 \\
\hline$z_2$ & 2 & $z_1, z_3$ & 4 \\
\hline$z_3$ & 2 & $z_1, z_4$ & 2 \\
\hline
\end{tabular}
\end{center}
\begin{center}
\begin{tabular}{|c|c|c|c|}
\hline Variable & Basis size & Variable & Basis size\\
\hline$z_2, z_3$ & 4& $z_1, z_2, z_4$ & 4\\ 
\hline$z_2, z_4$ & 2 & $z_1, z_3, z_4$ & 4 \\
\hline$z_3, z_4$ & 2 & $z_2, z_3, z_4$ & 4\\
\hline$z_1, z_2, z_3$ & 8 & $z_1, z_2, z_3, z_4$ & 8 \\
\hline
\end{tabular} 
\end{center}
The associated basis differential forms are:
\begin{center}
$$
\begin{array}{cccc}
 \left\langle e_1^{(4)}\right|=\left\langle d \mathbf{z}\right|
&  
\left\langle e_2^{(\mathbf{4})}\right|=\left\langle\frac{d \mathbf{z}}{z_1}\right|
&
\left\langle e_3^{(4)}\right|=\left\langle\frac{d \mathbf{z}}{z_2}\right| 
&
\left\langle e_4^{(4)}\right|=\left\langle\frac{d \mathbf{z}}{z_3}\right|\\\\
\left\langle e_5^{(4)}\right|=\left\langle\frac{d \mathbf{z}}{z_1 z_2}\right|
&
\left\langle e_6^{(4)}\right|=\left\langle\frac{d\mathbf{z}}{z_1 z_3 }\right|
&
\left\langle e_7^{(4)}\right|=\left\langle\frac{d \mathbf{z}}{z_2 z_3}\right|
&
\left\langle e_8^{(4)}\right|=\left\langle\frac{d \mathbf{z}}{z_1 z_2 z_3}\right|
\end{array}
$$
\end{center}
\paragraph{layer 4 set $z_1,z_2,z_3,z_4$}
\begin{center}
\begin{tabular}{|c|c|c|c|}
\hline Variable & Basis size & Variable & Basis size\\
\hline $0$ & 0 & $z_4$ & 0 \\
\hline$z_1$ & 0 & $z_1, z_2$ & 1 \\
\hline$z_2$ & 1 & $z_1, z_3$ & 1 \\
\hline$z_3$ & 1 & $z_1, z_4$ & 1 \\
\hline
\end{tabular}
\end{center}
\begin{center}
\begin{tabular}{|c|c|c|c|}
\hline Variable & Basis size &  Variable & Basis size\\
\hline$z_2, z_3$ & 3 & $z_1, z_2, z_4$ & 3 \\
\hline$z_2, z_4$ & 1 & $z_1, z_3, z_4$ & 3 \\
\hline$z_3, z_4$ & 1 & $z_2, z_3, z_4$ & 3 \\
\hline$z_1, z_2, z_3$ & 3 & $z_1, z_2, z_3, z_4$ & 7 \\
\hline
\end{tabular}
\end{center}
The associated differential forms are:
\begin{center}
$$
\begin{array}{cccc}
\left\langle e_1^{(4)}\right|=\left\langle\frac{d \mathbf{z}}{z_2}\right|&
\left\langle e_2^{(\mathbf{4})}\right|=\left\langle\frac{d \mathbf{z}}{z_3}\right|&
\left\langle e_3^{(4)}\right|=\left\langle\frac{d \mathbf{z}}{z_1 z_4}\right|&
\left\langle e_4^{(4)}\right|=\left\langle\frac{d \mathbf{z}}{z_2 z_3}\right|\\\\
\left\langle e_5^{(4)}\right|=\left\langle\frac{d \mathbf{z}}{z_1 z_2 z_4}\right|&
\left\langle e_6^{(4)}\right|=\left\langle\frac{d \mathbf{z}}{z_1 z_3 z_4}\right|&
\left\langle e_7^{(4)}\right|=\left\langle\frac{d \mathbf{z}}{z_1 z_2 z_3 z_4}\right|
\end{array}
$$
\end{center}
\subsection{Cut Box}
The following table shows the dimension of $\nu_{\Sigma}$, e.g. the dimensions of the twisted co-homology groups corresponding to each layer of the cut box fibration. 
The code used can be found at
\url{https://www.wolframcloud.com/obj/wfv651/Published/uregEd_BoxCutGrav.nb}
\begin{center}
\begin{tabular}{|c|c|}
\hline solve & $\nu$ \\
\hline$z_2$ & 2 \\
\hline$z_3$ & 2 \\
\hline$z_2, z_3$ & 4 \\
\hline
\end{tabular}
\end{center}
We choose the variable ordering from the inside out $z_1,z_2,z_3,z_4$.
The minimum dimension of the twisted core knowledge group at the individual layers of the fibration is
\begin{equation}
\nu_{\Sigma,2}=2\quad\nu_{\Sigma,2,3}=4\label{CXnu}
\end{equation}
again, every other arrangement would give either the same or a larger basis in each layer. Equation \eqref{nsolw} implies that for $\nu_{\Sigma,2}$ there are $2$ master integrals with poles at $z_2$, for $\nu_{\Sigma,2,3}$ there are $4$ master integrals with poles at either $z_2$ or $z_2$ or both,
\paragraph{Laya 1 set $z_2$:}
\begin{center}
\begin{tabular}{|c|c|}
\hline Variable & Basis size\\
\hline $0$ & 1 \\
\hline$z_2$ & 2 \\
\hline$z_3$ & 1 \\
\hline$z_2,z_3$ & 2\\
\hline
\end{tabular}
\end{center}
The associated basis differential forms are:
\begin{equation}
\left\langle e_1^{(4)}\right|=\left\langle d \mathbf{z}\right|\quad
\left\langle e_2^{(\mathbf{4})}\right|=\left\langle\frac{d \mathbf{z}}{z_2}\right|
\end{equation}
%\begin{align}
%\\
%\left\langle e_3^{(4)}\right|&=\left\langle\frac{d \mathbf{z}}{z_1 z_4}\right|\\
%\left\langle e_4^{(4)}\right|&=\left\langle\frac{d \mathbf{z}}{z_2 z_3}\right|\\
%\left\langle e_5^{(4)}\right|&=\left\langle\frac{d \mathbf{z}}{z_1 z_2 z_4}\right|\\ \left\langle e_6^{(4)}\right|&=\left\langle\frac{d \mathbf{z}}{z_1 z_3 z_4}\right|\\
%\left\langle e_7^{(4)}\right|&=\left\langle\frac{d \mathbf{z}}{z_1 z_2 z_3 z_4}\right|
%\end{align}\\\\
\paragraph{Laya 2 set $z_2,z_3$:}
\begin{center}
\begin{tabular}{|c|c|}
\hline Variable & Basis size\\
\hline $0$ & 1 \\
\hline$z_2$ & 2 \\
\hline$z_3$ & 2 \\
\hline$z_2, z_3$ & 4\\
\hline
\end{tabular}
\end{center}
The associated basis differential forms are:
\begin{equation}
\left\langle e_1^{(4)}\right|=\left\langle d \mathbf{z}\right|\quad
\left\langle e_2^{(\mathbf{4})}\right|=\left\langle\frac{d \mathbf{z}}{z_2}\right|\quad
\left\langle e_3^{(4)}\right|=\left\langle\frac{d \mathbf{z}}{z_3}\right|\quad
\left\langle e_3^{(4)}\right|=\left\langle\frac{d \mathbf{z}}{z_2 z_3}\right|
\end{equation}
\subsection{Cut Cross-Box}
The following table shows the dimension of $\nu_{\Sigma}$, e.g. the dimensions of the twisted co-homology groups corresponding to each layer of the fibration for the Cut cross-box. The code that was used can be found under \url{https://www.wolframcloud.com/obj/wfv651/Published/uregEdCrosBoxCutGrav.nb}
\begin{center}
\begin{tabular}{|c|c|}
\hline Variable & Basis size\\
\hline$z_1$ & 1\\
\hline$z_2$ & 2 \\
\hline$z_3$ & 2 \\
\hline$z_2, z_3$ & 4 \\
\hline
\end{tabular}
\end{center}
We choose the variable ordering from the inside out $z_1,z_2,z_3,z_4$.
The minimum dimension of the twisted core knowledge group at the individual layers of the fibration is
\begin{equation}
\nu_{\Sigma,2}=2\quad\nu_{\Sigma,2,3}=4\label{CXnu2}
\end{equation}
Equation \eqref{CXnu2} implies that for $\nu_{\Sigma,2}$ there are $2$ master integrals with poles at $z_2$, for $\nu_{\Sigma,2,3}$ there are $4$ master integrals with poles at either $z_2$ or $z_3$ both:
\paragraph{Laya 1 set $z_2$:}
\begin{center}
\begin{tabular}{|c|c|}
\hline Variable & Basis size\\
\hline $0$ & 1 \\
\hline$z_2$ & 2 \\
\hline$z_3$ & 1 \\
\hline$z_2, z_3$ & 2\\
\hline
\end{tabular}
\end{center}
The associated basis differential forms are:
\begin{equation}
\left\langle e_1^{(4)}\right|=\left\langle d \mathbf{z}\right|\quad\left\langle
e_2^{(\mathbf{4})}\right|=\left\langle\frac{d \mathbf{z}}{z_2}\right|
\end{equation}
%\begin{align}
%\\
%\left\langle e_3^{(4)}\right|&=\left\langle\frac{d \mathbf{z}}{z_1 z_4}\right|\\
%\left\langle e_4^{(4)}\right|&=\left\langle\frac{d \mathbf{z}}{z_2 z_3}\right|\\
%\left\langle e_5^{(4)}\right|&=\left\langle\frac{d \mathbf{z}}{z_1 z_2 z_4}\right|\\ \left\langle e_6^{(4)}\right|&=\left\langle\frac{d \mathbf{z}}{z_1 z_3 z_4}\right|\\
%\left\langle e_7^{(4)}\right|&=\left\langle\frac{d \mathbf{z}}{z_1 z_2 z_3 z_4}\right|
%\end{align}
\paragraph{Laya 2 set $z_2,z_3$:}
\begin{center}
\begin{tabular}{|c|c|}
\hline Variable & Basis size\\
\hline $0$ & 1 \\
\hline$z_2$ & 2\\
\hline$z_3$ & 2 \\
\hline$z_2, z_3$ & 4 \\
\hline
\end{tabular}
\end{center}
The associated basis differential forms are:
\begin{equation}
\left\langle e_1^{(4)}\right|=\left\langle d \mathbf{z}\right|\quad
\left\langle e_2^{(\mathbf{4})}\right|=\left\langle\frac{d \mathbf{z}}{z_2}\right|\quad
\left\langle e_3^{(4)}\right|=\left\langle\frac{d \mathbf{z}}{z_3}\right|\quad
\left\langle e_3^{(4)}\right|=\left\langle\frac{d \mathbf{z}}{z_3 z_2}\right|
\end{equation}
\section{Code overview, and set up.}
At the beginning of the numerical setups, all mathematical quantities necessary for the calculation of the multivariate intersection number are defined. The Baikov polynomial $u_{reg}(Z)$, a list of Baikov variables $\{z_1,...,z_n\}$, and $\omega$, defined as:
\begin{equation}
\omega(\mathbf{z})=d\logv{u(\mathbf{z})}
\end{equation}
In the code, due to the vectorial nature of $\omega$, this is done via a table containing the derivatives of the logarithm of the regularised Baikov polynomial with respect to the Baikov variables.
The code also contains a base list $\mathtt{phileftlist}$, which contains the individual bases of the different layers of the multivariate intersection number as well as their dual $\mathtt{phirightlist}$. The base list was previously obtained in section \ref{Dimension2}. Further, the code contains an inner base list $\mathtt{e}=\{\mathtt{1},\mathtt{1}/\mathtt{Z32}\}$ and its dual  $\mathtt{e=h}$. A distinction between the full base list and the inner base list is necessary since the first step of the calculation is to compute the intersection number of the innermost layer based solely on the inner base list and its dual. The code obviously also includes a differential form $\bra{\varphi}$ which corresponds to the differential form of the Feynman integral in the Baikov representation that one wants to calculate. In the code, this is creatively called  $\mathtt{bigexp}$. The differential form is combined with the base list $\mathtt{phileftlist}$ to form a new list $\mathtt{phileftlistuse}$. The code consists of two functions, $\mathtt{IntersecRec}$ and $\mathtt{DeqRes}$. The first function, $\mathtt{IntersecRec}$, defines in each layer the different intersection numbers of the quantities necessary to obtain the complete multivariate intersection number, The second function $\mathtt{DeqRes}$ in the intersection number code has the aim of counting the polls and calculating their residue. The code follows in its idea the intersection number residue formula Eq. \eqref{Res}. The initialisation of the code is based on the following intuition. Since the final intersection number is a matrix whose dimension is given by the size of the basis plus the differential form $\bra{\varphi}$ and the dual basis, it makes sense to define a 2D array called $\mathtt{result}$ with these dimensions. Each element of this array is first set to $0$. The array is then populated by two loops, the outer loop iterates over the rows of the array, while the inner loop iterates over the columns of the array. Inside the inner loop, the function $\mathtt{IntersecRec}[\mathtt{phileftlistuse}[[i]],\mathtt{phirightlist}[[j]], {\mathtt{e}}, {\mathtt{h}},\mathtt{omega}, \mathtt{vars}]$ is called. The result of the $\mathtt{IntersecRec}$ function call is assigned to the corresponding element in an array at position $(i, j)$. This part of the code essentially calculates the intersection number of the elements of $\mathtt{phileftlistuse}$ and $\mathtt{phirightlist}$ using the function $\mathtt{IntersecRec}$ and stores the results in the 2D array.
The result of the matrix  has in general the following form:
\begin{center}
$$ \mathtt{result}
=
\begin{bmatrix}
\braket{\varphi}{h_1}&\braket{\varphi}{h_2}& \braket{\varphi}{h_3}& \braket{\varphi}{h_4}\\
\braket{e_1}{h_1} & \braket{e_1}{h_2} & \braket{e_1}{h_3}& \braket{e_1}{h_4}\\
\braket{e_2}{h_1} & \braket{e_2}{h_2} & \braket{e_2}{h_3}& \braket{e_2}{h_4}\\
\braket{e_3}{h_1} & \braket{e_3}{h_2} & \braket{e_3}{h_3}& \braket{e_3}{h_4}\\
\braket{e_4}{h_1} & \braket{e_4}{h_2} & \braket{e_4}{h_3}& \braket{e_4}{h_4}
\end{bmatrix}
$$
\end{center}
It detailed description of the intersection number code can be found in the appendix, 

For our particular case, the box and the cross-box, we have decided to simplify the calculation by making a cut at the massless 
propagators $z_1=l^2$ and $z_4=\kl{l+q_1+q_2}$. It was also necessary to decompose the differential form $\bra{\varphi}$ into a monomial list. The details can be found in a notebook \url{https://www.wolframcloud.com/obj/wfv651/Published/multivar42ed.nb} that includes the code for the intersection number, as well as working examples. The results were compared with the intersection number code of Frellesvig, as well as \cite{Smirnov:2013dia} 
The simulations set up for the cut box reads as follows:
\begin{verbatim}
ur3 = 4^(5 - d) Z32^rho3[1]Z33^rho3[2] 
    (t (m1^4 t + m2^4 t + t (s + Z32)^2 - 2 m2^2(s t - (t - 2 Z32)
    (Z32 - Z33)) - 2 m1^2 (m2^2 t + s t + (t - 2 Z33) (Z32 - Z33))
    + 2 s t Z33 - 2 (2 s + t) Z32 Z33 + t Z33^2))^(1/2 (-5 + d))

vars = {Z32, Z33}; (*Defining Baikov variables*)

omega = Table[Simplify[D[Log[ur], vars[[i]]]],
        {i, Length[vars]}]; (*Defining Omega*)

phileftlist3 = {1, 1/Z32, 1/Z33, 1/(Z32*Z33)}; (*Defining the basis*)

phirightlist3 = phileftlist3; (*Defining the dual basis*)

N32 = 1/(2 (-2 + D)^2) ((-2 + D) (- (1/2) Z32 (2 (m1^2 + m2^2 - s) - Z33)
    + (m1^2 + m2^2 - s) (m1^2 + m2^2 - s - Z33)) - 
    (2 m1^2 - Z32) (2 m2^2 - Z33)) (-2 (2 m1^2 - Z32) (2 m2^2 - Z33) +
    (-2 + D) ((m1^2 + m2^2 - s) (m1^2 + m2^2 - s - Z32 - Z33) +
    (-m1^2 - m2^2 + s + Z32) (-m1^2 - m2^2 + s + Z33)));
    (*Numerator for the cut box*)

bigexp32 = N32/(Z32*Z33);

(*Defining the differential form varphi*)

phileftlistuse = Join[phileftlist, bigexp32];

e = {1, 1/Z32};

(*Defining the inner basis*)

h = e;

(*Defining the inner dual basis*)
\end{verbatim}
The simulations set up for the cut crosbox reads as follows:
\begin{verbatim}
ur4 = 4^(5 - d) Z42^rho4[1] Z43^rho4[2] (t (m1^4 t + m2^4 t + t 
    (s + t - Z42)^2 - 2 m2^2 (s t + (t - 2 Z42) (t - Z42 - Z43))
    -2 m1^2 (m2^2 t + s t + (t - 2 Z43) (t - Z42 - Z43))
    - 2 t (s + t) Z43 + 2 (2 s + t) Z42 Z43 + t Z43^2))^(1/2 (-5 + d))

vars4 = {X42, X43}; (*Defining Baikov variables*)

omega4 = 
  Table[Simplify[D[Log[u4], vars4[[i]]]],
  {i, Length[vars4]}];(*Defining Omega*)

phileftlist4 = {1, 1/X32, 1/X33, 1/(X32*X33)}; (*Defining the basis*)

phirightlist4 = phileftlist4;(*Defining the dual basis*)

N42 = 1/(2 (-2 + D)^2) ((-2 + D) (-(1/2) X42 (2 (m1^2 + m2^2 - u) - X43) 
    + (m1^2 + m2^2 - u) (m1^2 + m2^2 - u - X43)) - (2 m1^2 - X42) (2 m2^2 - X43)) 
    (-2 (2 m1^2 - X42) (2 m2^2 - X43) + (-2 + D) ((m1^2 + m2^2 - u) 
    (m1^2 + m2^2 - u - X42 - X43) + (-m1^2 - m2^2 + u + X42) 
    (-m1^2 - m2^2 + u + X43)));

bigexp42 = N42/(X42*X43)(*Defining the differential form varphi*);

phileftlistuse4 = Prepend[phileftlist4, bigexp4];

e4 = {1, 1/X32};

h4 = e4;

\end{verbatim}
The question the reader may now ask is how exactly does the $\mathtt{result}$ provide the main integral decomposition for the cut-box and the cut-cros-box. Recall the definition of thethe Feynman integrals decomposition formula for different subsectors in Eq. \ref{eqMIINTER2}. We can use Eq. \eqref{eqCof2} to calculate the coefficient of \eqref{eqMIINTER2} and see that all relevant quantities are contained in the $\mathtt{result}$ matrix, where the $1$ row correspond to the vector $\braket{\varphi}{h_j}$, while the $2$ to $5$ rows of the $\mathtt{result}$ matrix correspond to the $\mathbf{C}$ matrix. 
Therefore, we can simply extract the relevant quantities from the matrix and use Eq. \eqref{eqCof2} to multiply them together, to obtain the final result, we simply have to put the regulator equal $\rho_i=0$, and we obtain a vector containing the four coefficients of the  master integrals.% The question the reader may now ask is how exactly does the $\mathtt{result}$ provide the main integral decomposition for the cut-box and the cut-cros-box. Recall the definition of the master integral decomposition formula for different subsectors in Eq. \ref{eqMIINTER2}. We can use Eq. \eqref{eqCof2} to calculate the coefficient of \eqref{eqMIINTER2} and see that all relevant quantities are contained in the $\mathtt{result}$ matrix, where the $1$ row correspond to the vector $\braket{\varphi}{h_j}$, while the $2$ to $5$ rows of the $\mathtt{result}$ matrix correspond to the $\mathbf{C}$ matrix. Therefore, we can simply extract the relevant quantities from the matrix and use Eq. \eqref{eqCof2} to multiply them together, to obtain the final result, we simply have to put the regulator equal $\rho_i=0$ and we obtain a vector containing the four coefficients of the master integrals.
\section{Master integral decomposition of the gravity box and cross-box}\label{MIDGrav}
In this section, we are going through the coefficients that were obtained via thethe Feynman integrals decomposition formula and give further intuition for the master integral decomposition as it relates to scaring amplitudes.
Since we have performed a cut, the tadpole master integrals have disappeared. Consequently, we have also not obtained a coefficient for them. 
The first coefficient belongs to the bubble MI and has the following form, 
\begin{equation}
\begin{aligned}
c_{\mathsf{Buble},B}=&\frac{1}{16(-2+D)^2(-1+D)\left(4m_1^2-t\right)\left(-4m_2^2+t\right)}\\
& -\left(\left(128(-4+D)^2(-1+D) m_1^8+128(-1+D)\left(m_2^2-s\right)\left((-4+D) m_2^3-(-2+D) m_2 s\right)^2\right.\right. \\
& -64(-1+D)\left((50+D(-32+5 D)) m_2^6+(-78+(60-11 D) D) m_2^4 s+(-2+D) \times\right. \\
& \left.(-18+7 D) m_2^2 s^2-(-2+D) s^3\right) t+8((-168+D(292+D(-133+18 D))) \times \\
& m_2^4+(176+D(-336+(179-28 D) D)) m_2^2 s+2(-2+D) \times \\
& \left.(-1+D)(-12+5 D) s^2\right) t^2-2(-4+D)\left((20+D(-39+10 D)) m_2^2\right. \\
& +(-12+(23-8 D) D) s) t^3+(-4+D)^2 D t^4+64(-1+D) m_1^6 \times \\
& \left(4(42+5(-6+D) D) m_2^2-2(-4+D)(-8+3 D) s+(-50+(32-5 D) D) t\right) \\
& +8 m_1^4\left(1 6 ( - 1 + D ) \left(2(92+D(-56+9 D)) m_2^4-3(-2+D)(-22+7 D) m_2^2 s\right.\right. \\
& \left.+(-2+D)(-10+3 D) s^2\right)-4(-364+D(612+D(-277+38 D))) m_2^2 t \\
& \left.+8(-1+D)(78+D(-60+11 D)) s t+(-168+D(292+D(-133+18 D))) t^2\right) \\
& +2 m_1^2\left(6 4 ( - 1 + D ) \left(2(42+5(-6+D) D) m_2^6-3(-2+D)(-22+7 D) m_2^4 s\right.\right. \\
& \left.+4(-2+D)(-7+3 D) m_2^2 s^2-(-2+D)^2 s^3\right) \\
& -16\left((-364+D(612+D(-277+38 D))) m_2^4+(320+D(-600+(323-52 D) D)) \times\right. \\
& \left.+m 2^2 s+2(-2+D)(-1+D)(-18+7 D) s^2\right) t \\
& +\left((-368+2 D(328+D(-145+19 D))) m_2^2+(176+D(-336+(179-28 D) D)) s\right) t^2 \\
& \left.\left.+4+D)(20+D(-39+10 D)) t^3\right)\right)
\end{aligned}
\end{equation}
The second coefficient belongs to the triangle down MI and has the form: 
\begin{equation}
\begin{aligned}
c_{\triangledown,m_1,B}=&-\frac{1}{(-2+D)^2\left(4 m_1^2-t\right)}\left((-2+D) m_1^2+(-4+D) m_2^2-(-2+D) s\right) \\
& \left(8(-2+D) m_1^6-\left(m_2^2-s\right)\left(d\left(m_2^2-s\right)+2 s\right) t+2 m_1^2\right. \\
& \left(4(-2+D)\left(m_2^2-s\right)^2+(9-2 D) m_2^2 t+2(-2+D) s t\right) \\
& \left.+m_1^4\left(16(-4+D) m_2^2-(-2+D)(16 s+3 t)\right)\right)
\end{aligned}
\end{equation}
The third coefficient belongs to the  triangle up MI and has the form:
\begin{equation}
\begin{aligned}
c_{\triangle,m_2,B}=&-\frac{1}{(-2+D)^2\left(4 m_2^2-t\right)}\left((-2+D) m_2^2+(-4+D) m_1^2-(-2+D) s\right) \\
& \left(8(-2+D) m_2^6-\left(m_1^2-s\right)\left(d\left(m_1^2-s\right)+2 s\right) t+2 m_2^2\right. \\
& \left(4(-2+D)\left(m_2^2-s\right)^2+(9-2 D) m_2^2 t+2(-2+D) s t\right) \\
& \left.+m_2^4\left(16(-4+D) m_1^2-(-2+D)(16 s+3 t)\right)\right)
\end{aligned}
\end{equation}
The fourth coefficient is the scalar Box MI:
\begin{equation}
    c_{\square,s,B}=\left(\frac{1}{2} \left(-m_1^2-m_2^2+s\right)^2-\frac{2 m_1^2 m_2^2}{Q-2}\right)^2
\end{equation}
For the coss-Box. 
The first coefficient belongs to the bubble MI and has the form:
\begin{equation}
    \begin{aligned}
&c_{\mathsf{Buble},CB}\\
=& \frac{1}{16(-2+D)^2(-1+D)\left(4 m_1^2-t\right)\left(-4 m_2^2+t\right)} \\
& \left(128(-2+D)^2(-1+D) m_1^8+128(-2+D)^2(-1+D) m_2^2\left(m_2^2-s\right)^3\right. \\
& -64(-1+D)\left(m_2^2-s\right)\left((14+3(-4+D) D) m_2^4\right. \\
& \left.-4(-2+D)(-1+D) m_2^2 s+(-2+D)^2 s^2\right) t+8((-3+2 d) \\
& (-8+d(-4+3 d)) m_2^4-(16+d(32+d(-59+20 d))) m_2^2 s \\
& \left.+2(-2+D)(-1+D)(-12+7 d) s^2\right) t^2+2((-4+D) \\
& \left.(44+5 d(-9+2 d)) m_2^2+(-48+d(136+d(-103+24 d))) s\right) t^3 \\
& -(-4+D)^2(-2+D) t^4+8 m_1^4\left(1 6 ( - 1 + d ) \left(\left(-80+40 d-6 d^2\right) m_2^4\right.\right. \\
& \left.+(40+d(-28+3 d)) m_2^2 s+3(-2+D)^2 s^2\right) \\
& +4\left((-236+d(364+d(-159+22 d))) m_2^2+2(-1+D) \times\right. \\
& \left.(22+d(-24+7 d)) s) t+(-3+2 d)(-8+d(-4+3 d)) t^2\right) \\
& +2 m_1^2\left(-64(-1+D)\left(m_2^2-s\right)\left(2(6+(-6+D) d) m_2^4\right.\right. \\
& \left.-(-14+D)(-2+D) m_2^2 s-(-2+D)^2 s^2\right) \\
& +16\left((-236+d(364+d(-159+22 d))) m_2^4+(224\right. \\
& +d(-344+3(47-4 d) d)) m_2^2 s-2(-2+D) \\
& \left.(-1+D)(-6+5 d) s^2\right) t-4((-560+2 d(424+27(-7+D) d)) \\
& \left.m_2^2+(16+d(32+d(-59+20 d))) s\right) t^2+(-4+D) \\
& \left.(44+5 d(-9+2 d)) t^3\right)-64(-1+D) m_1^6\left(4(6+(-6+D) d) m_2^2\right. \\
& +24 s+14 t+3(-4+D) d(2 s+t))) \\
&
\end{aligned}
\end{equation}
The second coefficient belongs to triangle down MI and has the form: 
\begin{equation}
    \begin{aligned}
&c_{\triangledown,m_1,CB}\\=& \frac{1}{(-2+D)^2\left(4 m_1^2-t\right)}\left(d\left(m_1^2+m_2^2-s-t\right)+2\left(-m_2^2+s+t\right)\right) \\
& \left(8(-2+D) m_1^6-(-2+D) t\left(-m_2^2+s+t\right)^2+m_1^4\left(16(-4+D) m_2^2\right.\right. \\
& -16(-2+D) s+36 t-15 d t)+2 m_1^2\left(4(-2+D)\left(m_2^2-s\right)^2\right. \\
& \left.\left.+(21-8 d) m_2^2 t+(-17+8 d) s t+(-9+4 d) t^2\right)\right)
\end{aligned}
\end{equation}
The third coefficient belongs to the triangle MI up that has the form:
\begin{equation}
\begin{aligned}
c_{\triangle,m_2
,CB}=& \frac{1}{(-2+D)^2\left(4 m_1^2-t\right)}\left((-2+D) m_2^2+d\left(m_1^2-s-t\right)+2(s+t)\right) \\
& \left(8(-2+D) m_1^6+(-2+D) m_2^4\left(8 m_1^2-t\right)-(-2+D) t(s+t)^2\right. \\
& +m_1^4(-16(-2+D) s+36 t-15 d t)+2 m_1^2(s+t)(4(-2+D) s \\
& \quad+(-9+4 d) t)+2 m_2^2\left(8(-4+D) m_1^4+(-2+D) t(s+t)\right. \\
& \left.\left.+m_1^2(21 t-8((-2+D) s+d t))\right)\right)
\end{aligned}  
\end{equation}
The fourth coefficient is the scalar Box:
\begin{equation}
    c_{\square,u,CB}=\left(\frac{1}{2} \left(-m_1^2-m_2^2+u\right)^2-\frac{2 m_1^2 m_2^2}{D-2}\right)^2
\end{equation}
Now that we have obtained the coefficient via the intersection theory, we can use the last term in the the Feynman integrals decomposition formula Eq. \eqref{eqMIINTER2} to find the Master integral decomposition of the graviton box Eq. \ref{FIGB} and the coss--box Eq \ref{FIGCB}. Since master integrals, roughly speaking, form the bases of a vector space, they are scalar integrals whose numerator can then be understood as the coefficient of the base vector that depend on the scattering amplitude that is that is being decomposed. The numerator of the master integral is obtained by summing the individual coefficients of the different Feynman diagrams of the scattering amplitude that have the same basis vector.
\begin{center}
\begin{figure}[h]
\centering
\includegraphics[width=0.66\textwidth]{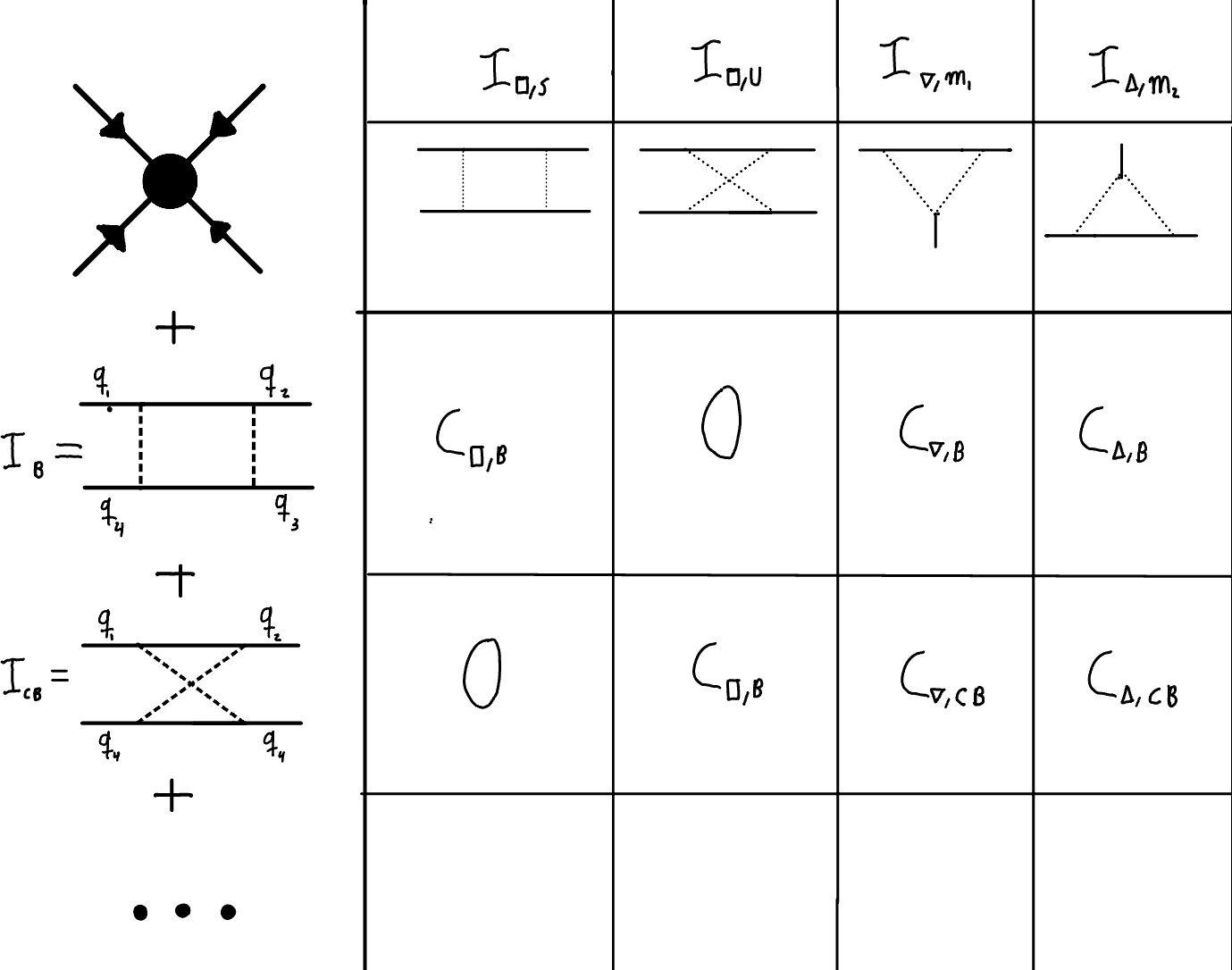}
\end{figure}
\end{center}
Since in the next section we will go into the soft or classical limit, we will not discuss the master integral of the bubble any further, since it is known to be zero in the soft limit. This leaves 4 master integrals, the master integral for the box Eq. \eqref{MIB}, the cross-box Eq \eqref{MICB}, the triangle down Eq. \eqref{MITD} and the triangle up Eq. \eqref{MITU},  \eqref{MITU}. \cite{PhysRevD.100} 
\begin{align}
    \mathcal{I}_{\square,s}(k)=&\int \frac{d^D l}{i \pi^{\frac{D}{2}}}\frac{1}{k^2} \frac{1}{(Q+k)^2}\frac{1}{\left(q_1+k\right)^2+m_1^2} \frac{1}{\left(q_4-k\right)^2+m_2^2}\label{MIB}\\
    \mathcal{I}_{\square,u}(k)=&\int \frac{d^D k}{ i\pi^{\frac{D}{2}}} \frac{1}{k^2} \frac{1}{(Q+k)^2} \frac{1}{\left(q_2+k\right)^2+m_1^2} \frac{1}{\left(q_4-k\right)^2+m_2^2}\label{MICB}\\    
    \mathcal{I}_{\triangledown,m_1}\left(k\right)=&\int\frac{d^D k}{ i\pi^{\frac{D}{2}}} \frac{1}{k^2} \frac{1}{(Q+k)^2} \frac{1}{\left(k+q_1\right)^2+m_1^2}\label{MITD}\\   \mathcal{I}_{\triangle,m_2}\left(k\right)=&\int\frac{d^D k}{ i\pi^{\frac{D}{2}}} \frac{1}{k^2} \frac{1}{(Q+k)^2} \frac{1}{\left(k+q_2\right)^2+m_2^2}\label{MITU} 
\end{align}

%\graphicspath{{./images/}}
\chapter{The Soft Limit}\label{Soft}
The section as largely based, on \cite{Cristofoli_2020} and \cite{PhysRevD.100}.
We will discuss the classical limit $\h\to0$ of the master integrals $\mathcal{I}_{\square, s}$, $\mathcal{I}_{\square,u}$, $\mathcal{I}_{\triangledown,m_1}$and $\mathcal{I}_{\triangle,m_2}$ and their respective coefficients, that is where the transferred momentum $Q=\kl{q_1+q_2}=\kl{q_3+q_4}=-\sqrt{t}$ is small. This corresponds to a classical long-range gravitational potential. In order to achieve this, we will borrow an idea originating in the methods of the regions.\\\\
 The methods of the regions generally allow for avoiding the explicit evaluation of the Feynman integral if one is only interested in the asymptotic series expansion and its behaviour in certain limit values.  The technique consists of splitting the domain of integration into sectors defined by suitable scaling relations. In our case, we consider the asymptotic evolution of the master integrals with the integration domain of the loop moments $k$ divided into the soft low-energy region where the integrated moment $k$ scales with $k\sim\mathcal{O}(Q)$ and, in general, a high-energy heart region where the integrated moment scales as $k\sim\mathcal{O}(1)$.\\ 

The non-asymptotic low-energy contribution in momentum space originating from the soft region leads to long-range effects in position space. Since we are only interested in the low-energy limit of the master integrals, we will not have a full discussion of the methods of the regions, hence, no hard region is evaluated. 
\section{Box and CrosBox}
As a reminder, the scaler box integral, the following form:
\begin{equation}
\mathcal{I}_{\square,s}(k)=\int \frac{d^D l}{i \pi^{\frac{D}{2}}}\frac{1}{k^2} \frac{1}{(Q+k)^2}\frac{1}{\left(q_1+k\right)^2+m_1^2} \frac{1}{\left(q_4-k\right)^2+m_2^2}\label{MIB2}\\
\end{equation}
with $Q^2=-|t|$
The scaler cross-box integral has the following form:
\begin{equation}
\mathcal{I}_{\square,u}(k)=\int \frac{d^D k}{ i\pi^{\frac{D}{2}}} \frac{1}{k^2} \frac{1}{(Q+k)^2} \frac{1}{\left(q_2+k\right)^2+m_1^2} \frac{1}{\left(q_4-k\right)^2+m_2^2}\label{MICB2}
\end{equation}
expanding the square of $\left(q_1+k\right)^2$ and $\left(q_4-k\right)^2$, in the Box and using the fact that $q_4^2=q_3^2=-m_2^2$ and $q_1^2=q_2^2=-m_1^2$ we get:
\begin{equation}
\mathcal{I}_{\square,s}(k)=\int \frac{d^D k}{i\pi^{\frac{D}{2}}} \frac{1}{\left(l^2+i\varepsilon\right)\left((Q+k)^2+i\varepsilon\right)\left(k^2-2 p_4 \cdot k+i\varepsilon\right)\left(k^2+2 p_1 \cdot k+i\varepsilon\right)}\label{SIntBox}
\end{equation}
and for the cross box we simply replace $q_1=q_2$ in 
eq. \eqref{SIntBox}, and get:
\begin{equation}
\mathcal{I}_{\square,u}(k)=\int \frac{d^D k}{i\pi^{\frac{D}{2}}} \frac{1}{\left(k^2+i\varepsilon\right)\left((Q+k)^2+i\varepsilon\right)\left(k^2-2 p_4 \cdot k+i\varepsilon\right)\left(k^2+2 p_2 \cdot k+i\varepsilon\right)}\label{MIBEXP12}
\end{equation}
In the centre of mass frame, we have:
\begin{center}
$
\begin{array}{ll}
q_1^\mu=\left(E_1(p), \vec{p}\right), & q_2^\mu=\left(E_1(p), \vec{p}^{\prime}\right), \\
q_4^\mu=\left(E_4(p),-\vec{p}\right), & q_3^\mu=\left(E_4(p),-\vec{p}^{\prime}\right)
\end{array} 
$
\end{center}

and we define
\begin{center}$
\begin{gathered}
p \equiv|\vec{p}|=\left|\vec{p}^{\prime}\right| \\
E_1(p) \equiv \sqrt{p^2+m_1^2},  E_4(p) \equiv \sqrt{p^2+m_2^2} \\
E_p \equiv E_1(p)+E_4(p),  \xi(p) \equiv \frac{E_1(p) E_4(p)}{E_p^2} \\
Q^\mu \equiv q_1^\mu+q_2^\mu,  \vec{Q} \equiv \vec{p}+\vec{p}^{\prime} .
\end{gathered}$
\end{center}
as previously stated, the Mandelstamvariables are: 
\begin{align}
s=-\kl{q_1+q_4}^2&=-\kl{q_2+q_3}^2\nonumber\\
u=-\kl{q_1+q_3}^2&=-\kl{q_2+q_4}^2\label{MSV2}\\
t=-\kl{q_1+q_2}^2&=-\kl{q_3+q_4}^2=-Q^2\nonumber
\end{align}
Furthermore, we introduce the variables
\begin{equation}
q_{\perp}=q_1-q_2=q_4-q_3, w=q_1+q_4=q_2+q_3\label{newvar}
\end{equation}
Eqations \eqref{newvar} allow us to rewrite Eq. \eqref{SIntBox}  as follows:
\begin{equation}
\mathcal{I}_{\square,s}(k)^{(1 s)}=\int \frac{d^D k}{ i\pi\hbar^{\frac{D}{2}}} \frac{\hbar^5}{k^2(k-Q)^2\left(k^2+\left(2 w-q_{\perp}-Q\right) \cdot k\right)\left(k^2-\left(q_{\perp}+Q\right) \cdot k\right)}\label{NVBox}
\end{equation}
The new variables in Eq. \eqref{newvar} satisfy, the following relations:
\begin{equation}
Q \cdot q_{\perp}=0=Q \cdot w,  q_{\perp} \cdot w=\left(m_1^2+m_2^2\right)-w^2\label{softrel}
\end{equation}
By nature of the classical limit, a soft region emerges, which is characterised by the scaling $k\sim\mathcal{O}\kl{\h}$ and thus $k\sim Q \ll q_{\perp}$, and a heart region, which is characterised by $k\sim\mathcal{O}\kl{1}$ and thus $k\sim q_{\perp}\gg Q$. In the low-energy regime of the soft limit, it follows that, the transferred momentum $Q$ vanishes, while the transferred wave vector $\frac{1}{\h}Q$ and the average momentum $\eh q_{\perp}$ of the massive particles stay fixed. This can be expressed as follows:
\begin{equation}
    Q \sim\mathcal{O}(\hbar),  q_{\perp} \sim\mathcal{O}(1),  Q\ll q_{\perp}\label{hlimC}
\end{equation}
 $\h\to0$, implies by the relation:
\begin{equation}
    -q_{\perp}^2=4m1^2+Q^2
\end{equation}
  This can be derived from Eq. \eqref{newvar}, \eqref{softrel}, and \eqref{hlimC} and the Mandelstam variables Eq. \eqref{MSV2}. The leading soft term can then be obtained by Taylor expanding the integrand of Eq. \eqref{NVBox} to leading order of $k\sim\mathcal{O}(\h)$, which implies $k\sim Q\ll q_{\perp}, w$. The leading soft term hen has the following form:
\begin{equation}
    \mathcal{I}_{\square,s}(k)^{(1 s)}=\int \frac{d^D k}{i \pi \hbar^{\frac{D}{2}}}\frac{\hbar^5}{k^2(k-Q)^2\left(\left(2 w-q_{\perp}\right) \cdot k\right)\left(-\left(q_{\perp} \cdot k\right)\right)}
\end{equation}
If we introduce the Feynman parameter $x$ for the two linear factors in the denominator, we obtain:
\begin{equation}
\mathcal{I}_{\square,s}(k)^{(1 s)}=\int_0^1 d x \int \frac{d^D k}{i\pi \hbar^{\frac{D}{2}}} \frac{\hbar^5}{k^2(k-Q)^2\left(\left(2 x w-q_{\perp}\right) \cdot k\right)^2}
\end{equation}
Since $2xw-q_{\perp}$ is orthogonal to $Q$ we can apply the following Identity: 
If two vectors are orthogonal $a\cdot b=0$ it holds:
\begin{align}
I_{\perp}\left(a^2, b^2\right) & =\int \frac{d^D k}{i \pi^{\frac{D}{2}}\left(k^2+i\varepsilon\right)^{\lambda_1}\left((a-k)^2+i\varepsilon\right)^{\lambda_2}(2 b \cdot k+i\varepsilon)^{\lambda_3}} \\
& =\frac{\Gamma\left(\lambda_1+\lambda_2+\frac{\lambda_3-D}{2}\right) \Gamma\left(\frac{\lambda_3}{2}\right)}{2 \Gamma\left(\lambda_1\right) \Gamma\left(\lambda_2\right) \Gamma\left(\lambda_3\right) \Gamma\left(D-\lambda_1-\lambda_2-\lambda_3\right)} \frac{\Gamma\left(\frac{D-\lambda_3}{2}-\lambda_1\right) \Gamma\left(\frac{D-\lambda_3}{2}-\lambda_2\right)}{\left(a^2\right)^{\lambda_1+\lambda_2+\frac{\lambda_3-D}{2}}\left(-b^2\right)^{\frac{\lambda_3}{2}}} .
\end{align}
This gives us:
\begin{equation}
\mathcal{I}_{\square,s}(k)^{(1 s)}=\frac{\Gamma\left(\frac{D-4}{2}\right)^2 \Gamma\left(\frac{6-D}{2}\right)}{2 \Gamma(D-4)} \frac{1}{\hbar}\left(\frac{Q^2}{\hbar^2}\right)^{\frac{D-6}{2}} \int_0^1 \frac{d x}{-\left(x w-\frac{1}{2} q_{\perp}\right)^2+i\varepsilon}
\end{equation}
where we have reinstated the $+i\varepsilon$. The route of the polynomials is
\begin{equation}
\left(x-\frac{q_{\perp}}{2w}\right)^2+i\varepsilon
\end{equation}
are then:
\begin{equation}
x_{ \pm}=\frac{w\cdot q_{\perp}}{2 w^2} \pm\frac{1}{2} \sqrt{\left(\frac{w \cdot q_{\perp}}{w^2}\right)^2-\frac{q_{\perp}^2}{Q^2}}\label{rut}
\end{equation}
%which can be rewritten as:
%\begin{equation}
%x_{ \pm}=\frac{m_1^2-q_1 \cdot q_4 \p m_\sqrt{\left(q_1 \cdot q_4\right)^2-\left(m_1 m_2\right)^2}}{m_1^2+m_2^2+2 q_1 \cdot q_4} \p m_i \epsilon\label{rut}
%\end{equation}
The integral then takes a form
\begin{equation}
\mathcal{I}_{\square,s}(k)^{(1 s)}=\frac{i \Gamma\left(\frac{D-4}{2}\right)^2 \Gamma\left(\frac{6-D}{2}\right)}{2
\Gamma(D-4)} \frac{1}{\hbar}\left(\frac{Q^2}{\hbar^2}\right)^{\frac{D-6}{2}}\int_0^1 \frac{d x}{-w(x-x_{+})(x-x_{-})+i\varepsilon}
\end{equation}
Performing a partial fraction decomposition gives us the following result 
\begin{align}
\mathcal{I}_{\square,s}(k)^{(1 s)}=&\frac{i \Gamma\left(\frac{D-4}{2}\right)^2 \Gamma\left(\frac{6-D}{2}\right)}{2\Gamma(D-4)} \frac{1}{\hbar}\left(\frac{Q^2}{\hbar^2}\right)^{\frac{D-6}{2}}\int_0^1\frac{dx}{w(x_{+}-x_{-})}\kl{\frac{1}{(x-x_{-})}-\frac{1}{(x-x_{+})}}\\
\mathcal{I}_{\square,s}(k)^{(1 s)}=&\frac{i \Gamma\left(\frac{D-4}{2}\right)^2 \Gamma\left(\frac{6-D}{2}\right)}{2\Gamma(D-4)} \frac{1}{\hbar}\left(\frac{Q^2}{\hbar^2}\right)^{\frac{D-6}{2}}\frac{1}{w^2\left(x_{+}-x_{-}\right)}\logv{\frac{\left(1-x_{+}\right) x_{-}}{\left(1-x_{-}\right) x_{+}}}
\end{align}
whtis
\begin{equation}
(x_{+}-x_{-})=\frac{2\sqrt{(q_1\cdot q_4)^2-(m_1m_2)^2}}{(q_1+q_4)}, w=(q_1+q_4)=\sqrt{s}
\end{equation}
To leading order in $\h$ we have $q^2_{\perp}=\frac{4m^2_2}{\h^2}-q^2\sim\frac{4m_2^2}{\h^2}$ and using the relations in Eq.$\eqref{softrel}$ appropriately,
we get: 
\begin{equation}
\mathcal{I}_{\square,s}(k)^{(1 s)}=\frac{i \Gamma\left(\frac{D-4}{2}\right)^2 \Gamma\left(\frac{6-D}{2}\right)}{2 \Gamma(D-4)} \frac{1}{\hbar}\left(\frac{Q^2}{\hbar^2}\right)^{\frac{D-6}{2}}\frac{\logv{\frac{q_1\cdot q_4}{m_1m_2}+\sqrt{\kl{\frac{q_1\cdot q_4}{m_1m_2}}-1}}+i\pi}{\sqrt{\kl{q_4\cdot q_1}-m_1m_2}}
\end{equation}
which we can rewrite by the not so well-known identity  $\cosh^{-1}{(x)}=\logv{x+\sqrt{x^2-1}}$ as
\begin{equation}
\mathcal{I}_{\square,s}(k)^{(1 s)}=\frac{i \Gamma\left(\frac{D-4}{2}\right)^2 \Gamma\left(\frac{6-D}{2}\right)}{2\Gamma(D-4)} \frac{1}{\hbar}\left(\frac{Q^2}{\hbar^2}\right)^{\frac{D-6}{2}}\frac{-\cosh(-\frac{q_1\cdot q_2}{m_1m_2})+i\pi}{\sqrt{\kl{q_1\cdot q_4}-m_1m_2}}
\end{equation}
The transverse box is obtained by $q_1\to q_2$, which corresponds to $q_1\cdot q_4\to -q_1\cdot q_4$ up to $\mathcal{O}(\h^2)$, the real part of the root analogous to Eq. \eqref{rut} then no longer falls, between $0$ and $1$, the resulting integral and is:\begin{equation}
\mathcal{I}_{\square,s}(k)^{(1 s)}=\frac{i \Gamma\left(\frac{D-4}{2}\right)^2 \Gamma\left(\frac{6-D}{2}\right)}{2 \Gamma(D-4)} \frac{1}{\hbar}\left(\frac{Q^2}{\hbar^2}\right)^{\frac{D-6}{2}}\frac{\cosh(-\frac{q_1\cdot q_4}{m_1m_2})}{\sqrt{\kl{q_1\cdot q_4}-m_1m_2}}
\end{equation}
Combining the leading box and cross box gives us:
\begin{equation}
\mathcal{I}^{(1 s)}_{\square, u}+\mathcal{I}_{\square, u}^{(1 s)}=\frac{\Gamma\left(\frac{D-4}{2}\right)^2 \Gamma\left(\frac{6-D}{2}\right)}{2\Gamma(D-4)} \frac{-\pi}{\sqrt{\left(q_1 \cdot q_4\right)^2-m_1^2 m_2^2}}\left(\frac{Q^2}{\hbar^2}\right)^{\frac{D-6}{2}}
\end{equation}
So the first order soft limit expansion of the box and cross contributions of the master integral decomposition, of the box and cross box feynman diagram contribution of the 2PM amplitude takes the following form:
\begin{equation}
    i\mathcal{M}^{(1s)}_B=-\frac{\pi}{2} \frac{4i \kappa^4 c_{\square}^{1s}(q^2)}{\sqrt{\left(k_1 k_2\right)^2-m_1^2 m_2^2}} \frac{\Gamma^2\left(\frac{D}{2}-2\right) \Gamma\left(3-\frac{D}{2}\right)}{\Gamma(D-4)}\left(\frac{Q^2}{\h^2}\right)^{\frac{D}{2}-3}\label{M1B}
\end{equation}
As shown in \cite{Bjerrum-Bohr:2019kec} and \cite{Cheung:2020gyp}, the second or next-order term of the soft expansion of Eq. \eqref{MIB2} and Eq. \eqref{MICB2} also contributes to the box and cross-box contributions of the master integral decomposition of the box and cross-box contributions of the 2PM scattering amplitude. Therefore, we will now perform the second order expansion of Eq. \eqref{MIB2} in the soft limit.
The general Taylor expansion for the soft region reads:
\begin{equation}
\mathcal{I}_{\square}^{(s)}=\sum_{n, m=1}^{\infty} \int \frac{d^D l}{(2 \pi)^D} \frac{\left(q \cdot l-l^2\right)^{m+n-2}}{l^2(q-l)^2\left[\left(2 Q-q_{\perp}\right) \cdot l\right]^m\left(-q_{\perp} \cdot l\right)^n}
\end{equation}
and the Feynman parametrist version reads, then:
\begin{equation}
\mathcal{I}_{\square}^{(s)}=\sum_{n, m=1}^{\infty} \frac{\Gamma(m+n)}{\Gamma(m) \Gamma(n)} \int_0^1 d x x^{m-1}(1-x)^{n-1} \int \frac{d^D k}{i\pi^{\frac{D}{2}}} \frac{\left(p \cdot k-k^2\right)^{m+n-2}}{k^2(p-k)^2\left[\left(2 x w-q_{\perp}\right) \cdot k\right]^{m+n}}
\end{equation}
The next two leading order term, $m=1,n=2$ and $m=2, n=1$ than reads:
\begin{equation}
\mathcal{I}_{\square,s}^{(2 s)}(k)=\int_0^1 d x \int \frac{d^D k}{i\pi^{\frac{D}{2}}} \frac{p \cdot k-k^2}{k^2(p-k)^2\left[\left(2 x w-q_{\perp}\right) \cdot k\right]^3} .
\end{equation}
we can split the integral and find that the second term in the numerator gives a scalar integral:
\begin{equation}
\int \frac{d^D k}{i\pi^{\frac{D}{2}}}\frac{-k^2}{k^2(Q-k)^2\left[\left(2w-q_{\perp}\right) \cdot k\right]^3}=\int\frac{d^D k}{i\pi^{\frac{D}{2}}} \frac{1}{k^2\left[\left(2 w-q_{\perp}\right) \cdot k\right]^3}=0
\end{equation}
where in the second step we performed a variable change from $k\to Q-k$ and realised that it is equal to 0.
\begin{equation}
\mathcal{I}_{\square}^{(2, s)}=\int_0^1 d x \int \frac{d^D k}{i\pi^{\frac{D}{2}}} \frac{Q\cdot k}{k^2(Q-k)^2\left[\left(2 x w-\mathrm{q}_{\perp}\right) \cdot l\right]^3}
\end{equation}
which can be evaluated with the help of the fact that, if $a\cdot b=0$, than: 
\begin{align}
I_{\perp}\left(a^2, b^2\right) & =\int \frac{d^D k}{i \pi^{\frac{D}{2}} \left(k^2+i\varepsilon\right)^{\lambda_1}\left((a-k)^2+i\varepsilon\right)^{\lambda_2}(2 b \cdot k+i\varepsilon)^{\lambda_3}} \\
& =\frac{\Gamma\left(\lambda_1+\lambda_2+\frac{\lambda_3-D}{2}\right) \Gamma\left(\frac{\lambda_3}{2}\right)}{2 \Gamma\left(\lambda_1\right) \Gamma\left(\lambda_2\right) \Gamma\left(\lambda_3\right) \Gamma\left(D-\lambda_1-\lambda_2-\lambda_3\right)} \frac{\Gamma\left(\frac{D-\lambda_3}{2}-\lambda_1\right) \Gamma\left(\frac{D-\lambda_3}{2}-\lambda_2\right)}{\left(a^2\right)^{\lambda_1+\lambda_2+\frac{\lambda_3-D}{2}}\left(-b^2\right)^{\frac{\lambda_3}{2}}}
\end{align}
which gives for our integral:
\begin{equation}
    \mathcal{I}_{\square,s}^{(2 s)}(k)=-\frac{ \Gamma\left(\frac{5-D}{2}\right) \Gamma\left(\frac{D-3}{2}\right)^2}{4\Gamma(D-4)}\left(\frac{Q^2}{\hbar^2}\right)^{\frac{D-5}{2}} \int_0^1 \frac{d x}{\left[-\left(x w-\frac{q_{\perp}}{2}\right)^2+i\varepsilon\right]^{\frac{3}{2}}}
\end{equation}
The integral can now be evaluated and gives
\begin{equation}
\int_0^1 \frac{d x}{\left[\left(x-x_{+}\right)\left(x-x_{-}\right)\right]^{\frac{3}{2}}}=-\frac{4}{\left(x_{+}-x_{-}\right)^2}\left[\frac{x_{+}+x_{-}}{2 \sqrt{x_{+} x_{-}}}+\frac{\left(1-x_{+}\right)+\left(1-x_{-}\right)}{2 \sqrt{\left(1-x_{+}\right)\left(1-x_{-}\right)}}\right]
\end{equation}
To simplify the expression, we use the following identities:
\begin{align}
\left(x_{+}-x_{-}\right)^2 & =\frac{4}{s}\left[\left(q_1 \cdot q_2\right)^2-m_1^2 m_2^2\right]+\left(-\hbar^2 Q^2\right)\\
x_{+}+x_{-}&=1+\frac{m_2^2-m_1^2}{s} \\
x_{+} x_{-}&=\frac{m_2^2}{s}+\frac{-\hbar^2 Q^2}{4 s}\\
\left(1-x_{+}\right)\left(1-x_{-}\right)&=\frac{m_1^2}{s}+\frac{-\hbar^2 Q^2}{4 s}\\  
\end{align}
next two leading order them, we find to leading order in $\h$:
\begin{equation}
    \mathcal{I}_{\square,s}^{(2 s)}(k)=-\frac{ \Gamma\left(\frac{5-D}{2}\right) \Gamma\left(\frac{D-3}{2}\right)^2}{4\Gamma(D-4)}\left(\frac{Q^2}{\hbar^2}\right)^{\frac{D-5}{2}} \frac{\hbar^3\left[s\left(\frac{1}{m_1}+\frac{1}{m_2}\right)+\left(m_1^2-m_2^2\right)\left(\frac{1}{m_1}-\frac{1}{m_2}\right)\right]}{\left(q_1 \cdot q_4\right)^2-m_1^2 m_2^2}
\end{equation}
In order to obtain the full next two leading-order expressions, we must also include the cross-box or $u$-channel that is obtained by replacing $q_1$ by $q_2$ or $q_1\to q_2\to -q_1\cdot q_2$, it takes the form:
\begin{equation}
    \mathcal{I}_{\square,u}^{(2 s)}(k)=-\frac{ \Gamma\left(\frac{5-D}{2}\right) \Gamma\left(\frac{D-3}{2}\right)^2}{4\Gamma(D-4)}\left(\frac{Q^2}{\hbar^2}\right)^{\frac{D-5}{2}}\frac{\hbar^3\left[u\left(\frac{1}{m_1}+\frac{1}{m_2}\right)+\left(m_2^2-m_1^2\right)\left(\frac{1}{m_2}-\frac{1}{m_1}\right)\right]}{\left(q_3 \cdot q_1\right)^2-m_1^2 m_2^2}
\end{equation}
using the fact that $2q_1\cdot q_4-2q_3\cdot q_4=-Q^2$, $q_1^2=m_1=q_2^2$, and $q_4^2=m_2^2=q^2_4$, we are able to combine both expressions $\mathcal{I}_{\square,s}^{(2 s)}(k)$ and $\mathcal{I}_{\square,u}^{(2 s)}(k)$ to: 
\begin{equation}
\mathcal{I}_{\square,s}^{(2 s)}(k)+\mathcal{I}_{\square,u}^{(2 s)}(k)=\frac{\Gamma\left(\frac{5-D}{2}\right) \Gamma\left(\frac{D-3}{2}\right)^2}{2 \Gamma(D-4)}\left(\frac{Q^2}{\hbar^2}\right)^{\frac{D-5}{2}} \frac{m_1+m_2}{\left(q_1 \cdot q_2\right)^2-m_1^2 m_2^2}    
\end{equation}
So the second-order soft limit expansion of the box and cross contributions of the master integral decomposition, of the box and cross-box feynman diagram contribution of the 2PM amplitude takes the following form:
\begin{equation}
    i \mathcal{M}^{(2s)}_B=\frac{ 2i\kappa^4 c_{\square}^{(2s)}(k^2)\kl{m_1+m_2}}{\left(k_1 k_2\right)^2-m_1^2 m_2^2} \frac{\Gamma\left(\frac{5-D}{2}\right) \Gamma^2\left(\frac{D-3}{2}\right)}{\Gamma(D-4)}\left(\frac{Q^2}{\h^2}\right)^{\frac{D-5}{2}}\label{M2B}
\end{equation}
Combining Eq. \eqref{M1B} and Eq. \eqref{M2B} gifs:
\begin{align}
i\mathcal{M}_B=&
-\frac{\pi}{2} \frac{4i \kappa^4 c_{\square}^{1s}(k^2)}{\sqrt{\left(k_1 k_2\right)^2-m_1^2 m_2^2}} \frac{\Gamma^2\left(\frac{D}{2}-2\right) \Gamma\left(3-\frac{D}{2}\right)}{\Gamma(D-4)}\left(\frac{Q^2}{\h^2}\right)^{\frac{D}{2}-3}\\
&+\frac{ 2i\kappa^4 c_{\square}^{1s}(k^2)\kl{m_1+m_2}}{\left(k_1 k_2\right)^2-m_1^2 m_2^2} \frac{\Gamma\left(\frac{5-D}{2}\right) \Gamma^2\left(\frac{D-3}{2}\right)}{\Gamma(D-4)}\left(\frac{Q^2}{\h^2}\right)^{\frac{D-5}{2}}\nonumber 
\end{align}
This agrees with the results for the box, and Cross-box contribution to the master integral basis of the 2PM contributions, as presented in 
\cite{Bjerrum-Bohr:2019kec} \cite{Cheung:2020gyp} also we could do not a full discussion on classical and super classical terms, this has already been done elsewhere, and we have achieved, what set out to achieve. 
\section{Triangle Master Integrates}
%The triangle integral takes the following form:

%together with $$$, which we can obtain as usual by $$.
\begin{equation}
\mathcal{I}_{\triangledown,m_1}\left(k\right)=\int\frac{d^D k}{ i\pi^{\frac{D}{2}}} \frac{1}{k^2} \frac{1}{(Q+k)^2} \frac{1}{\left(k+q_1\right)^2+m_1^2}\label{MITD2}
\end{equation}
%again, we introduce the %momentu m_transfer $Q=q_1+q_2$, and the addition of variable $q_{\perp}=q_1-q_2$, that obey together the property,$q\cdot q_{\perp}$.The classical limit consists in letting $\h\to0$ such that the momentu m_transfer $q$ vanishes, while the transferred wave-vector $\frac{1}{\h}q$ and the average momentu m_$1/2q_{\perp}$ of the massive particles stay fixed. We identify this situation with the mathematical fact that: \begin{equation}     q \si m_\mathcal{O}(\hbar),  q_{\perp} \si m_\mathcal{O}(1),  q \ll q_{\perp} \end{equation} fro m_the relation \begin{equation} -q_{\perp}^2=4 m_2^2+q^2 \end{equation}It follows that in the  limit, the mass $m_2\neq0$.  In general, one employs an expansion into two different regions to obtain an asymptotic approximation of the integral Eq. \eqref{NV3Ange} in the classical limit $\h\to0$. This method consists in splitting the integration over the momenta $l$ into a soft region characterised by the scaling $l\sim\mathcal{O}(\h)$ and hence $l\si m_q\ll q_{\perp}$, and a hard region in which $l\sim\mathcal{O}(1)$ and hence $l\si m_q_{\perp}\ll q$. this article, we are however only confirmed with the soft region:
If we combine as before equations \eqref{newvar},\eqref{softrel}  and Eq.\eqref{hlimC}, we get:
\begin{equation}
\mathcal{I}_{\triangledown,m_1}\left(k\right)=\int\frac{d^D k}{i\pi\hbar^{\frac{D}{2}}} \frac{\hbar^5}{\left(k^2+i\varepsilon\right)\left((Q-k)^2+i\varepsilon\right)\left(k^2-\left(q_{\perp}+Q\right) \cdot k+i\varepsilon\right)}
\end{equation}
The first contributions is for the soft region has the following form:
\begin{align}\mathcal{I}_{\triangledown,m_1}^{(1 s)}\left(k\right)&=\int\frac{d^D k}{(2 \pi \hbar)^D} \frac{\hbar^5}{\left(k^2+i\varepsilon\right)\left((Q-k)^2+i\varepsilon\right)\left(-q_{\perp} \cdot k+i\varepsilon\right)}\label{3Ange1Contr}\\\mathcal{I}_{\triangledown,m_1}\left(k\right)^{(2 s)}&=\int\frac{d^D k}{(2 \pi \hbar)^D} \frac{\hbar^5\left(-k^2+q \cdot k\right)}{\left(k^2+i\varepsilon\right)\left((Q-k)^2+i\varepsilon\right)\left(-q_{\perp} \cdot k+i\varepsilon\right)^2}\label{3Ange2Contr},
\end{align}
%in principle, can, then the integral be extended to the whole regen. The dimensional space in both regions in view of the fact that the error $$thus introduced always, always take the for m_of a scaler's integral and is therefore identically vanishing in dimensional regularisation: to leading order, for instance:$$
%by means of the above expression. We have reduced the proble m_to the evaluation of simpler Feynman integrals, which can be directly calculated introducing Feynman permit tries Asian and exploring the orthogonality between $$and $$.
The leading contribution of Eq. \eqref{3Ange1Contr} to the soft region can be obtained if we recognise that $Q$ and $q_{\perp}$ are orthogonal, then we can use the following integral identity: if vectors $a$ and $b$ are orthogonal, then $a\cdot b=0$ and:
\begin{align}
I_{\perp}\left(a^2, b^2\right) & =\int \frac{d^D k}{i \pi^{\frac{D}{2}}\left(k^2+i\varepsilon\right)^{\lambda_1}\left((a-l)^2+i\varepsilon\right)^{\lambda_2}(2 b \cdot k+i\varepsilon)^{\lambda_3}} \\
& = \frac{\Gamma\left(\lambda_1+\lambda_2+\frac{\lambda_3-D}{2}\right) \Gamma\left(\frac{\lambda_3}{2}\right)}{2 \Gamma\left(\lambda_1\right) \Gamma\left(\lambda_2\right) \Gamma\left(\lambda_3\right) \Gamma\left(D-\lambda_1-\lambda_2-\lambda_3\right)} \frac{\Gamma\left(\frac{D-\lambda_3}{2}-\lambda_1\right) \Gamma\left(\frac{D-\lambda_3}{2}-\lambda_2\right)}{\left(a^2\right)^{\lambda_1+\lambda_2+\frac{\lambda_3-D}{2}}\left(-b^2\right)^{\frac{\lambda_3}{2}}}
\end{align}
if the vectors $a$ and $b$ are orthogonal $a\cdot b=0$. 
This gives for the integral in Eq. \eqref{3Ange1Contr}:
\begin{equation}
\mathcal{I}_{\triangledown,m_1}^{(1 s)}\left(k\right)=\frac{\Gamma\left(\frac{D-3}{2}\right)^2 \Gamma\left(\frac{5-D}{2}\right)}{2 m_1 \Gamma(D-3)}\left(\frac{Q^2}{\hbar^2}\right)^{\frac{D-5}{2}}\label{SMITD}
\end{equation}
Hence, the first-order expansion in the soft limit of the triangle down contribution to the master Integral basis of the box and cross-box diagram contributions of the 2 PM scattering amplitude, takes the following form 
\begin{equation}
    i\mathcal{M}^{1s}_{\triangledown,m1}=
    \frac{2i\kappa^4c_{\triangledown}(m_1)\Gamma\left(\frac{D-3}{2}\right)^2 \Gamma\left(\frac{5-D}{2}\right)}{m_1 \Gamma(D-3)}\left(\frac{Q^2}{\hbar^2}\right)^{\frac{D-5}{2}}
\end{equation}
The $\mathcal{I}_{\triangle,m_2}^{(1 s)}\left(k\right)$ term, can now be obtained simply by switching $m_1\to m_2$, therefore, it has form
\begin{equation}
\mathcal{I}_{\triangle,m_2}^{(1 s)}\left(k\right)= \frac{\Gamma\left(\frac{D-3}{2}\right)^2 \Gamma\left(\frac{5-D}{2}\right)}{2 m_2\Gamma(D-3)}\left(\frac{Q^2}{\hbar^2}\right)^{\frac{D-5}{2}}\label{SMITU}
\end{equation}
For the first order term in the soft limit expansion of the triangle up contribution to the master integral basis of the box and cross box diagram contribution of the 2 PM scattering amplitude has the form: 
\begin{equation}
    \mathcal{M}^{1s}_{\triangle,m1}=
    \frac{2i\kappa^4 c_{\triangle}(m_1)\Gamma\left(\frac{D-3}{2}\right)^2 \Gamma\left(\frac{5-D}{2}\right)}{m_2 \Gamma(D-3)}\left(\frac{Q^2}{\hbar^2}\right)^{\frac{D-5}{2}}
\end{equation}
\section{Master Integral Coefficients}
Now that we have obtained the master integrals in the classical limit, all that remains is to determine the master integral coefficients in the classical limit for the complete master integral representation of the box and crossbox contribution of the 2PM amplitude in the classical limit. This is done quite quickly by neglecting the transmitted momentum $Q=\kl{q_1+q_2}=\kl{q_3+q_4}=-\sqrt{t}$ or $t=-\kl{q_1+q_2}^2=-\kl{q_3+q_4}^2$.
The triangle down MI coefficient that has the form:
\begin{equation}
\begin{aligned}
c_{\triangledown,m_1,B}^{(1 s)}=&-\frac{1}{(-2+D)^2} 2\left((-2+D) m_1^2+(-4+D) m_2^2-(-2+D) s\right) \\
& \left((-2+D) m_1^4+(-2+D)\left (m_2^2-s\right)^2+2 m_1^2 \right. \\
& \left.\left((-4+D) m_2^2-(-2+D) s\right)\right)\label{ctd1sB}
\end{aligned}
\end{equation}
The triangle up MI coefficient that has the form:
\begin{equation}
\begin{aligned}
c_{\triangle,m_2,B}^{(1 s)}=&-\frac{1}{(-2+D)^2} 2\left((-2+D) m_2^2+(-4+D) m_1^2-(-2+D) s\right) \\
& \left((-2+D) m_2^4+(-2+D)\left(m_1^2-s\right)^2+2 m_2^2 \right. \\
& \left.\left((-4+D) m_1^2-(-2+D) s\right)\right)\label{ctu1sB}
\end{aligned}
\end{equation}
The scalar Box MI coefficient is
\begin{equation}
    c_{\square,s,B}^{(1 s)}=\left(\frac{1}{2} \left(-m_1^2-m_2^2+s\right)^2-\frac{2 m_1^2 m_2^2}{D-2}\right)^2
\end{equation}
The coefficients for the master integral basis of the gravity cross box are as follows. 
The triangle down MI coefficient that has the form:
\begin{equation}
\begin{aligned}
c_{\triangledown,m_1,CB}^{(1 s)}=& \frac{1}{(-2+d)^2}\left(2\left(-2 m_2^2+d\left(m_1^2+m_2^2-s\right)+2 s\right)\right. \\
& \left((-2+d) m_1^4+(-2+d)\left(m_2^2-s\right)^2\right. \\
& \left.\left.+2 m_1^2\left((-4+d) m_2^2-(-2+d) s\right)\right)\right)\label{ctd1sCB}\end{aligned}
\end{equation}
The triangle up MI coefficient that has the form:
\begin{equation}
\begin{aligned}
c_{\triangle,m_2,CB}^{(1 s)}=& \frac{1}{(-2+d)^2}\left(2\left(-2 m_1^2+d\left(m_2^2+m_1^2-s\right)+2 s\right)\right. \\
& \left((-2+d) m_2^4+(-2+d)\left(m_1^2-s\right)^2\right. \\
& \left.\left.+2 m_2^2\left((-4+d) m_1^2-(-2+d) s\right)\right)\right)\label{ctu1sCB}
\end{aligned}
\end{equation}
The scalar cross-box MI coefficient is:
\begin{equation}
    c_{\square,u,CB}^{(1 s)}=\left(\frac{1}{2} \left(-m_1^2-m_2^2+u\right)^2-\frac{2 m_1^2 m_2^2}{D-2}\right)^2
\end{equation}
To obtain the numerator of the downward triangle master integral we follow the Intuition section \ref{MIDGrav}, and we combine $c_{\triangledown,m1,B}$, Eq \eqref{ctd1sB}, with $c_{\triangledown,m1,CB}$, Eq \eqref{ctd1sCB}. The numerator of the downward triangle master integral has than following form:
\begin{equation}
\begin{aligned}
N_{\triangledown}^{1s}=
& \frac{4\kappa^2}{(-2+d)^2}\left(4\left(m 1^2+m_2^2\right) \right. \\
& \left((-2+d) m_1^4+(-2+d)\left(m 2^2-s\right)^2\right. \\
& \left.\left.+2 m_1^2\left((-4+d) m_2^2-(-2+d) s\right)\right)\right)
\end{aligned}
\end{equation}
To obtain the numerator of the triangle up master integral, we combine $c_{\triangle,m1,B}$, Eq \eqref{ctu1sB}, with $c_{\triangle,m1,CB}$, Eq \eqref{ctu1sCB} and obtain the numerator of the triangle master integral, it has the following form for:
\begin{equation}
\begin{aligned}
N_{\triangle}^{1s}=&\frac{4\kappa^2}{(-2+d)^2}\left(4\left(m_2^2+m_1^2\right) \right. \\
& \left((-2+d) m_2^4+(-2+d)\left(m_1^2-s\right)^2\right. \\
& \left.\left.+2 m_2^2\left((-4+d) m_1^2-(-2+d) s\right)\right)\right)
\end{aligned}
\end{equation}
The numerator of the box master integral is
\begin{equation}
    N_{\square,s,B}^{(1 s)}=4\kappa^2\left(\frac{1}{2} \left(-m_1^2-m_2^2+s\right)^2-\frac{2 m_1^2 m_2^2}{D-2}\right)^2
\end{equation}
The numerator of the cross-box master integral is 
\begin{equation}
    N_{\square,u,CB}^{(1 s)}=4\kappa^        2\left(\frac{1}{2} \left(-m_1^2-m_2^2+u\right)^2-\frac{2 m_1^2 m_2^2}{D-2}\right)^2
\end{equation}
\newpage

\chapter{Summary and Outlook}
We began our journey in chapter \ref{GRnQFT}  with the analysis of the gauge theory of spin-2 gravitons. We established that the gauge-theoretical description of gravitons is a quantum field-theoretical manifestation of the principle of general covariance known from classical general relativity. Not surprisingly, gravitons in the lower energy limit can be described like other quantum fields, this comes from the fact that the space-time curvature is small and special relativity applies. This allowed us to formulate of the Feynman rules for gravity and therefore the formulation of the $2\to2$  2 PM scattering amplitude that consists of box, crossbox and triangle contributions. In this publication, we ignored the triangle contributions. We will be treated with the methods of intersection theory in a forthcoming paper where we will calculate the full $2 \rightarrow 2$ 2 PM scattering amplitude.\\\\
It is common knowledge that despite the great success of Feynman calculus, the deep mathematical structure underlying it is poorly understood. Lately, the community has become aware of the mathematical framework of twisted co-homology to evaluate the Feynman integral. It was originally developed to study hypergeometric functions. In this work we gave an introduction to the fundamental aspects of the theory as applied to the box cross-box contributions of the 2PM scattering process.\\\\
First, we introduced the essential ideas of co-homology, in the chapter on the univariate intersection number,chapter \ref{ISTU}. There we saw for the first time that, Feynman integrals in Baikov representation can naturally be interpreted as the pairing between a twisted co-cycle and a twisted cycle. Furthermore, we introduced the notion of a twisted (co)-homology group, including the corresponding group elements, twisted cycles and twisted co-cycles. From here, we continued our discussion and introduced the intersection number as the pairing between elements of the twisted co-homology group and their dual. We obtained the Feynman integrals decomposition formula formula Eq. \eqref{eqMIINTER} and the master decomposition formula \eqref{MDF}. We saw that the intersection number acts as a glorified scalar product in the space of Feynman integrals, which allows us to obtain the coefficients of the master integral by projecting the Feynman integral onto the master integral\\\
In the chapter \ref{Intersection_Theory_II}, on multivariate intersection theory, we have generalised the concept developed in chapter \ref{ISTU} to multiple integration variables and presented a recursive algorithm for evaluating multivariate co-homology intersection numbers. We have also presented an extensive example which we hope has given the reader a practical understanding of the mechanisms underlying the computation of the multivariate intersection number.\\\\
In chapter \ref{GravBox} we presented the underlying framework that allows us to obtain the intersection number of any given Feynman integral in Baikov representation. We presented codes that allow us to calculate the dimension $\nu$ of each fibre and find its basis elements and their size. We used this code for the case of the box and cross-box as well as their cut version. we obtained their minimum base elements and their size. Further, we obtained the coefficients of the master integrals bases of the cut gravity box and cross-box. This was done via the code presented in \ref{INC}, which implements the algorithm presented in chapter \ref{Intersection_Theory_II}, in mathematica.\\\\
Having successfully obtained the complete Master integral basis for both the box and cross-box contributions to the 2PM scattering amplitude, we proceeded to evaluate the Master integrals and their coefficients within the classical limit. Notably, the numerator of the box and cross-box contributions to the Master integral basis acquired through intersection theory, aligns with previously established results, \cite{PhysRevD.100}. This compellingly demonstrates the applicability of intersection theory to the post-Minkowskian expansion and scattering amplitudes involving gravity in a general sense, contingent upon the ability to represent individual Feynman diagrams in the Baikov representation.\\\

In our forthcoming publication, we intend to compute the full 2PM scattering amplitude utilizing intersection theory, aiming for consistency with prior findings in \cite{Cristofoli_2020}\cite{PhysRevD.100}. An intriguing avenue for future exploration involves delving into the 3PM, which encompasses two-loop Feynman diagrams, to determine whether intersection theory continues to be a reliable tool. We anticipate that our work will contribute to the field, encouraging more comprehensive scrutiny of this subject. It is a privilege to be useful.

\chapter*{Appendices}
\addcontentsline{toc}{chapter}{Appendices} 
\setcounter{section}{0}
\renewcommand\thesection{\Alph{section}}
This section gives an overview about the different Mathematica codes that were used in this thesis. 
\section{dimension code}\label{dimensioncode}
This Mathematica code is used to calculate the dimension $\nu$  of each fibration: 
\begin{verbatim}
Do[
(*loop over each value of'k' within the range of {2,Length[subsets3}*)

subsets3 = Subsets[ver3];
 
 (* All subsets for the Baikov variables var
 are defint via. They correspond to the different possible bases
 for each layer of the fibration of the manifold.*)
currentvars3 = subsets3[[k]];
(*take the kth element from the list "subsets3" and store it 
in "currentvars3"*)
remvars3 = Join[Complement[ver3, currentvars3],
{s, t, d, m1, m2, rho3[1], rho3[2], rho3[3], rho3[4]}];
(*Create a list "remvars3" by joining the elements of'ver3' not 
contained in "currentvars3", the rest of the Baikov variables, with additionally
%the Mandelstam variables, as well as the masses and the regulates
"{s,t,d,m1,m2,rho3[1],rho3[2],rho3[3],rho3[4]}"*)
numrules3 = Table[remvars3[[i]] -> 1/Prime[23 + 45*i], {i, Length[remvars3]}];
(*Create replacement rules where each variable in "remvars3" is 
replaced by a corresponding fraction. "1/Prime[23+45*i]" ,where "i" is the 
index of the variable in "remvars3", this done to speed up the 
computation, as Mathematica treats letters, differently than numbers*)

dlistdo3 = Table[Simplify[D[Log[ureg3c /. numrules3],
currentvars3[[j]]]], {j, 1, Length[currentvars3]}];

(*omega is calculated using a table. In order to do this, we apply the
replacement rule "numrules3" to "ureg3", then take the logarithm of the
resulting expression and calculate its derivative.*)

SOLdo3 = Solve[dlistdo3 == 0, currentvars3];

(*The system of equations of omega "dlistdo3 == 0" is solved for the variable
in "currentvars3", the result is then stored in "SOLdo3"*).

lenSOLdo3 = Length[SOLdo3]; 

The length of the solution list is "SOLdo3" obtained, it is in stored in "lenSOLdo3".

Print[k, "/", Length[subsets3], "   ", currentvars3, "  ", lenSOLdo3];

(*The current value of "k", the total length of "subsets3", the current subset of the variable "currentvars3" and the number of solutions found in "lenSOLdo3" are determined, and the loop ends.*) 

, {k, 2, Length[subsets3]}];

(*Iterate over values of "k" from 2 to the 
length of "subsets3"*)
\end{verbatim}

\section{Basis size code}\label{base_size_code}

This Mathematica code is used to compute the basis size, of each fibration layer: 
\begin{verbatim}
Do[

currentvars3 = Take[ver3, ii];  

(*The ii-th element is taken from the list ver4 and stored in currentvars3.
This currentvars3  represents the subset of variables that are being analyzed 
in the current iteration*)

Print["\nFor set ", currentvars3, " :"]; 

(*Print a message indicating the current set *)

  remvars3 = Join[Complement[ver3, currentvars3], 
  {s, t, d, m1, m2, rho4[1], rho4[2], rho4[3], rho4[4]}]; 

(*A list "remvars4" is created by combining two lists:
the elements in the list "ver4" that are not included in "currentvars3", 
i.e. the rest of the Baikov variables and the Mandelstam variables,
as well as the masses and the regulates
"{s, t, d, m1, m2, rho3[1], rho3[2], rho3[3], rho3[4]}".*)

umrules3 = Table[remvars3[[i]] -> 1/Prime[23 + 45*i], {i, Length[remvars3]}]; 

(*A replacement rule numrules3 is constructed using a table.
The rule corresponds to replacing each element in the list "remvars3" with 
"1/Prime[23 + 45*i]" where "i" is the index of the variable that is being replaced
in "remvars3"*)

Do[currentset3 = subsets3[[k]]; 
    
(*"currentset3" is defined as the list of all possible subsets
of the Baikov variable called "ver3". Each subset corresponds to a possible
basis choice of the lay of the fibration that corresponds to its length.*)

ureg3 = Product[currentset3[[i]]^rho3[i],

{i, 1, Length[currentset4]}]*B3^((d - 5)/2);

(*The regulated Baikov polynomial "ureg3" is calculated via the product
of the regulators "rho3[i]:i\in[1,Length[ver3]]" of the current subset
of Baikov variables "currentset3" and the Baikov polynomial "B3"*)

ureg32 = ureg3 /. numrules3;  

(*The replacement rules from "numrules3" are applied to "ureg3",
this is essentially done, as before, to speed up the calculation*)

dlistdo3 = Table[Simplify[D[Log[ureg32], currentvars3[[j]]]], {j, 1, Length[currentvars3]}];  

(*$\omega$ in the code "dlistdo3" is calculated using a table. 
In order to do this, we apply the replacement rule "numrules3" to "ureg33",
then take the logarithm of the resulting expression and calculate its derivative.*)

SOLdo3 = Solve[dlistdo3 == 0, currentvars3]; 

(*"SOLdo" is obtained by solving the system of equations of $\omega$ e.g
"dlistdo3 == 0" for the variables in
"currentvars3"*) 

lenSOLdo3 = Length[SOLdo3];  

(*The number of solutions found in "SOLdo3" in other words the length of "SOLdo3"
is stored in "lenSOLdo3"*)

Print[k, "/", Length[subsets3], "   ", currentset3, "  ", lenSOLdo3];

(*The progress and results of the current iteration are printed,
displaying the value of "k", the total number of subsets,
the subset of variables in "currentset3", and the number of solutions found 
for those variables*)

, {k, 1, Length[subsets3]}];  (* Loop through "subsets4" *)

, {ii, 1, 4}];  (* Outer loop with 'ii' from 1 to 4 *)
\end{verbatim}
\section{Intersection Number Code}\label{INC}
This Mathematica code is used to compute the Intersection number:
\begin{verbatim}
IntersecRec = Function[{phiL, phiR, elist, hlist, omegalist, varlist,

Module[{print, depth, var, e, h, nu, omega, eremaining, remaining,
omegaremaining, varsremaining, result, Cmatrix, CInverse,phiLred, 
phiRred, omegaarg, Omegamatrix},

(*The IntersecRec function computes an intersection number
have based on the input parameters and involves recursive calculations.*)

    print = True;
    Print["GO"];

(*This enables debugging output. The "GO" message is printed,
to indicate that the function is being executed.*)

    depth = Length[varlist];
 
(*The variable depth is calculated as the length of "varlist",
which indicates the depth of the recursion*)

    If[Length[elist] != depth - 1,
     Print["2 WARNING: Depth = ", depth, " but Length[elist] = ", 
      Length[elist]]];

(*An if statement checks whether the length of "elist" is one less than depth,
and prints a warning if not*

    If[Length[hlist] != depth - 1,
     Print["3 WARNING: Depth = ", depth, " but Length[hlist] = ", 
      Length[hlist]]];

(*A warning is printed if the length of "hlist" does not match the depth minus one.*)

    If[Length[omegalist] != depth,
     Print["4 WARNING: Depth = ", depth, " but Length[omegalist] = ", 
      Length[omegalist]]];

(*A warning is printed if the length of "omegalist" does not match the depth.*)

var = varlist[[-1]];
omega = omegalist[[-1]]; 

(*The outermost variable "var" and the outermost "omega" are
extracted from "varlist" and omegalist.*) 

    If[depth == 1, result = DeqRes[var, {{omega}}, {{1}}, {phiL}, {phiR}];
    
(*If the depth is 1, the recursion has reached its innermost level,
the "DeqRes" function is called directly with specific inputs to compute the result.
 If the depth is greater than 1:*)

    e = elist[[-1]];
    h = hlist[[-1]];
    nu = Length[e];
    
(*The innermost basis "e" and the innermost dual basis "h" are extracted from "elist" and
"hlist".*)

(*The dimension "nu", is determined as the length of "e".
If[Length[h] != nu,
  Print["6 WARNING: Length[h] = ", Length[h], " but Length[e] = ", 
   nu]];
   
    eremaining = elist[[;; -2]];
    hremaining = hlist[[;; -2]];
    omegaremaining = omegalist[[;; -2]];
    varsremaining = varlist[[;; -2]];
    
(*Lists and variables are defined based on
the remaining elements of
elist, hlist, omegalist, and varlist.*)

(*In the following are the individual intersection numbers  via the recursive function
"IntersecRec". First is "Cmatrix" calculated. In the code, this is done as follows:*)

Cmatrix = Table[Table[IntersecRec[e[[i]], h[[j]], eremaining, remaining, 
        omegaremaining, varsremaining], {j, nu}], {i, nu}];
CInverse = Inverse[Cmatrix];

phiLred = Table[IntersecRec[phiL, h[[i]], eremaining, remaining, 
    omegaremaining, varsremaining], {i, nu}];
phiLred = phiLred . CInverse;

omegaarg = Table[D[e[[i]], var] + omega*e[[i]], {i, nu}];
 
Omegamatrix = Table[Table[IntersecRec[omegaarg[[i]], h[[j]], eremaining, hremaining, 
     omegaremaining, varsremaining], {j, nu}], {i, nu}];
     
Omegamatrix = Omegamatrix.CInverse;

result = DeqRes[var, Omegamatrix, Cmatrix, phiLred, phiRred];];
   result = Factor[result];
   
   (*If[print,Print["IntersecRec:   result = ",result]]*);
   result]];

   DeqRes = Function[{varin, Omega, CMatrix, phiN, phiVN},
   
   Module[{polesol, aaa1, eks2, eks3, poles, aa, nu, max, pole, 
     varrule, jac, phiNuse, phiNVuse, omegause, psiansatz, zeroeks, 
     zeroser, zerolist, solvevars, eqs, zerosol, psieks, Expiseks, 
     contribution, result, rulp, Fzerolist, DOM, psi, 
     CMatrixuse},
     
(*DeqRes is a function that takes five input parameters:
varin, Omega, CMatrix, phiN, and phiVN.*)

(*Pole extraction*)

   aaa1 = Map[Denominator[Factor[#]] &, Omega/aa];

(*The denominators of the expressions in “omega/aa” are factorised and extracted to “aaa1”.*)

eks2 = Table[Apply[List, aaa1[[i]]], {i, Length[aaa1]}];

(*The denominators are converted to lists*)

eks3 = Apply[Times, Union[Flatten[eks2]] /. {aa -> 1}];

(*Unique denominators are multiplied *)

polesol = Solve[eks3 == 0, varin] // Union;

(*The equation “eks3==0” is solved for “varin” to find the poles*)

   poles = Append[Table[polesol[[i, 1, 2]], {i, Length[polesol]}], Infinity];

 result = 0;
max = 4;
nu = Length[phiN];


   Do[pole = poles[[i]];
     varrule = 
      If[pole === Infinity, {varin -> 1/y, 
        jac -> -1/y^2}, {varin -> y + pole, jac -> 1}];
     phiNuse = phiN*jac /. varrule;
     
(*Define a coordinate transformation based on whether the pole
is at a finite location or at infinity.*) 

     phiNVuse = phiVN*jac /. varrule;
     omegause = Omega*jac /. varrule;
     CMatrixuse = CMatrix /. varrule;(*I added this*)
     DOM = omegause;

(*The transformations are applied to “phiN”, “phiVN”, “Omega”, and} “CMatrix”.*) 

     psiansatz = Table[Sum[psi[i, j]*y^i, {i, -2, max}], {j, 1, nu}];

(*Construct a polynomial series “psiansatz” for the function “psi”,
using powers of the local coordinates of $y$.*)

   zeroeks = D[psiansatz, y] + psiansatz.DOM - phiNuse;

(*onstructing the intersection number differential equations*)

     zeroser = Series[zeroeks, {y, 0, max - 1}];

(*Series expansion “zeroser” of the differential equation “zeroeks”
is computed around the local coordinates of “y”.
 
     zerolist = Table[Coefficient[zeroser, y, i], {i, -3, max - 1}];

(*Coefficients of the series expansion “zeroser” are extracted and stored in “zerolist”*)

     Fzerolist = Flatten[zerolist];
     solvevars = Union[Cases[psiansatz, psi[__], All]];

(*Variables to be solved “psi[i, j]}” are extracted from “psiansatz”,
and stored in “solvevars”*)

     eqs = Thread[Fzerolist == 0];

(* A system of equations “eqs” is created by equating coefficients to 0.*)

     zerosol = Solve[eqs, solvevars];

(*The system of equations is solved for the variables in  “solvevars”,
and the solutions are stored in “zerosol”.*)

     psieks = psiansatz /. zerosol[[1]];

(*The solutions “zerosol” are applied to the polynomial ansatz “psiansatz”, 
and the resulting expression is stored in “psieks”.*) 

     rulp = {psi[__] :> 0};

(*A rule “rulp” is defined to set “psi[i, j]” to 0*)

(*Computing the expression inside the residue formula*)

Expiseks = Normal[Series[psieks . CMatrixuse . phiNVuse, {y, 0, 2}]] /. rulp;

     contribution = Residue[Expiseks, {y, 0}];
     result += contribution;
     , {i, 1, Length[poles]}];

(*Residue of “Expiseks” is calculated at $y=0$ and stored in contribution*)
  
   result]];
\end{verbatim}

\chapter*{Acknowledgments}
First and foremost, I would like to express my gratitude to my advisors, Poul Henrik Damgaard and Hjalte Frellesvig, for their patience during the writing of my Master's thesis.
During this time, I was able to learn a lot from you, which was not only useful for my Master's thesis. But it will also be helpful in my future.\\\\
I think that I often pushed you both to the limits of a supervising team and your patience, for example when I repeatedly spent the nights learning and researching and in the morning could and knew nothing at all. \\\\ 
I also want to extend my heartfelt thanks to my parents for unwaveringly supporting my education journey. They've fought for me and consistently placed their trust in my abilities, enabling me to pursue my dreams. They have sacrificed so much to provide me with the opportunity to study physics.\\\\
Additionally, I'd like to acknowledge and appreciate every teacher and government official who once doubted my potential and told me that I would never achieve more than a middle school diploma, let alone study physics. Your scepticism served as a driving force for my determination and brought me to this point. You have, in a way, been the wind beneath my wings of ambition. This achievement is dedicated to every dyslexic child, especially those in Germany who may not have had the same level of support I did. I hope this work inspires them to persevere in their own battles.\\\\
Furthermore, I want to express my gratitude to Waltraud Voelter for providing the dictation software, Dragon NaturallySpeaking, which allowed me to create this work through voice recognition. I also want to thank Washington Taylor for providing me with an extension that enabled me to dictate complex mathematical equations effortlessly. I'd also like to express my gratitude to  Ralf Metzler, who has consistently offered me valuable guidance and bolstered my confidence with his unwavering belief in my abilities.

\newpage
%\backmatter
\bibliography{quellen}
\bibliographystyle{unsrtnat}
\newpage
\begin{center}
\section*{Statutory declaration
}    
\end{center}
I herewith declare that I have comprised and written this thesis myself without any other than the cited sources and aids. Sentences or passages taken quoted literally or in spirit from other works are marked as such; other references with regard to the statement and the scope are identified by full details of the publications concerned. A thesis in the same or similar form has not been submitted to any examination board and has not been published. This thesis has not been used, either as a whole or in excerpts, in any other examination or as coursework. 
\vfill
Copenhagen,
\today\hspace{4mm}\\\\\includegraphics[height=2\baselineskip]{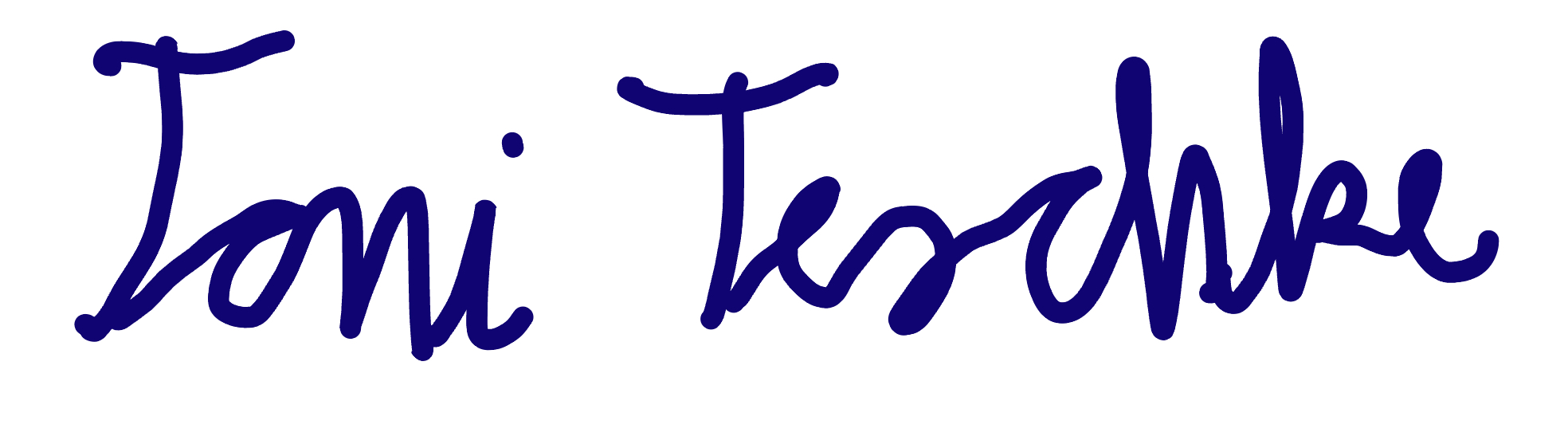}\\
\noindent Toni, Teschke
\end{document}